\documentclass[10pt,aps,prd,onecolumn,showpacs,superscriptaddress,longbibliography,nofootinbib]{revtex4-2}
\usepackage{mathrsfs,amsmath,amsthm,latexsym,amssymb,amsfonts,epsfig,cancel,enumerate,graphicx,txfonts,diagbox}
\usepackage[utf8]{inputenc}  
\usepackage[T1]{fontenc}     
\usepackage{lmodern}         
\usepackage[percent]{overpic}
\usepackage{multirow}
\usepackage[T1]{fontenc}
\usepackage[utf8]{inputenc}
\usepackage[colorlinks=true,linkcolor=blue,citecolor=blue,urlcolor=blue]{hyperref}
\usepackage{orcidlink}

\usepackage{xcolor}

\usepackage{booktabs}
\usepackage{graphicx}
\definecolor{navy}{RGB}{0,0,150}
\usepackage{appendix}
\allowdisplaybreaks

\newcommand{\RGUI}{Department of Physics, The Assam Royal Global University, Guwahati-781035, Assam, India}
\newcommand{\UCCC}{Centro de Investigaci\'{o}n en Ciencias del Espacio y F\'{i}sica Te\'{o}rica (CICEF), Universidad Central de Chile, La Serena 1710164, Chile}
\newcommand{\UCCB}{Programa de P\'os-Gradua\c c\~ao em F\'{\i}sica \& Coordena\c c\~ao do Curso de F\'{\i}sica -- Bacharelado, Universidade Federal do Maranh\~{a}o, 65085-580 S\~{a}o Lu\'{\i}s, Maranh\~{a}o, Brazil}

\begin{document}
	\baselineskip=16pt
	
	\title{Charged Black Holes in KR-gravity Surrounded by Perfect Fluid Dark Matter}
	
	\author{Faizuddin Ahmed\orcidlink{0000-0003-2196-9622}}
	\email{faizuddinahmed15@gmail.com}
	\affiliation{\RGUI}
	
	\author{Mohsen Fathi\orcidlink{0000-0002-1602-0722}}
	\email{mohsen.fathi@ucentral.cl}
	\affiliation{\UCCC} 
	
	\author{Edilberto O. Silva\orcidlink{0000-0002-0297-5747}}
	\email{edilberto.silva@ufma.br}
	\affiliation{\UCCB}
	
		\begin{abstract}
		In this work, we systematically investigate the null geodesics of electrically charged black holes in a gravitational framework that incorporates Lorentz violation induced by a background Kalb–Ramond (KR) field, in the presence of perfect-fluid dark matter. The properties of the photon sphere, black hole shadow, and photon trajectories are analyzed in detail. Furthermore, to explore the combined effects of Lorentz violation and dark matter on the motion of neutral test particles, we examine the innermost stable circular orbit (ISCO) in this spacetime. In addition, the epicyclic frequencies of test particles are studied to gain further insight into the dynamical behavior of particle motion around these black holes. The main analytical results are complemented by a phenomenological QPO analysis, a thermodynamic investigation, and a discussion of the sparsity of Hawking radiation, allowing us to connect optical, dynamical, and thermodynamic signatures within a single framework.\\
		
		{\bf Keywords:} charged Kalb-Ramond black holes; null geodesics; photon sphere; black hole shadow; QPOs; fundamental frequencies

	\end{abstract}

	\maketitle

	%\tableofcontents

	\section{Introduction}\label{sec:1}
	
	The shadow of a black hole constitutes one of the most striking observational signatures of the strong-field regime of gravity, arising from the extreme bending, capture, and scattering of null geodesics in the vicinity of the event horizon \cite{Synge1966,Luminet1979}. For a spherically symmetric black hole, the shadow boundary is uniquely determined by the photon sphere. Due to the high degree of spacetime isometry, every light ray constituting this boundary follows a circular orbit at a fixed radius, which can be effectively mapped onto an equatorial trajectory.  From an observational perspective, the morphology of the black hole shadow-particularly its angular size, shape, and possible distortions-encodes detailed information about the underlying spacetime geometry. In the case of rotating black holes, for instance, frame-dragging effects introduce asymmetries that lead to characteristic deformations of the shadow silhouette. Consequently, precise measurements of shadow properties provide a powerful diagnostic tool for determining key black hole parameters such as mass, spin, and the presence of additional charges or deviations from the Kerr solution \cite{Bardeen1973a,Chandrasekhar1984}. For a comprehensive overview of the theoretical framework and recent advances in black hole shadow physics, including analytical methods, numerical techniques, and observational prospects, we refer the reader to Refs. \cite{Cunha2018,Perlick2022,Chen2023}.

	The theoretical frameworks developed so far have acquired remarkable significance in light of recent observational breakthroughs achieved by the Event Horizon Telescope (EHT). The first horizon-scale images of the supermassive black holes M87$^{*}$ \cite{EHTL1,EHTL4,EHTL6} and Sgr A$^{*}$ \cite{EHTL12,EHTL14,EHTL16,EHTL17} have provided unprecedented empirical support for general relativity in the strong-field regime, while simultaneously enabling precise constraints on key physical parameters such as black hole mass, spin, and potential deviations from standard Kerr geometry. The visual depiction of these black holes not only confirmed the predictions of GR but also provided new opportunities for investigating the distribution of matter and electromagnetic phenomena in the vicinity of black holes. These landmark observations have substantially advanced theoretical modeling of black hole shadows. In particular, recent studies have progressed beyond idealized vacuum spacetimes to incorporate increasingly realistic astrophysical environments, including the effects of accretion disks, magnetized plasma flows, and surrounding dark matter distributions \cite{Gralla2019,Zeng2025}. Such environmental factors can significantly alter photon trajectories and, consequently, modify the observed shadow morphology. These developments provide a crucial framework for investigating more intricate and potentially exotic shadow features, which may serve as sensitive probes of new physics and possible deviations from classical general relativity.
	
	Lorentz-violating extensions of gravity sourced by antisymmetric tensor backgrounds have attracted growing attention as a theoretically motivated arena in which departures from local Lorentz invariance can be studied in strong gravitational fields. In this context, the Kalb-Ramond (KR) field \cite{KalbRamond1974} provides a natural geometric source of anisotropy, and the corresponding black-hole solutions offer a useful laboratory for investigating how Lorentz violation modifies both particle dynamics and observable optical signatures. It can be described as a self-interacting second-rank antisymmetric tensor field, and the KR modification is closely related to heterotic string theory \cite{Gross1985}. Due to the non-minimal coupling of the tensor field with the Ricci scalar, Lorentz symmetry violation may arise in this framework \cite{Altschul2010}. 
	
	Black hole solutions in KR gravity, in both uncharged and charged cases, have been studied in the literature. For instance, investigations include particle dynamics and gravitational weak lensing \cite{Atamurotov2022,AhmedSakalliAlBadawi2026IJGMMP}, thermodynamic properties and shadow analysis \cite{Yang2023, Liu2024, Duan2024, AhmedAlBadawiSakalli2026MPLA,AlBadawiAhmedSakalli2025}, and extensions involving a cloud of strings \cite{AhmedSilva2025, AhmedSilva2025b}. Other studies include classical tests such as the perihelion precession of Mercury \cite{Lessa2020}, orbital dynamics and phase transition \cite{SucuSakalli2025}, quasinormal modes and greybody factors in the presence of a global monopole \cite{Baruah2025}, geodesic structure in the presence of global monopole \cite{Belchior2025,FathiOvgun2025}, particle dynamics and shadow in string cloud backgrounds \cite{AhmedarXiv2026}, thermodynamic properties in nonlinear electrodynamics (ModMax) with a global monopole \cite{Ahmed2026} and in the presence of perfect fluid dark matter \cite{Sohan2025,Shokhzod2025}. The charged KR solution considered in this work inherits this structure through the parameter $\ell$, which measures the effective strength of the Lorentz-violating background, while the PFDM sector introduces an additional environmental contribution controlled by $\lambda$. The coexistence of these ingredients makes the spacetime particularly suitable for testing how fundamental modifications of the gravitational sector and astrophysical matter distributions jointly affect the near-horizon phenomenology of compact objects.
	
	The presence of dark matter has become a cornerstone of modern astrophysics, primarily motivated by a wide range of observational anomalies that cannot be explained by luminous matter alone. Early evidence arose from studies of the rotation curves of spiral and elliptical galaxies, which revealed mass distributions extending well beyond their visible components \cite{Rubin1980,Peebles1982}. Subsequent analyses have shown that the dominant contribution to galactic mass budgets originates from a non-luminous component, with estimates suggesting that nearly ninety percent of a typical galaxy’s total mass may be attributed to dark matter \cite{Persic1996}. Within this broader astrophysical framework, compact objects are not expected to exist in isolation. In particular, black holes located at galactic centers are likely embedded within extended dark matter halos \cite{EHTL1,EHTL6}. This realization has stimulated increasing efforts to incorporate dark matter effects into models describing physical processes in the vicinity of galactic cores \cite{Sofue2013, Boshkayev2019, Konoplya2022}. Among the various approaches, modeling dark matter as a perfect fluid has attracted considerable attention in the literature (see, e.g., \cite{Kiselev2003, Hendi2020, Liang2023}).
	
	Quasi-periodic oscillations (QPOs) have been widely detected in the X-ray emission of neutron star and black hole binary systems in astrophysics, providing a powerful probe of strong-field gravity and accretion physics in extreme environments. The presence of quasi-periodic variability in the observed X-ray flux suggests coherent or semi-coherent physical processes in the inner accretion flow, such as disk oscillations, relativistic orbital motion, or nonlinear resonances in the accretion disk. These mechanisms offer valuable tools for investigating the fundamental properties of spacetime in the strong-gravity regime. Twin-peak (3:2) QPOs have been observed in several black hole systems, motivating theoretical models that link these frequencies to oscillatory modes of accretion disks and relativistic epicyclic motion. The first systematic identification of timing variability and early QPO-like features in X-ray binary systems was reported in observational studies, including \cite{Angelini1989}. Subsequent progress in timing astronomy was significantly advanced by long-term monitoring missions, particularly the Rossi X-ray Timing Explorer (RXTE), which provided high-time-resolution observations of numerous black hole transients and neutron star binaries, thereby making it a cornerstone of QPO research \cite{Verbunt1993, Belloni2012}.
	
	QPOs in X-ray binaries typically appear in two main categories: low-frequency (LF) QPOs (approximately 0.1--30 Hz) and high-frequency (HF) QPOs (approximately 40--450 Hz). These characteristic frequencies are generally interpreted as being inversely proportional to the mass of the compact object, consistent with relativistic timescales near the innermost stable circular orbit (ISCO). HF QPOs, in particular, are of great importance because they are potentially associated with orbital and epicyclic frequencies of matter in the inner accretion disk and may provide constraints on the mass and spin of black holes and neutron stars. The current challenge in QPO research is to identify the correct physical mechanism responsible for their origin. Competing models include geodesic resonance models, relativistic precession models, diskoseismology, and magnetohydrodynamic oscillation scenarios. These models are actively used to test general relativity and its extensions in the strong-field regime, as well as to probe the structure of the innermost accretion flow and the ISCO radius.
	
	Recent theoretical studies have extended QPO modeling into modified gravity and non-vacuum black hole spacetimes, including the effects of dark matter halos, external electromagnetic fields, and alternative gravity theories \cite{Rayimbaev2022EPJC, Rayimbaev2022CQG, Qi2023EPJC, Rayimbaev2023EPJC, Rayimbaev2022Dark, Rayimbaev2023Galaxies, Rayimbaev2021, Rayimbaev2022, Murodov2023, Rayimbaev2023, Rayimbaev2024, Shermatov2025}. These models show that QPO frequencies depend sensitively on spacetime geometry and surrounding fields, making them effective observational tools for testing strong-gravity physics. In the geodesic oscillation framework, QPO frequencies are directly associated with orbital and epicyclic frequencies of test particles in curved spacetime \cite{Gao2021}. This approach provides a natural relativistic interpretation of HF QPOs and links observational timing features to fundamental properties of black hole spacetimes.

	The foundation of black hole thermodynamics was established by Bekenstein and Hawking \cite{Bekenstein1973,Hawking1975}. Subsequently, the thermodynamic properties of black holes and the formulation of laws of black hole mechanics, analogous to the classical laws of thermodynamics, were developed in \cite{Bardeen1973,Sciama1976,Davies1978,Landsberg1992}. These early formulations primarily describe stationary and isolated systems. However, realistic black holes exist in dynamical environments where gravitational radiation and matter flux continuously perturb the spacetime geometry. In such non-equilibrium situations, the global event horizon is inadequate as a physical boundary because of its teleological nature. To address this limitation and describe real-time evolution, modern approaches employ quasi-local concepts such as dynamical and trapping horizons \cite{Hayward1994,Ashtekar2004}. These frameworks define the black hole boundary in terms of the local expansion of null geodesic congruences, thereby enabling a consistent treatment of time-dependent surface gravity and entropy flux. Thermodynamics of black hole solutions in general relativity as well as modified gravity theories has widely been investigated in the literature (see, Refs. \cite{Yang2023, Liu2024, Duan2024, AhmedSilva2025, AhmedSilva2025b, Ahmed2026, Sohan2025,Shokhzod2025, AhmedAlBadawiSakalli2026MPLA,AlBadawiAhmedSakalli2025} and related references therein).
	
	The main objective of the present work is to provide a unified analysis of a charged black hole in KR-gravity surrounded by perfect fluid dark matter, emphasizing how Lorentz violation and the dark-matter environment modify optical, dynamical, and thermodynamic observables. More specifically, we study the null geodesic structure through the effective potential, the photon sphere, the shadow radius, the effective radial force, and the corresponding photon trajectories. We then turn to the timelike motion of neutral test particles, focusing on the ISCO, the epicyclic frequencies, and the associated QPO phenomenology, including a Bayesian comparison with observational data from representative black-hole sources. Finally, we investigate the horizon thermodynamics, the local and global stability conditions, and the sparsity of Hawking radiation, thereby assembling a broad phenomenological portrait of the charged KR+PFDM spacetime.

	\section{Charged BH Surrounded by PFDM in KR-gravity }
	
	The modified Einstein equations in Lorentz-violating gravity with a background KR field, in the presence of perfect-fluid dark matter, are given by the following equations. These equations incorporate contributions from the electromagnetic field, the Kalb--Ramond (KR) field, and the dark matter sector, thereby providing a consistent framework to study the interplay 
	between matter sources and modified gravitational dynamics \cite{Zheng2024}:
	\begin{align}
		G^{\mu\nu}&= T^{\mu\nu}_{\rm EM}+T^{\mu\nu}_{\rm KR}+T^{\mu\nu}_{\rm DM},\label{aa1}
	\end{align}
	Here, $T^{\mu\nu}_{\rm EM}$ is the energy-momentum tensor of the electromagnetic field, derived as
	\begin{align}
		T^{\text{ EM }}_{\mu \nu }= & {} 2 F_{\mu \alpha } F_{\nu }{}^{\alpha }- \frac{1}{2} g_{\mu \nu }F^{\alpha \beta } F_{\alpha \beta } +\eta \left( 8 B^{\alpha \beta } B_{\nu }{}^{\gamma } F_{\alpha \beta } F_{\mu \gamma }-g_{\mu \nu } B^{\alpha \beta } B^{\gamma \rho } F_{\alpha \beta } F_{\gamma \rho } \right) ,\label{aa2}
	\end{align}
	$ T^{\mu\nu}_{\rm KR}$  is the effective energy-momentum tensor of the KR field, given by 
	\begin{align}
		T^{\text{ KR }}_{\mu \nu }= & {} \frac{1}{2} H_{\mu \alpha \beta } H_{\nu }^{\alpha \beta }-\frac{1}{12} g_{\mu \nu } H^{\alpha \beta \rho } H_{\alpha \beta \rho }+2V' B_{\alpha \mu }B^{\alpha }{}_\nu-g_{\mu \nu }V+ \xi _2 \Bigg [\frac{1}{2} g_{\mu \nu } B^{\alpha \gamma } B^{\beta }{}_{\gamma }R_{\alpha \beta } - B^{\alpha }{}_{\mu } B^{\beta }{}_{\nu }R_{\alpha \beta } \nonumber \\{} & {} - B^{\alpha \beta } B_{\nu \beta } R_{\mu \alpha }\!-\!B^{\alpha \beta } B_{\mu \beta } R_{\nu \alpha }\!+\!\frac{1}{2} \nabla _{\alpha }\nabla _{\mu }\left( B^{\alpha \beta } B_{\nu \beta }\right)+\frac{1}{2} \nabla _{\alpha }\nabla _{\nu }\left( B^{\alpha \beta } B_{\mu \beta }\right) -\frac{1}{2}\nabla ^{\alpha }\nabla _{\alpha }\left( B_{\mu }{}^{\gamma }B_{\nu \gamma } \right) \nonumber \\{} & {} -\frac{1}{2} g_{\mu \nu } \nabla _{\alpha }\nabla _{\beta }\left( B^{\alpha \gamma } B^{\beta }{}_{\gamma }\right) \Bigg ]. \label{aa3}
	\end{align}
	And $T^{\mu\nu}$ is the energy-momentum of perfect fluid dark matter is given by \cite{MHL2012}
	\begin{equation}
		T^{\text{DM}}_{\,\,\mu\nu} = \mathrm{diag}\left(-\mathcal{E}_{DM},\, P_{r\,DM},\, P_{\theta\,DM},\, P_{\phi\,DM}\right)
		=\mbox{diag}\left(\frac{\lambda}{8\pi r^3},\,\frac{\lambda}{8\pi r^3},\,-\frac{\lambda}{16\pi r^3},\,-\frac{\lambda}{16\pi r^3}\right)\label{aa4}
	\end{equation}
	where $\lambda$ is a constant that takes real values, called the PFDM parameter.
	
	The equation of motion for the KR field and the modified Maxwell equation are obtained as \cite{Zheng2024}
	\begin{align}
		\nabla ^{\alpha }H_{\alpha \mu \nu }\!+\!3\xi R_{\alpha [\mu }B^{\alpha }{}_{\nu ]}\!-\!6 V' B_{\mu \nu } \!-\!12 \eta B^{\alpha \beta } F_{\alpha \beta } F_{\mu \nu }\!=\!0.,\label{aa5}\\
		\nabla ^{\nu }\left( F_{\mu \nu } + 2\eta B_{\mu \nu }B^{\alpha \beta } F_{\alpha \beta }\right) =0.\label{aa6}
	\end{align}
	
	Therefore, a static and spherically symmetric spacetime describing a charged black hole in Lorentz-violating gravity in the presence of PFDM is given by the following line element:
	\begin{equation}
		ds^2 = -f(r)\,dt^2 + \dfrac{dr^2}{f(r)} + r^2 (d\theta ^2 + \sin ^2{\theta }\,d\varphi ^2),\label{metric}
	\end{equation}
	where the lapse function is given by
	\begin{align}
		f(r)= \frac{1}{1-\ell} - \frac{2M}{r}+\frac{Q^2}{(1-\ell)^2 r^2} + \frac{\lambda }{r} \ln \frac{r}{|\lambda |},\label{function}
	\end{align}
	where $M$ denotes the black hole mass, $\ell \equiv \xi_2 b^2/2$ is the Lorentz-violating parameter characterizing the strength of Lorentz symmetry breaking, and $Q$ represents the electric charge.
	
	\begin{figure}[ht!]
		\centering
		\includegraphics[width=0.4\linewidth]{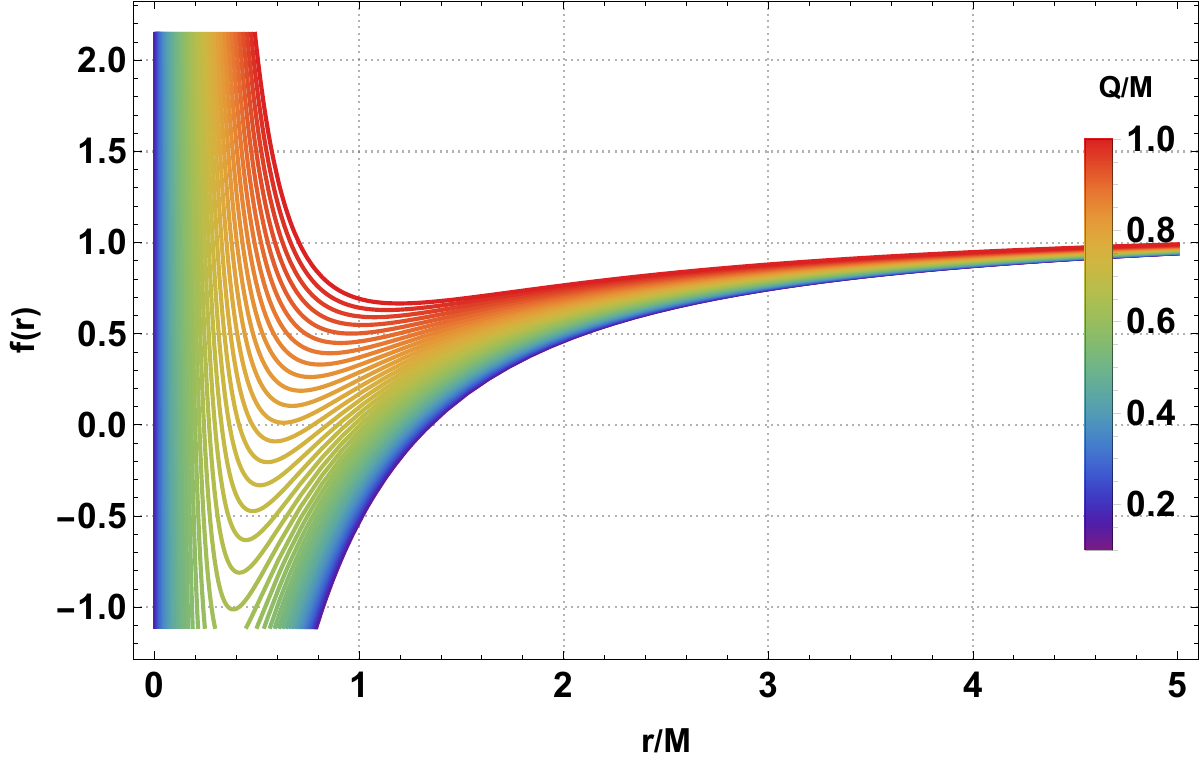}\qquad
		\includegraphics[width=0.4\linewidth]{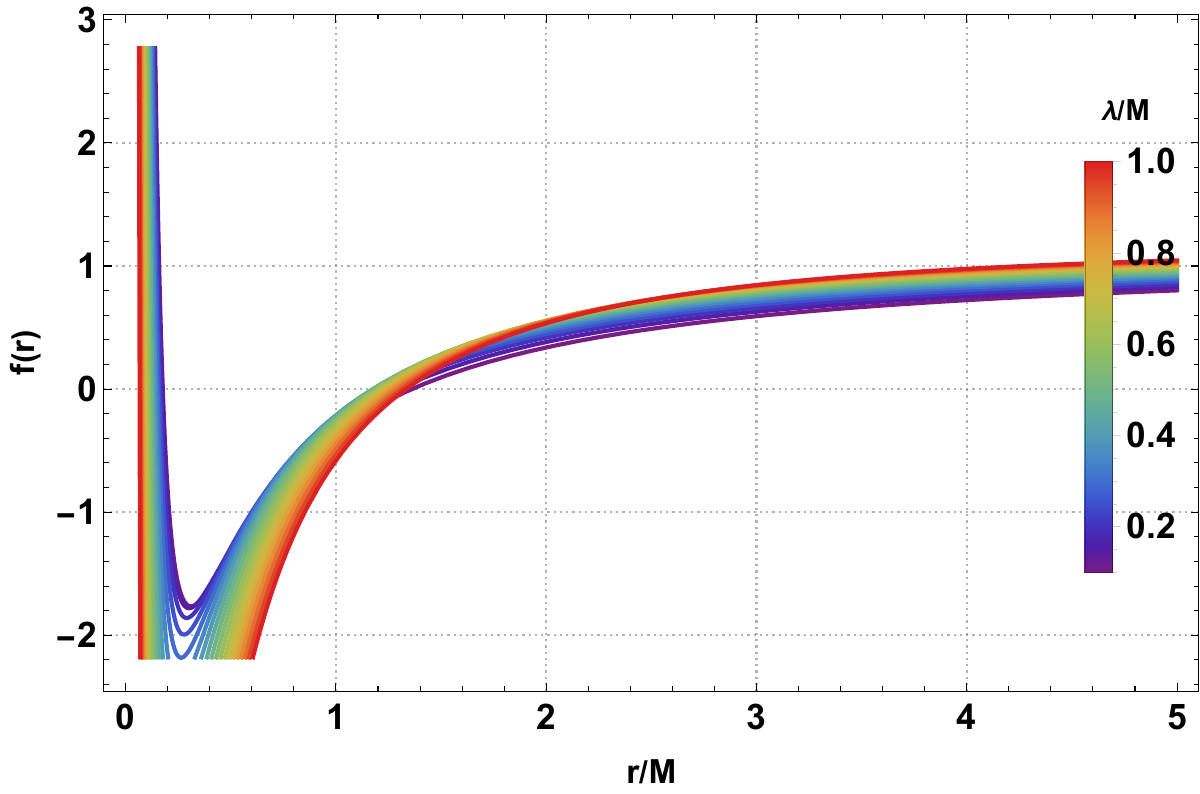}\\
		(i) $\lambda/M=0.5,\,\ell=0.1$ \hspace{6cm} (ii) $Q/M=0.5,\,\ell=0.1$\\
		\hfill\\
		\includegraphics[width=0.4\linewidth]{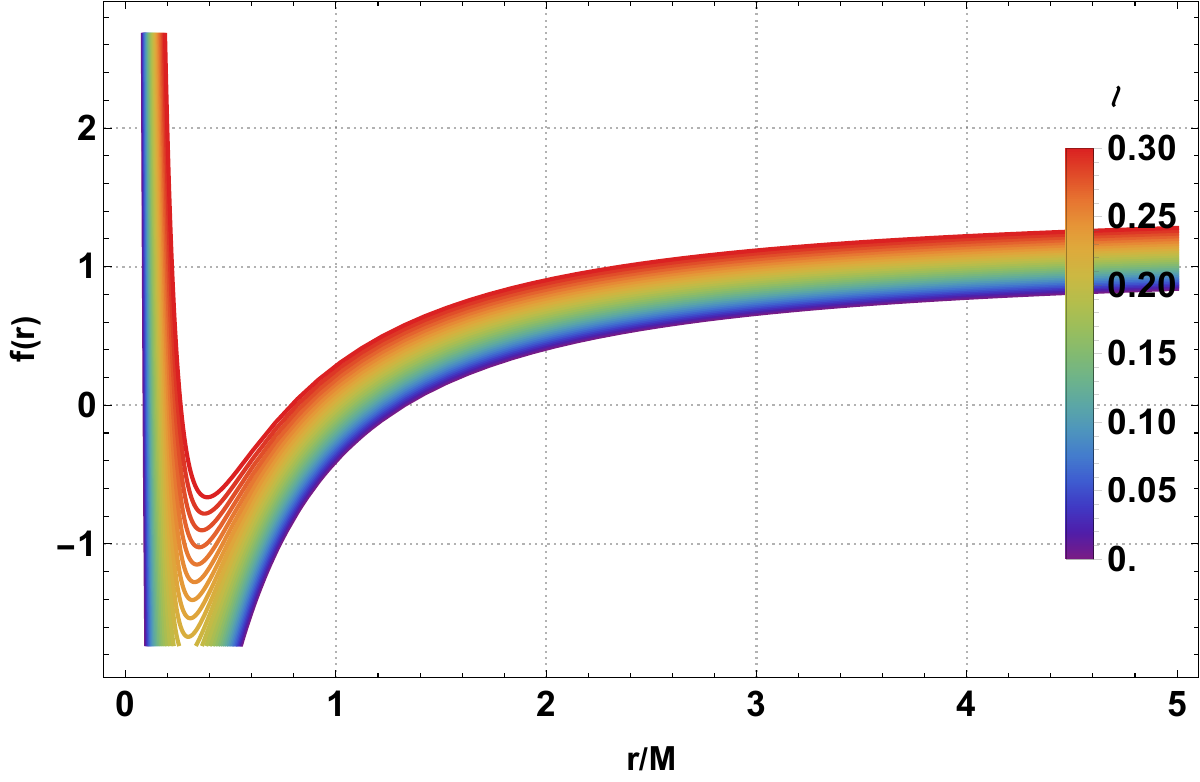}\\
		(iii) $\lambda/M=0.5=Q/M$
		\caption{Behavior of the lapse function $f(r)$ by varying the charge $Q$, PFDM parameter $\lambda$, and KR-field parameter $\ell$. }
		\label{fig:function}
	\end{figure}
	
	The behavior of the lapse function is illustrated in Figure \ref{fig:function} as a function of the dimensionless radial distance by varying parameters $\{Q, \lambda, \ell\}$. From this Figure, we observe that the event horizon exists provided the electric charge must have small values $Q < 1$.
	
	The limiting cases are summarized as follows:
	\begin{itemize}
		\item For $Q=0$, corresponding to the absence of electric charge, the lapse function reduces to
		\begin{equation}
			f(r)= \frac{1}{1-\ell} - \frac{2M}{r}+ \frac{\lambda }{r} \ln \frac{r}{|\lambda |},\label{function-2}
		\end{equation}
		In this case, the BH spacetime describes an uncharged solution in KR gravity surrounded by PFDM \cite{Sohan2025,Shokhzod2025}.
		
		\item In the limiting case $\lambda \to 0$, the lapse function becomes
		\begin{equation}
			f(r)= \frac{1}{1-\ell} - \frac{2M}{r}+\frac{Q^2}{(1-\ell)^2\,r^2},\label{function-3}
		\end{equation}
		This corresponds to a charged BH solution in KR gravity \cite{Zheng2024}.
		
		\item When both $Q=0$ and $\lambda \to 0$, i.e., in the absence of electric charge and surrounding matter, the lapse function simplifies to
		\begin{equation}
			f(r)= \frac{1}{1-\ell} - \frac{2M}{r}.\label{function-4}
		\end{equation}
		This solution represents the Schwarzschild BH spacetime in KR gravity \cite{Yang2023}.
		
		\item For $\ell=0$, corresponding to the absence of Lorentz-violating effects, the lapse function reduces
		\begin{align}
			f(r)=1- \frac{2M}{r}+\frac{Q^2}{r^2} + \frac{\lambda }{r} \ln \frac{r}{|\lambda |},\label{function-5}
		\end{align}
		In that limiting case, the solution represents Reissner-Nordstrom BH surrounded by PFDM \cite{MHL2012,Sadeghi2024}.
	\end{itemize}

	\section{Null Geodesics}
	
	Null geodesics describe how photons move in curved spacetime around black holes, showing key features such as the photon sphere, black hole shadow, and light paths near the event horizon. Recent observational images of M87* and Sagittarius A* have increased significant interest in both general relativity and alternative gravity models, offering new ways to test predictions about strong gravity and photon behavior.
	
	For simplicity, and due to the spherical symmetry of the spacetime, the geodesic motion can be restricted to the equatorial plane defined by $\theta=\pi/2$ without loss of generality. The photon's motion is described by the null geodesic as
	\begin{equation}
		ds^2=0\Longrightarrow -f(r) \dot t^2+\frac{1}{f(r)} \dot r^2+r ^2 \dot \phi^2=0\,,\label{bb1}
	\end{equation}
	where dot represents derivative with respect to an affine parameter.
	
	As the space-time is static and spherically symmetric, there are two conserved quantities for the temporal coordinate $t$ and the angular coordinate $\phi$. There are given by
	\begin{equation}
		\mathrm{E}=f(r) \dot t\quad,\quad \mathrm{L}=r^2 \sin^2 \theta \dot \phi\,,\label{bb2} 
	\end{equation}
	where $\mathrm{E}$ is the conserved energy and $\mathrm{L}$ is the conserved angular momentum.
	
	With these, the equation of motion for photon particles is given by
	\begin{equation}
		\dot r^2=\mathrm{E}^2-V_{\rm eff}(r),\label{bb3}
	\end{equation}
	where $V_{\rm eff}$ is the effective potential of the null geodesics system and is given by
	\begin{equation}
		V_{\rm eff}(r)=\frac{\mathrm{L}^2}{r^2}\,f(r).\label{bb4}
	\end{equation}
	
	\begin{figure}[ht!]
		\centering
		\includegraphics[width=0.4\linewidth]{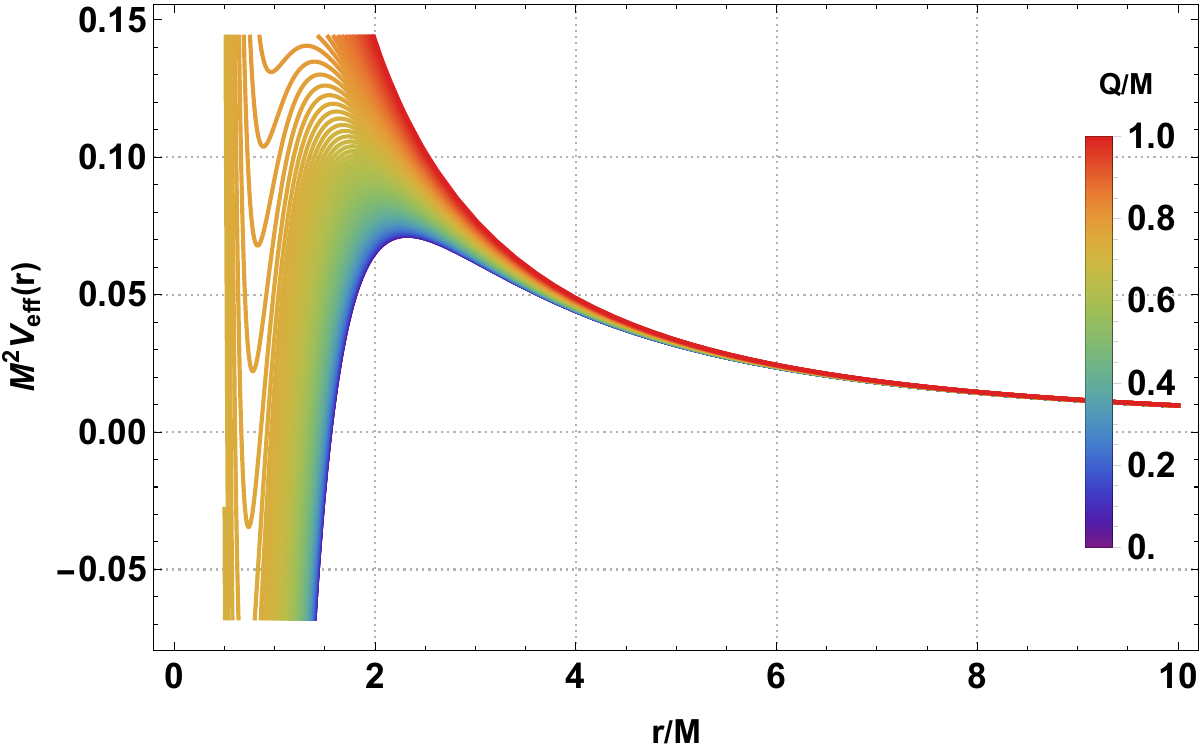}\qquad
		\includegraphics[width=0.4\linewidth]{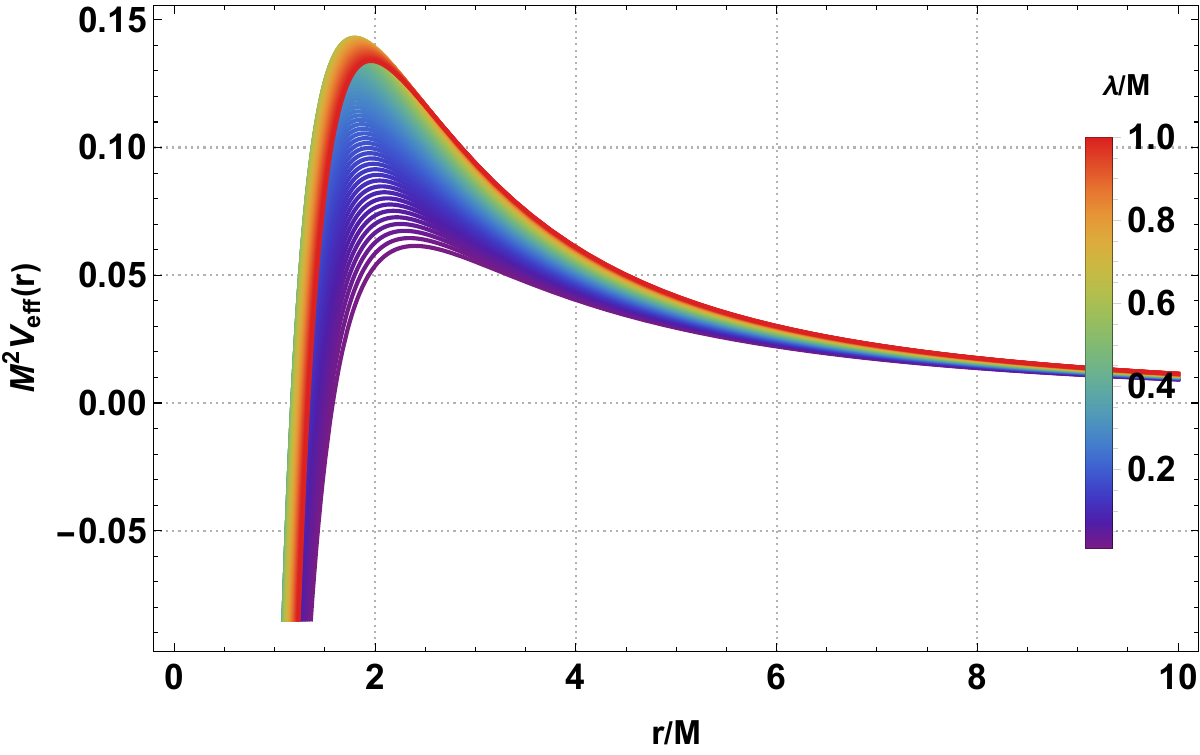}\\
		(i) $\lambda/M=0.5,\,\ell=0.1$ \hspace{6cm} (ii) $Q/M=0.5,\,\ell=0.1$\\
		\hfill\\
		\includegraphics[width=0.4\linewidth]{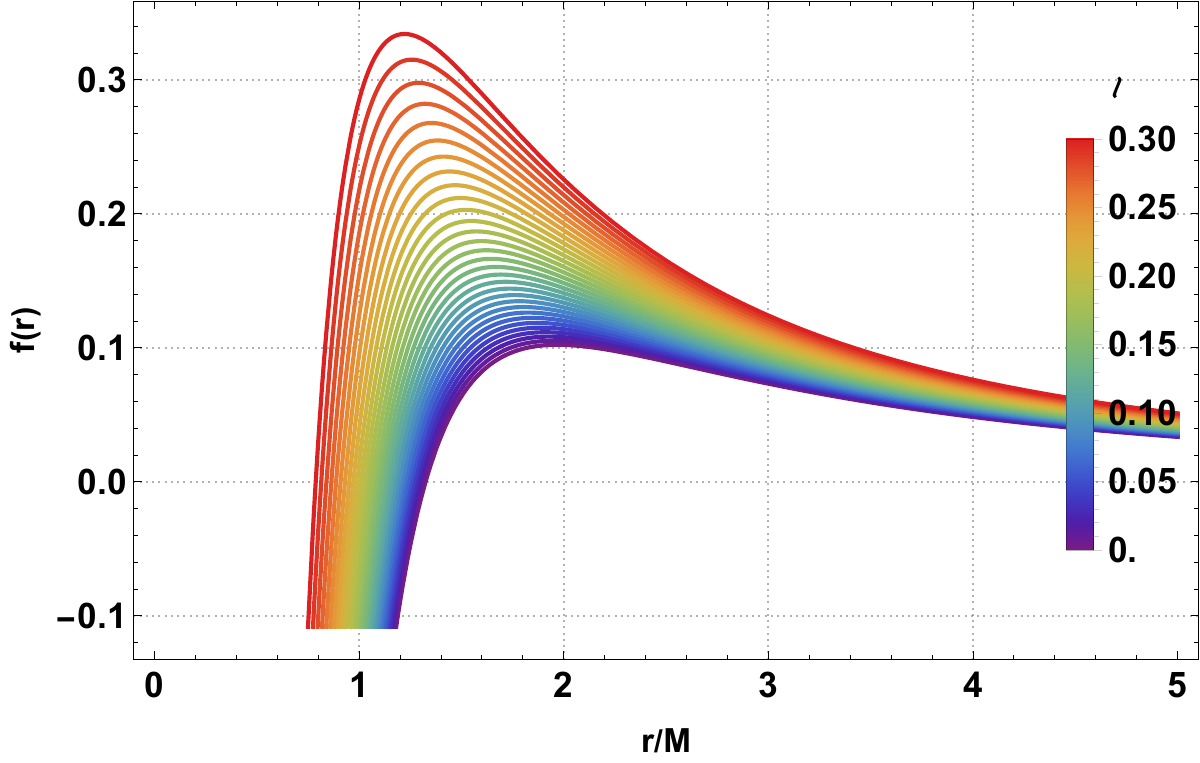}\\
		(iii) $\lambda/M=0.5=Q/M$
		\caption{Behavior of the effective potential governing the photon dynamics by varying the charge $Q$, PFDM parameter $\lambda$, and KR-field parameter $\ell$.}
		\label{fig:potential}
	\end{figure}
	
	This effective potential plays an important role in determining the optical features, such as the photon sphere, black hole shadow, photon trajectories, and the effective force experienced by massless photons, which we shall discuss below. 
	
	The behavior of the effective potential is illustrated in Fig.~\ref{fig:potential} as a function of the dimensionless radial coordinate for different values of the parameters $\{Q, \lambda, \ell\}$. It is observed that the height of the effective potential barrier increases as these parameters increase. This indicates that the gravitational field becomes stronger, leading to a more pronounced confinement of photon trajectories near the black hole. Consequently, the unstable circular photon orbits are associated with higher potential peaks, reflecting the combined influence of the electric charge, dark matter parameter, and Lorentz-violating effects on the spacetime geometry.
	
	\begin{center}
		{\bf A.\,Circular Orbits: Photon Sphere}
	\end{center}
	
	For circular null orbits, the conditions $\dot r=0$ and $\ddot r=0$ must be satisfied. These conditions using Eq. (\ref{bb3}) results the following relations:L
	\begin{equation}
		\mathrm{E}^2=V_{\rm eff}=\frac{\mathrm{L}^2}{r^2}\,f(r)\quad,\quad \partial_r V_{\rm eff}(r)=0.\label{cc1}
	\end{equation}
	
	Using the given potential in (\ref{bb4}) into the above relation (\ref{cc1}), we find the following polynomial relation in $r$ as,
	\begin{equation}
		\frac{1}{1-\ell} - \frac{3M}{r} + \frac{2 Q^2}{(1-\ell)^2 r^2} + \frac{3 \lambda}{2 r} \ln \frac{r}{|\lambda|} - \frac{\lambda}{2 r}=0.\label{cc2}
	\end{equation}
	The exact solution of the above equation gives us the photon sphere radii $r_{\rm ph}$. Noted that an exact solution is quite impossible due to the logarithmic function. However, one can obtain numerical results for the photon sphere radii by selecting suitable values for various geometric parameters.
	
	\begin{table}[ht!]
		\centering
		\begin{tabular}{|c|c|c|c|c|c|c|}
			\hline
			$\ell$ & $Q/M (\downarrow) \backslash \lambda/M (\rightarrow)$ & 0.1 & 0.2 & 0.3 & 0.4 & 0.5 \\
			\hline
			0.1 & 0.1 & 2.3114 & 2.1397 & 2.0465 & 1.9998 & 1.9836 \\
			0.1 & 0.2 & 2.2837 & 2.1116 & 2.0188 & 1.9732 & 1.9581 \\
			0.1 & 0.3 & 2.2361 & 2.0630 & 1.9711 & 1.9272 & 1.9143 \\
			0.1 & 0.4 & 2.1657 & 1.9909 & 1.9003 & 1.8591 & 1.8497 \\
			0.1 & 0.5 & 2.0674 & 1.8896 & 1.8007 & 1.7638 & 1.7599 \\
			\hline
			-0.1 & 0.1 & 2.7988 & 2.5614 & 2.4234 & 2.3450 & 2.3060 \\
			-0.1 & 0.2 & 2.7802 & 2.5424 & 2.4045 & 2.3267 & 2.2884 \\
			-0.1 & 0.3 & 2.7487 & 2.5100 & 2.3724 & 2.2955 & 2.2586 \\
			-0.1 & 0.4 & 2.7034 & 2.4633 & 2.3261 & 2.2506 & 2.2156 \\
			-0.1 & 0.5 & 2.6427 & 2.4006 & 2.2638 & 2.1903 & 2.1580 \\
			\hline
			-0.2 & 0.1 & 3.0399 & 2.7680 & 2.6062 & 2.5108 & 2.4594 \\
			-0.2 & 0.2 & 3.0243 & 2.7519 & 2.5902 & 2.4952 & 2.4445 \\
			-0.2 & 0.3 & 2.9979 & 2.7247 & 2.5631 & 2.4688 & 2.4191 \\
			-0.2 & 0.4 & 2.9601 & 2.6857 & 2.5242 & 2.4310 & 2.3828 \\
			-0.2 & 0.5 & 2.9101 & 2.6338 & 2.4725 & 2.3807 & 2.3346 \\
			\hline
		\end{tabular}
		\caption{Numerical values of $r_s/M$ for varying $Q/M$, $\lambda/M$, and $\ell$.}
		\label{tab: photon-radius}
	\end{table}
	
	\begin{figure}[ht!]
		\centering
		\includegraphics[width=0.4\linewidth]{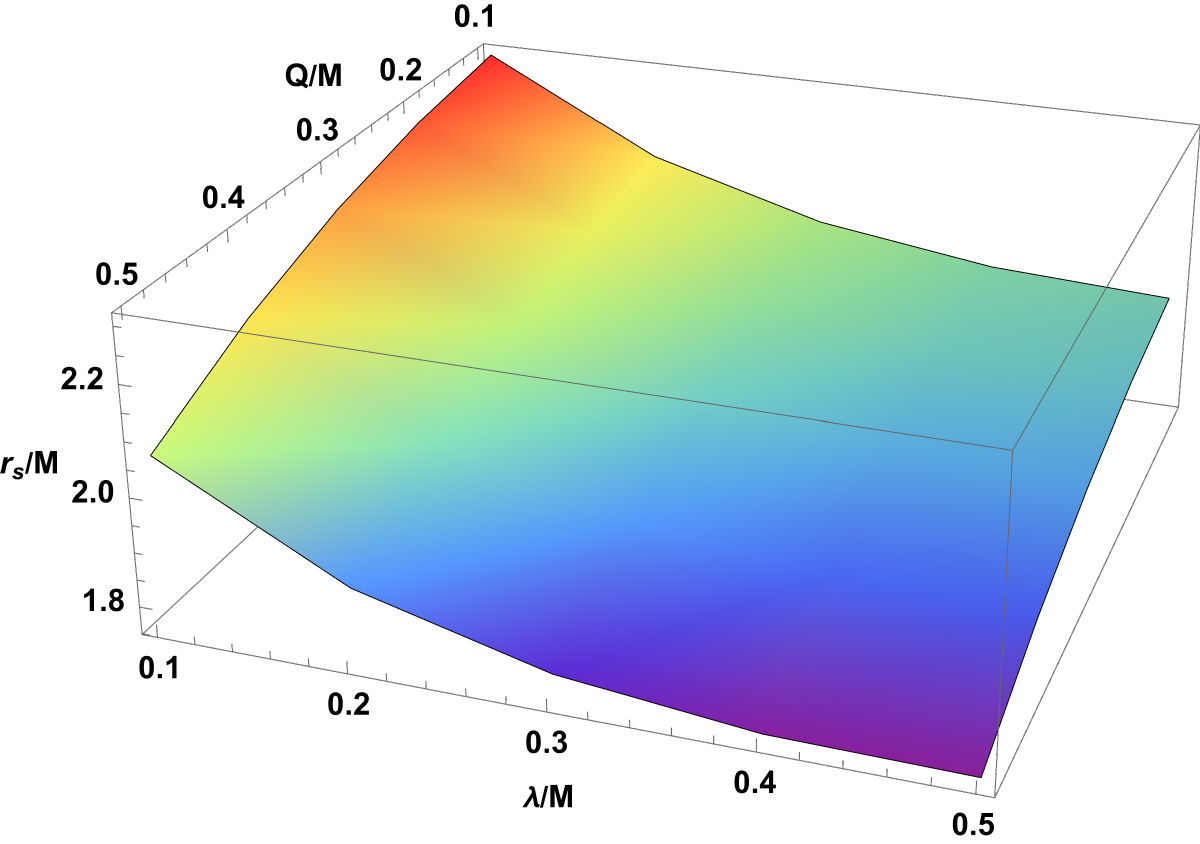}\qquad
		\includegraphics[width=0.4\linewidth]{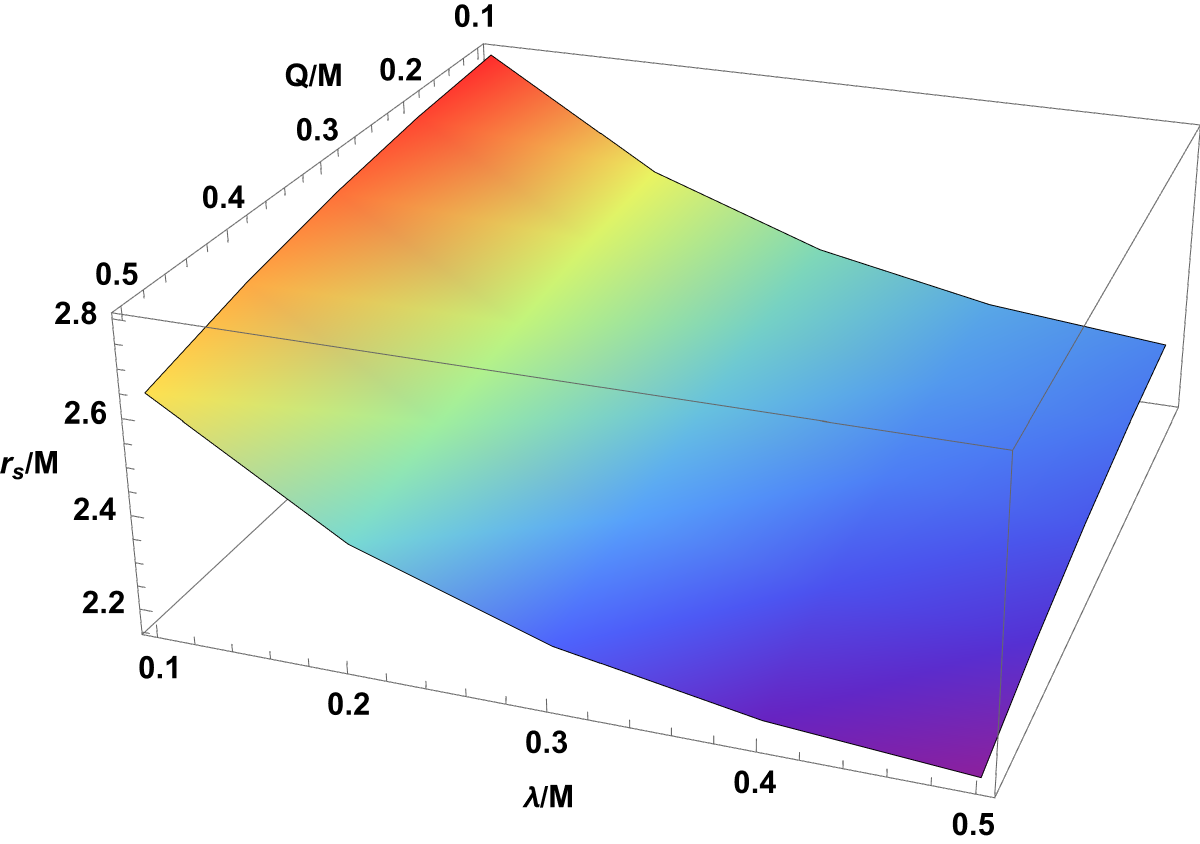}\\ 
		(i) $\ell=0.1$ \hspace{6cm} (ii) $\ell=-0.1$\\
		\hfill\\
		\includegraphics[width=0.4\linewidth]{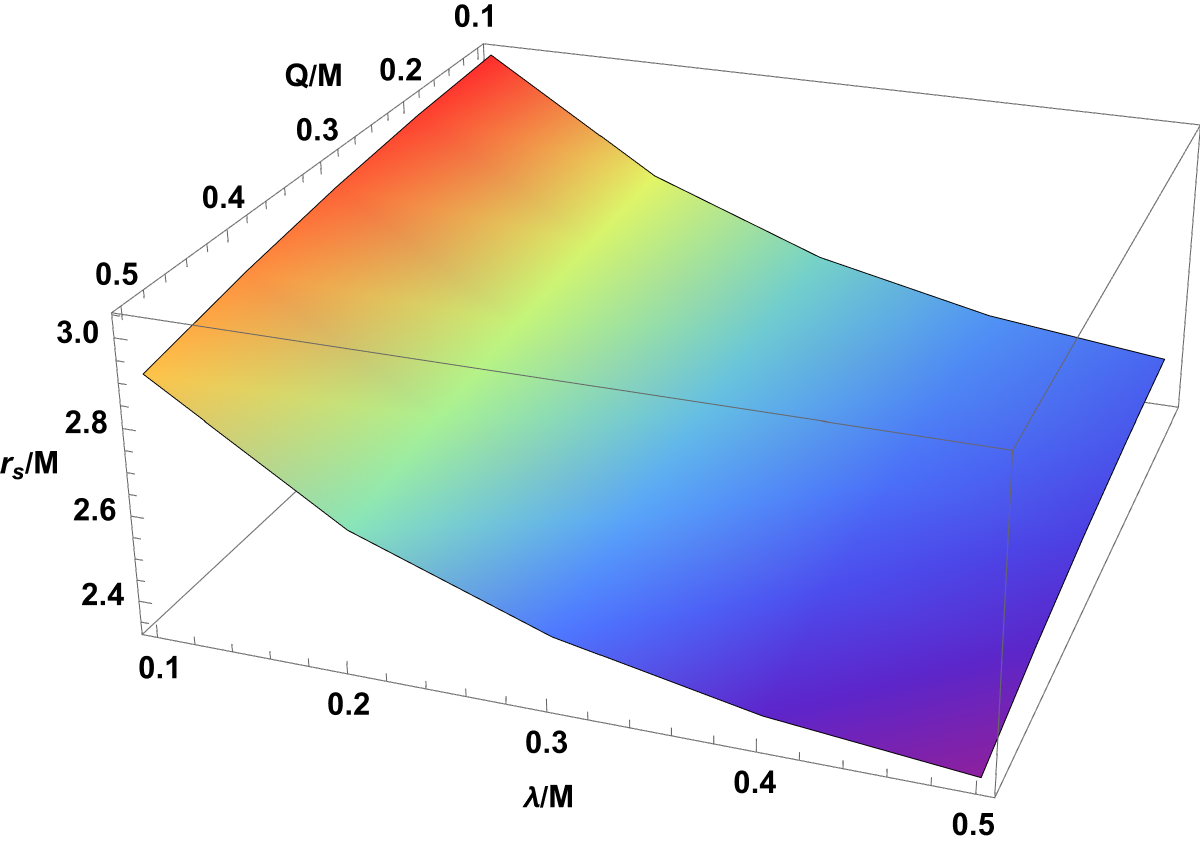}\\
		(iii) $\ell=-0.2$\\
		\caption{Three-dimensional visualization of the photon sphere radius $r_s$ with $\{Q,\lambda\}$ parameter space for three distinct values of the KR-field parameter $\ell$.}
		\label{fig:photon}
	\end{figure}

	In Table \ref{tab: photon-radius}, we present the numerical values of the photon sphere radius $r_{s}$ for different choices of the electric charge $Q$, PFDM parameter $\lambda$, and Lorentz-violating parameter $\ell$. From the table, it is evident that for fixed values of $\ell$ and $Q$, the radius $r_s$ decreases as $\lambda$ increases. A similar decreasing trend in $r_s$ is observed with increasing $Q$, indicating that both the electric charge and PFDM parameter tend to reduce the photon sphere radius. 
	
	In contrast, decreasing the Lorentz-violating parameter $\ell$ leads to an increase in $r_s$, suggesting that Lorentz-violating effects act to shrink the photon sphere. Furthermore, Fig.~\ref{fig:photon} illustrates a three-dimensional visualization of the photon sphere radius as a function of $Q$ and $\lambda$ for different values of $\ell$, providing a more comprehensive view of how these parameters jointly influence the photon sphere structure.
	
	\begin{center}
		{\bf B.\, Shadow Radius}
	\end{center}
	
	The black hole shadow is the apparent dark region formed by photons being captured by the event horizon, as seen by a distant observer. Its size and shape are determined by the spacetime geometry and are strongly influenced by parameters such as charge, dark matter distribution, and Lorentz-violating effects.
	
	For the selected space-time (\ref{metric}), the given lapse function at large distances behaves as,
	\begin{equation}
		\lim_{r \to \infty} f(r) =\frac{1}{1-\ell} \neq 1 \label{condition}
	\end{equation}
	indicating that the metric is not asymptotically flat.
	
	This feature warrants emphasis because in non-asymptotically flat geometries, the inferred shadow size depends not only on the photon sphere but also on the observer's position and normalization. For that reason, throughout the present analysis, we keep the observer at a finite radial location $r_0$ and interpret $R_{\rm sh}$ as an effective observable radius for a static distant observer, rather than as the asymptotic celestial radius familiar from asymptotically flat black-hole spacetimes.
	
	Thus, for a static distant observer located at $r_0$, the shadow radius is given by \cite{Perlick2022}
	\begin{equation}
		R_{\rm sh}=r_s\,\sqrt{\frac{f(r_0)}{f(r_{\rm ph})}}=r_{\rm ph}\sqrt{\frac{\frac{1}{1-\ell} - \frac{2M}{r_0}+\frac{Q^2}{(1-\ell)^2 r^2_0} + \frac{\lambda }{r_0} \ln \frac{r_0}{|\lambda |}}{\frac{1}{1-\ell} - \frac{2M}{r_s}+\frac{Q^2}{(1-\ell)^2 r^2_s} + \frac{\lambda }{r_s} \ln \frac{r_s
				}{|\lambda |}}}.\label{cc8}
	\end{equation}
	
	\begin{table}[ht!]
		\centering
		\begin{tabular}{|c|c|c|c|c|c|c|}
			\hline
			$\ell$ & $Q/M (\downarrow) \backslash \lambda/M (\rightarrow) $ & 0.1 & 0.2 & 0.3 & 0.4 & 0.5 \\
			\hline
			0.1 & 0.1 & 3.6494 & 3.3627 & 3.1903 & 3.0886 & 3.0346 \\
			0.1 & 0.2 & 3.6168 & 3.3290 & 3.1568 & 3.0560 & 3.0033 \\
			0.1 & 0.3 & 3.5609 & 3.2710 & 3.0991 & 2.9999 & 2.9496 \\
			0.1 & 0.4 & 3.4787 & 3.1855 & 3.0141 & 2.9175 & 2.8709 \\
			0.1 & 0.5 & 3.3652 & 3.0666 & 2.8957 & 2.8032 & 2.7624 \\
			\hline
			-0.1 & 0.1 & 4.3364 & 3.9648 & 3.7296 & 3.5807 & 3.4910 \\
			-0.1 & 0.2 & 4.3152 & 3.9425 & 3.7071 & 3.5586 & 3.4697 \\
			-0.1 & 0.3 & 4.2792 & 3.9046 & 3.6690 & 3.5211 & 3.4335 \\
			-0.1 & 0.4 & 4.2276 & 3.8501 & 3.6140 & 3.4672 & 3.3815 \\
			-0.1 & 0.5 & 4.1587 & 3.7772 & 3.5405 & 3.3952 & 3.3121 \\
			\hline
			-0.2 & 0.1 & 4.6653 & 4.2522 & 3.9857 & 3.8128 & 3.7048 \\
			-0.2 & 0.2 & 4.6477 & 4.2336 & 3.9668 & 3.7941 & 3.6867 \\
			-0.2 & 0.3 & 4.6180 & 4.2021 & 3.9348 & 3.7626 & 3.6561 \\
			-0.2 & 0.4 & 4.5755 & 4.1570 & 3.8891 & 3.7176 & 3.6125 \\
			-0.2 & 0.5 & 4.5195 & 4.0973 & 3.8285 & 3.6578 & 3.5547 \\
			\hline
		\end{tabular}
		\caption{Numerical values of shadow radius $R_{sh}/M$ for varying $Q/M$, $\lambda/M$, and $\ell$. Here $r_0/M=10$.}
		\label{tab:shadow-radius}
	\end{table}
	
	\begin{figure}[ht!]
		\centering
		\includegraphics[width=0.4\linewidth]{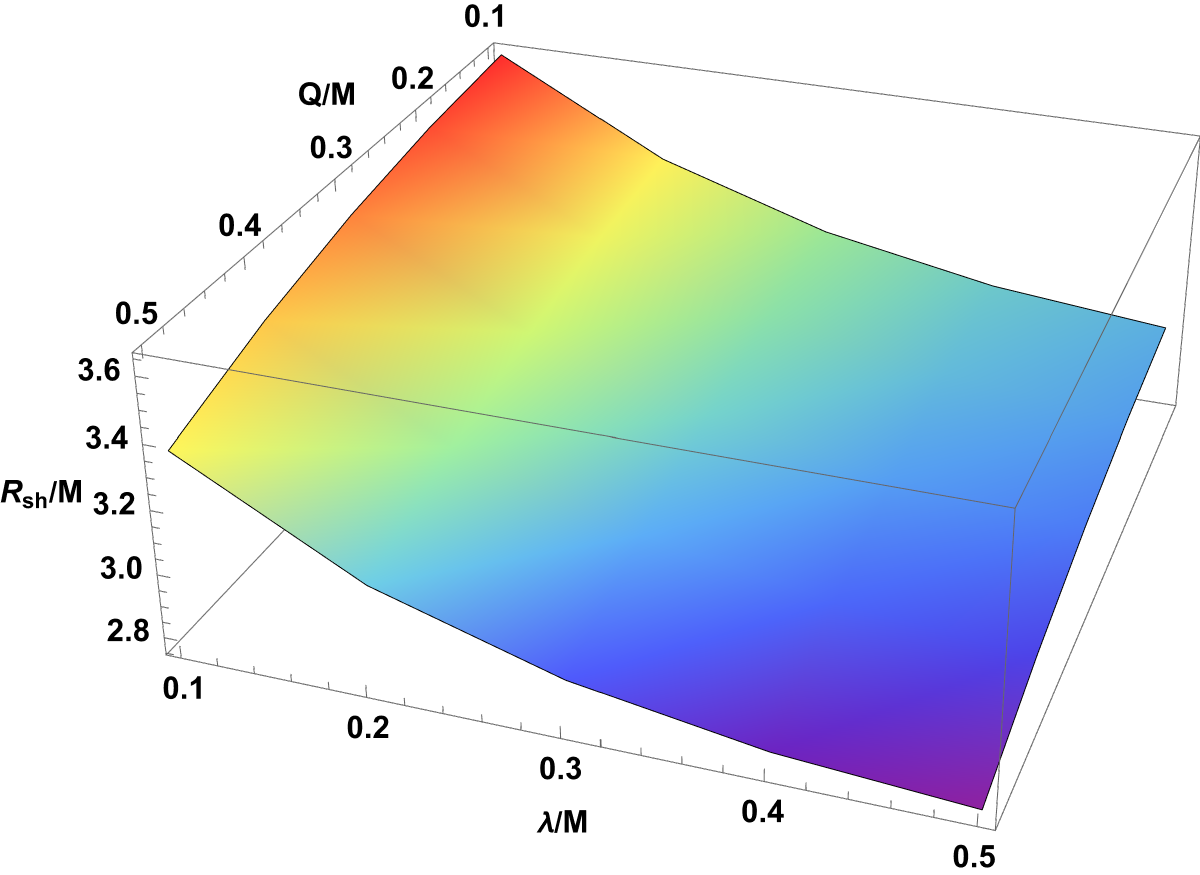}\qquad
		\includegraphics[width=0.4\linewidth]{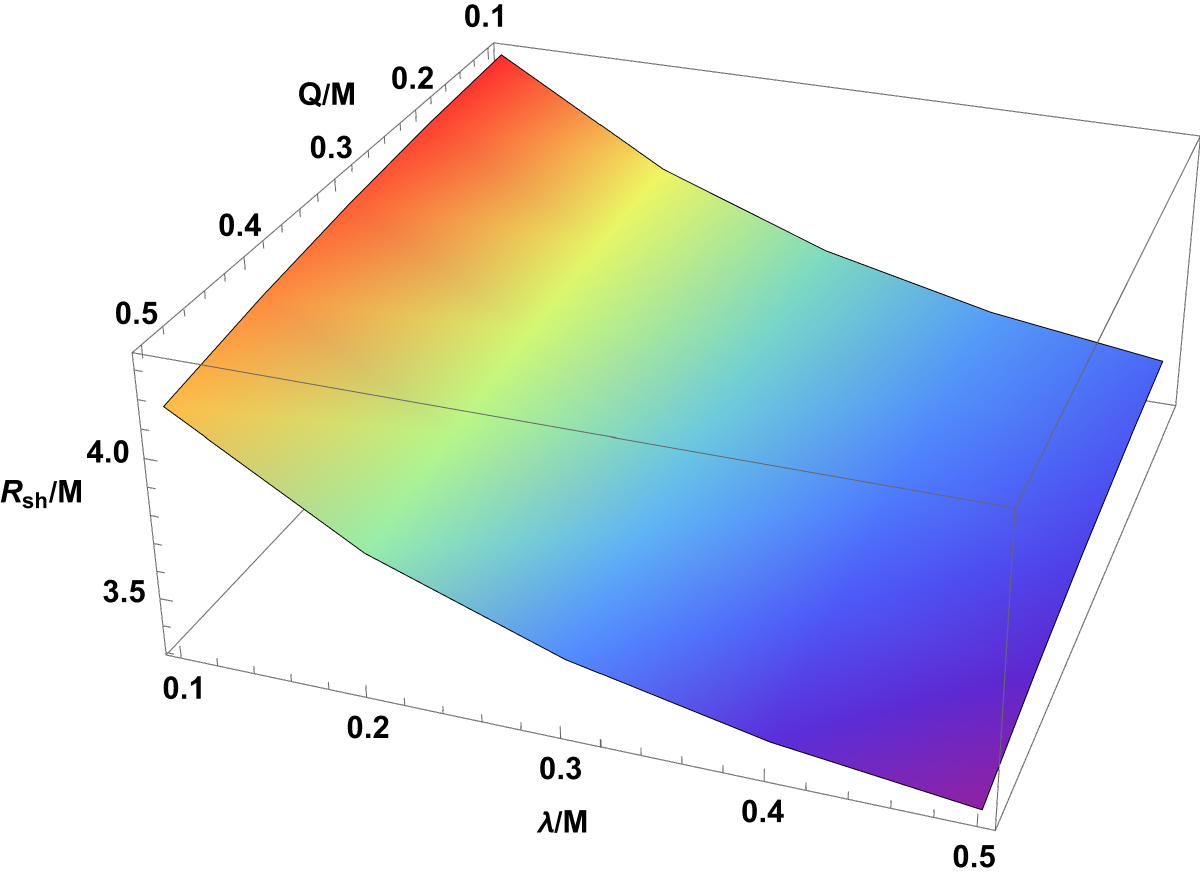}\\ 
		(i) $\ell=0.1$ \hspace{6cm} (ii) $\ell=-0.1$\\
		\hfill\\
		\includegraphics[width=0.4\linewidth]{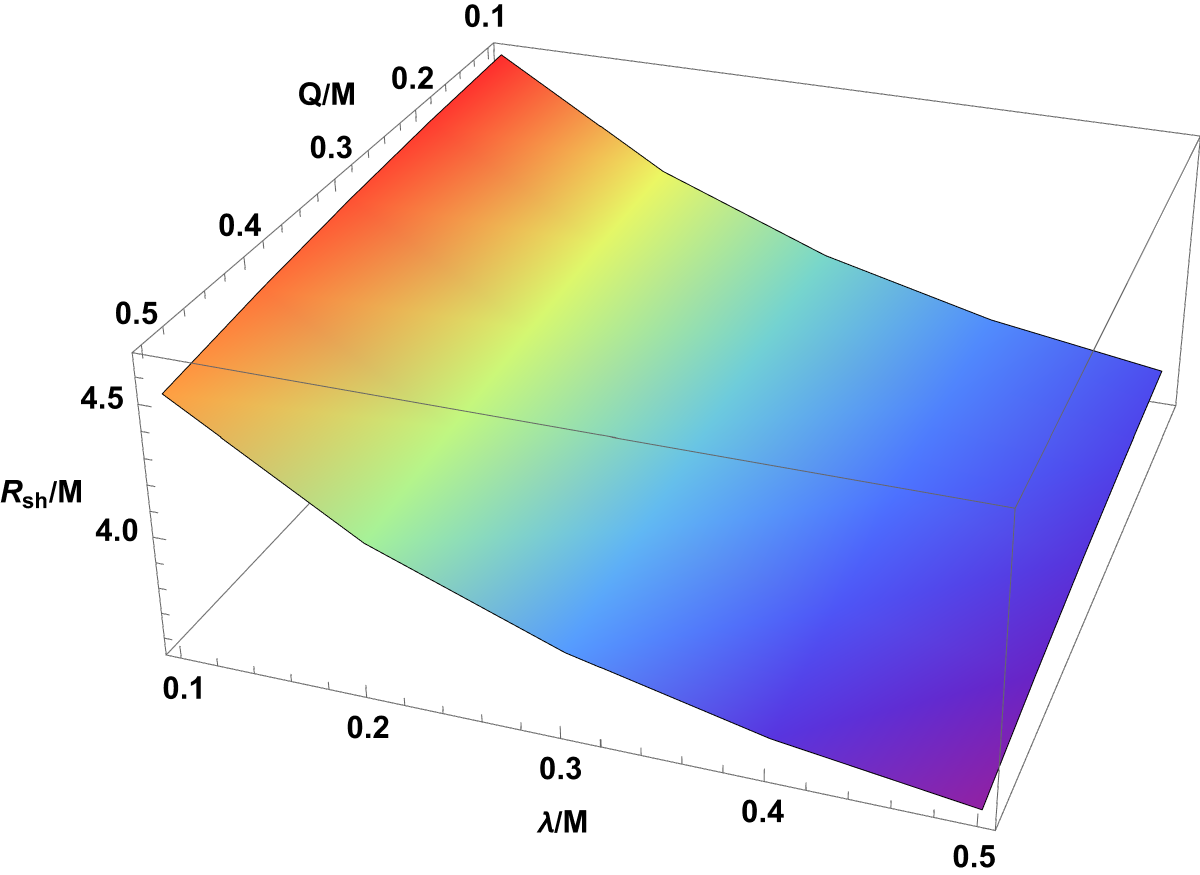}\\
		(iii) $\ell=-0.2$\\
		\caption{Three-dimensional visualization of the shadow radius $R_{\rm sh}$ with $\{Q,\lambda\}$ parameter space for three distinct values of the KR-field parameter $\ell$.}
		\label{fig:shadow}
	\end{figure}

	In Table \ref{tab:shadow-radius}, we present the numerical values of the shadow radius $R_{\rm sh}$ for various choices of the electric charge $Q$, PFDM parameter $\lambda$, and Lorentz-violating parameter $\ell$. Similar to the behavior of the photon sphere radius, it is observed that increasing $Q$ and $\lambda$ leads to a decrease in the shadow radius. This indicates that both the electric charge and the dark matter parameter contribute to shrinking the apparent size of the black hole shadow.
	
	In contrast, decreasing the Lorentz-violating parameter $\ell$ results in an increase in $R_{\rm sh}$, implying that Lorentz-violating effects tend to enlarge the shadow size. Furthermore, Fig.~\ref{fig:shadow} presents a three-dimensional visualization of the shadow radius as a function of $Q$ and $\lambda$ for different values of $\ell$, offering a more comprehensive understanding of how these parameters collectively influence the shadow structure.
	
	\begin{center}
		{\bf C.\, Effective Radial Force}
	\end{center}
	
	The effective radial force experienced by photon particles in the gravitational field is determined by the negative gradient of the effective potential, which governs the motion and stability of photon trajectories. This force indicates whether photons are attracted toward or repelled from a given radial location and plays a crucial role in defining the photon sphere and the black hole shadow. Mathematically, it is expressed as
	\begin{equation}
		\mathrm{F}_{\rm ph}=-\frac{1}{2} \partial_r V_{\rm eff}(r).\label{dd1}
	\end{equation}
	where $V_{\rm eff}(r)$ is the effective potential for photon motion. The points where $F_{\rm eff}(r)=0$ correspond to circular photon orbits, with stability/instability determined by the sign of the second derivative of the potential.  
	
	Substituting potential (\ref{bb4}), we find the following expression for the force experienced by the photons as,
	\begin{equation}
		\mathrm{F}_{\rm ph}=\frac{\mathrm{L}^2}{r^3}\left(\frac{1}{1-\ell} - \frac{3M}{r} + \frac{2 Q^2}{(1-\ell)^2 r^2} + \frac{3 \lambda}{2 r} \ln \frac{r}{|\lambda|} - \frac{\lambda}{2 r}\right).\label{dd2}
	\end{equation}
	
	\begin{figure}[ht!]
		\centering
		\includegraphics[width=0.4\linewidth]{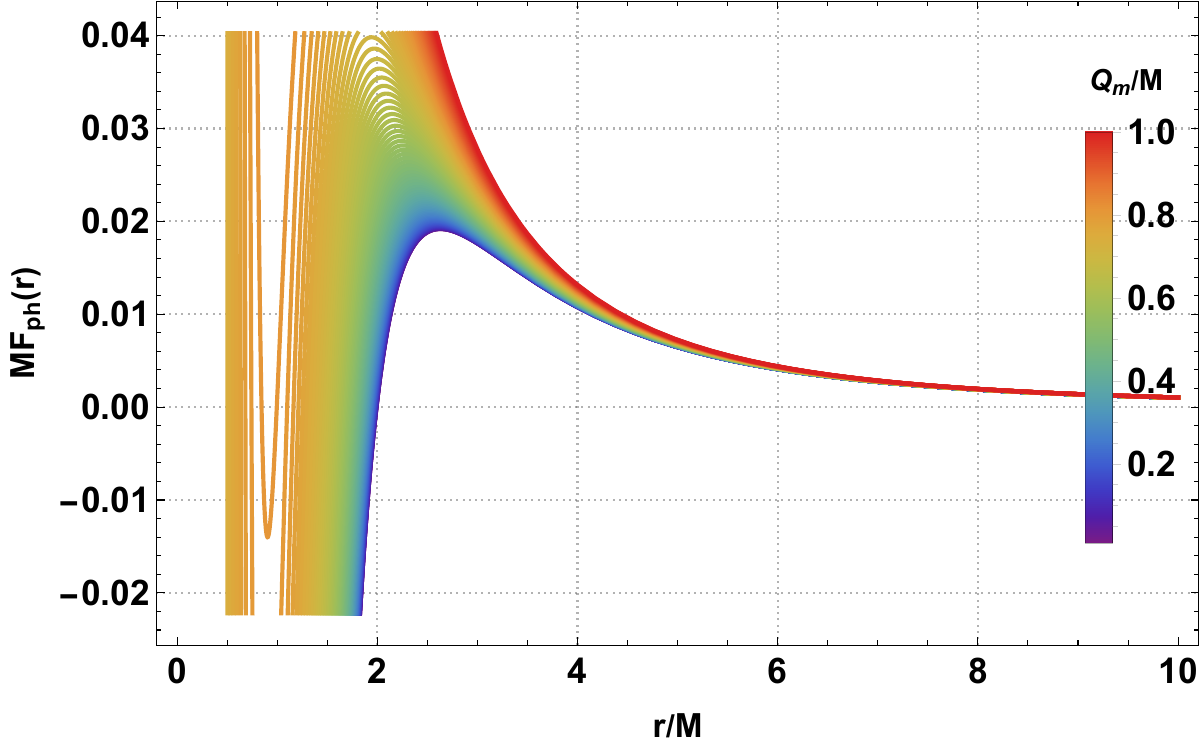}\qquad
		\includegraphics[width=0.4\linewidth]{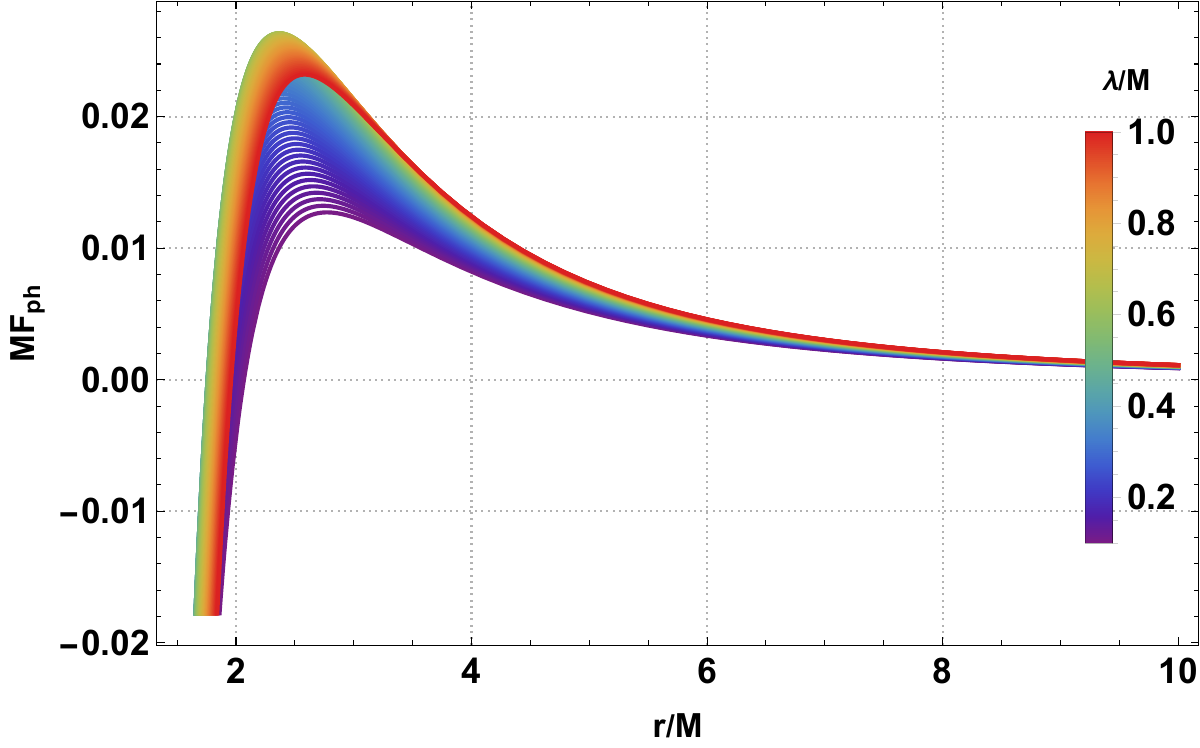}\\
		(i) $\lambda/M=0.5,\,\ell=0.1$ \hspace{6cm} (ii) $Q/M=0.5,\,\ell=0.1$\\
		\hfill\\
		\includegraphics[width=0.4\linewidth]{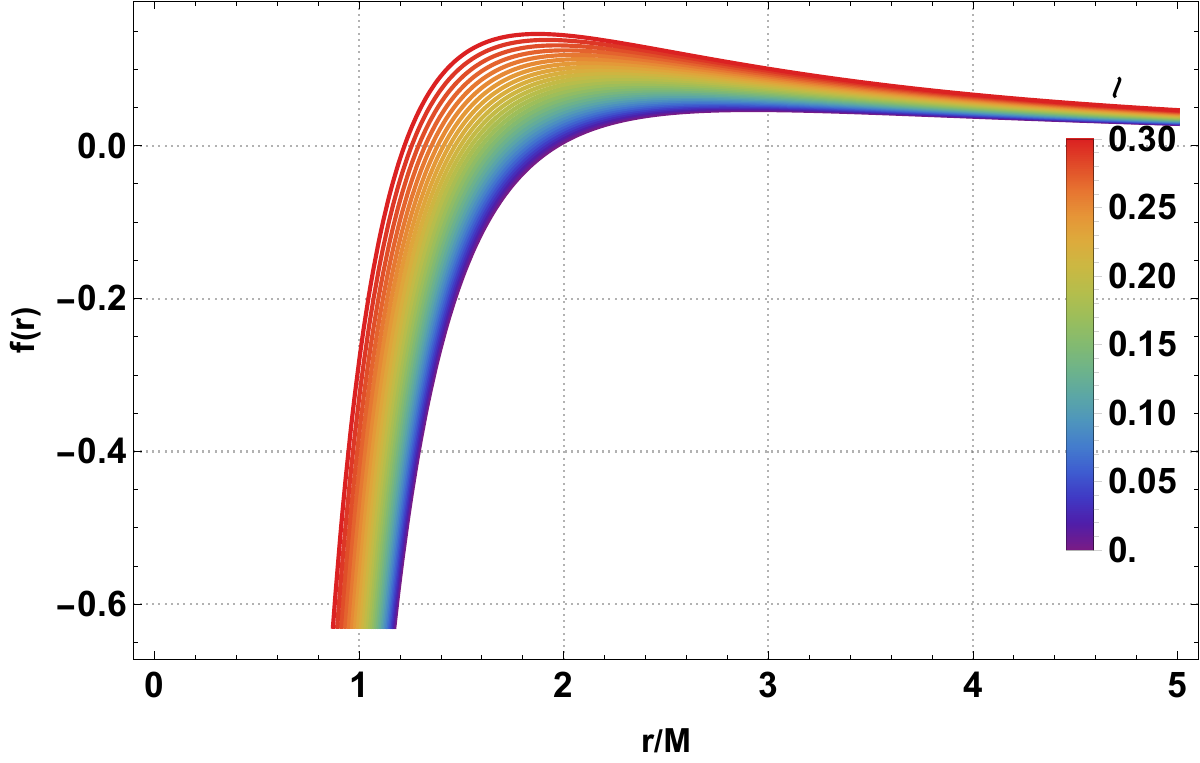}\\
		(iii) $\lambda/M=0.5=Q/M$
		\caption{Behavior of the effective force experienced by photon particles by varying the charge $Q$, PFDM parameter $\lambda$, and KR-field parameter $\ell$.}
		\label{fig:force}
	\end{figure}
	
	From the above expression, we have seen that several geometric parameters $\{Q,\,\lambda,\,\ell\}$ that alter the space-time curvature influence the magnitude of force experienced by photon particles.

	The behavior of the effective radial force is illustrated in Fig.~\ref{fig:force} as a function of the dimensionless radial coordinate for different values of the parameters $\{Q, \lambda, \ell\}$. It is observed that the magnitude of the force increases with increasing values of these parameters, indicating a stronger gravitational pull that more tightly confines photons near the black hole. As a result, the unstable circular photon orbits are located closer to the event horizon, reflecting the combined influence of electric charge, dark matter, and Lorentz-violating effects on photon dynamics.
	
	\begin{figure}[ht!]
		\centering
		\includegraphics[width=0.32\linewidth]{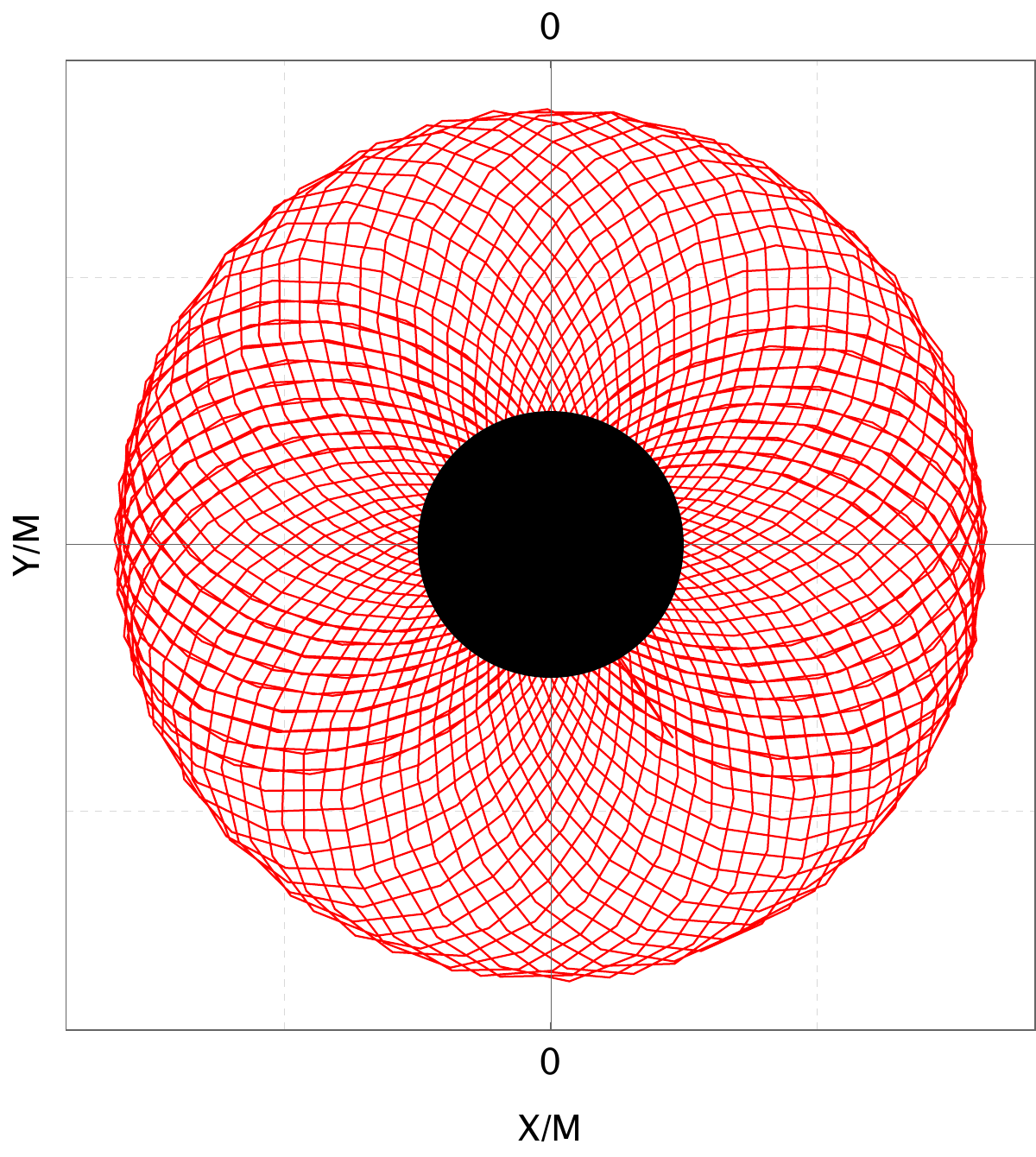}\quad
		\includegraphics[width=0.32\linewidth]{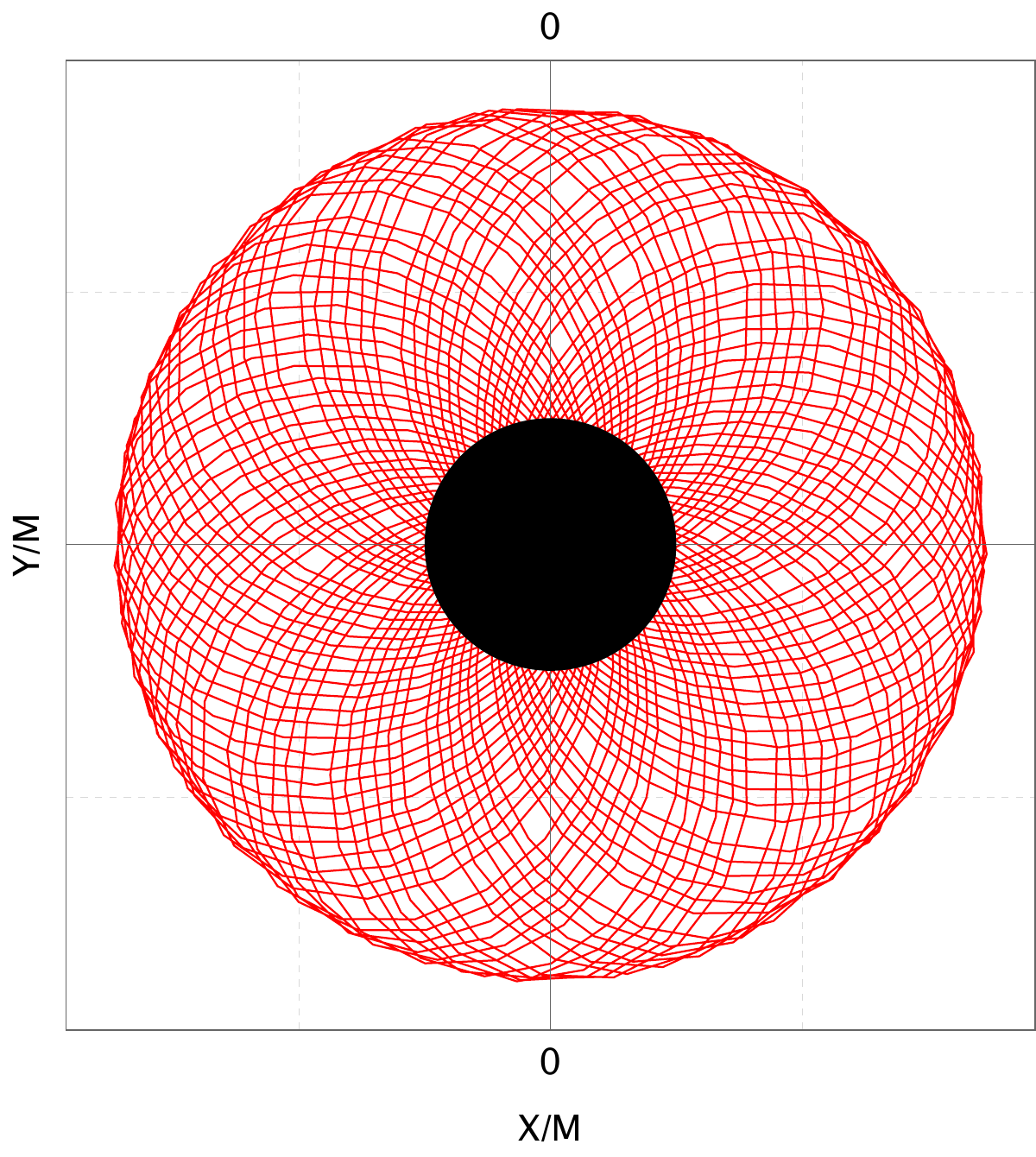}\quad
		\includegraphics[width=0.32\linewidth]{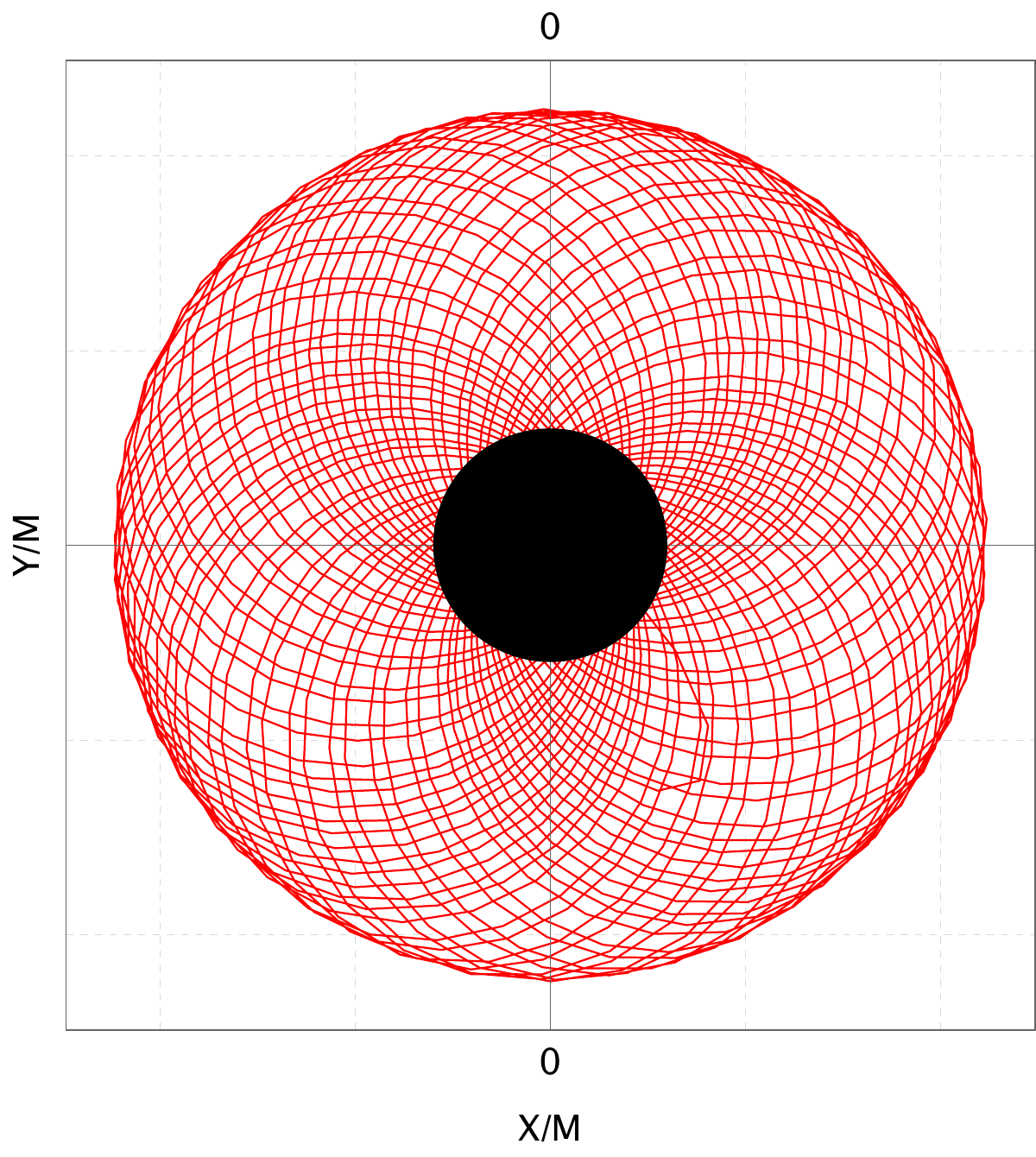}\\
		(i) $\ell=-0.2$ \hspace{4cm} (ii) $\ell=0.0$ \hspace{4cm} (iii) $\ell=0.2$\\
		\caption{Photon trajectories in the $(X, Y)$-plane for different values of the Lorentz-violating parameter $\ell$. The black hole parameters are fixed at $\lambda/M=-0.1$ and $Q/M=0.5$. The figure illustrates how variations in $\ell$ affect the bending of photon paths, showing that smaller values of $\ell$ increase the curvature of trajectories and bring unstable circular orbits closer to the black hole, thereby influencing the size and shape of the photon sphere and the corresponding black hole shadow.}
		\label{fig:trajectory-1}
	\end{figure}
	
	\begin{figure}[ht!]
		\centering
		\includegraphics[width=0.32\linewidth]{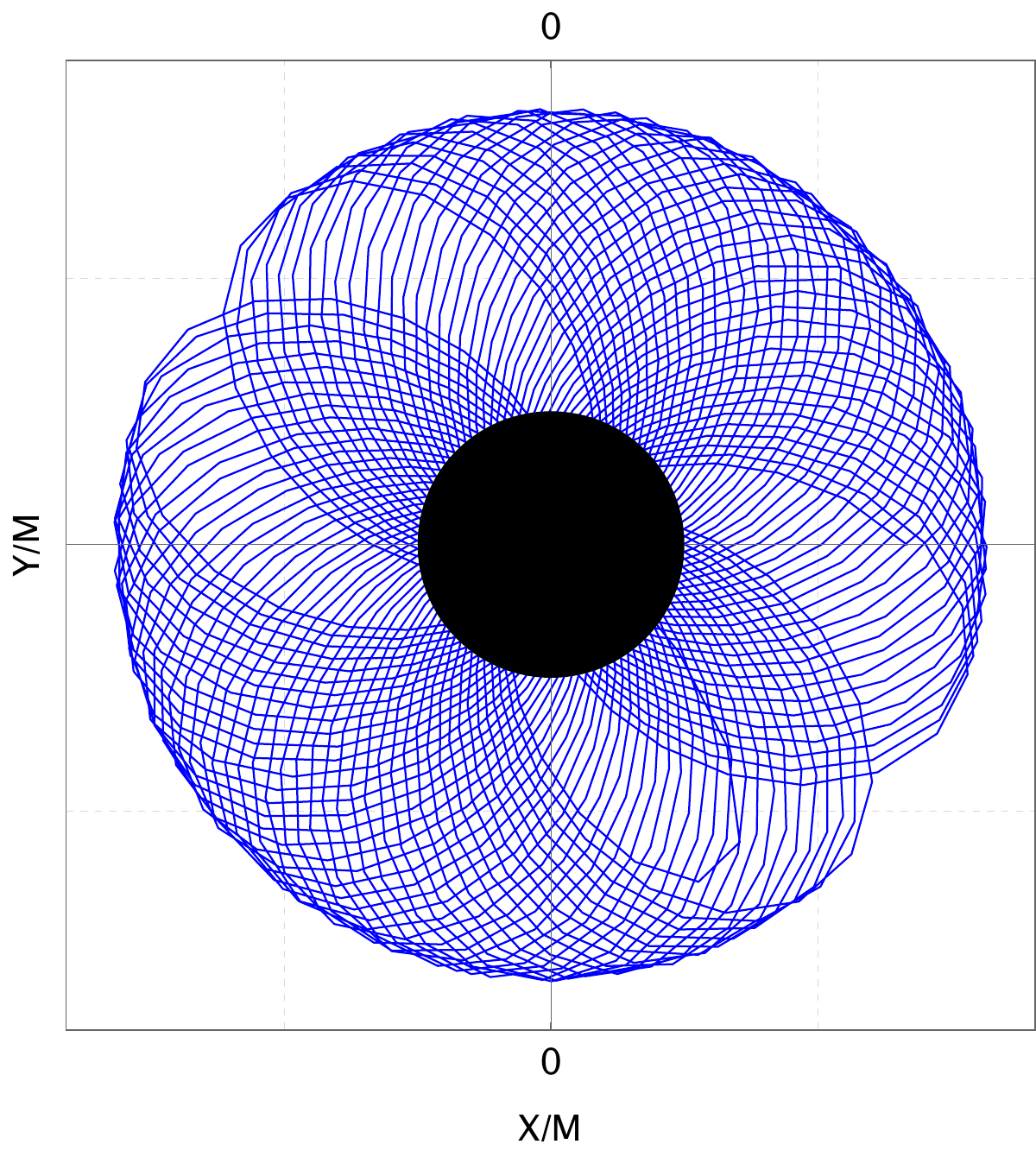}\quad
		\includegraphics[width=0.32\linewidth]{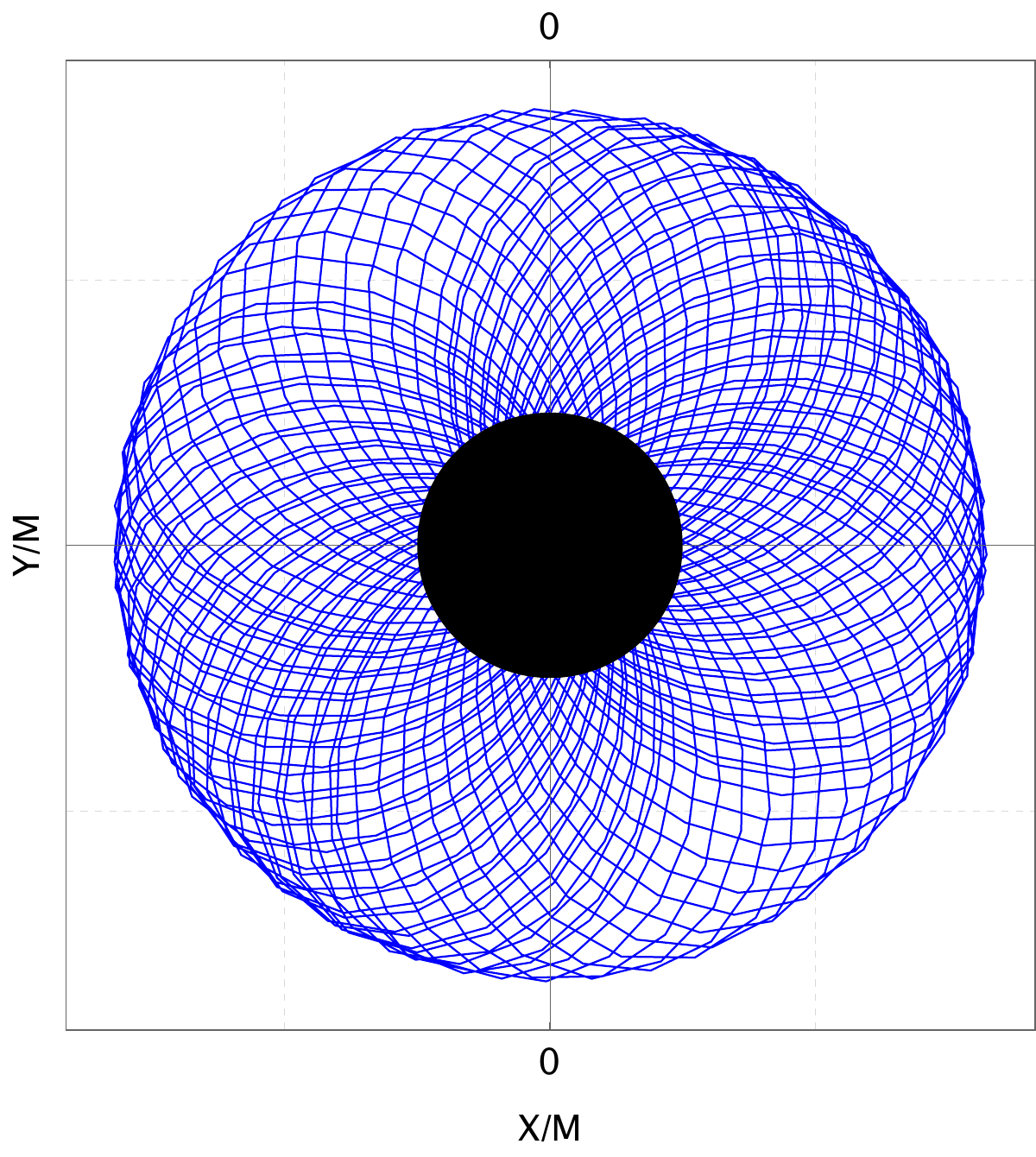}\quad
		\includegraphics[width=0.32\linewidth]{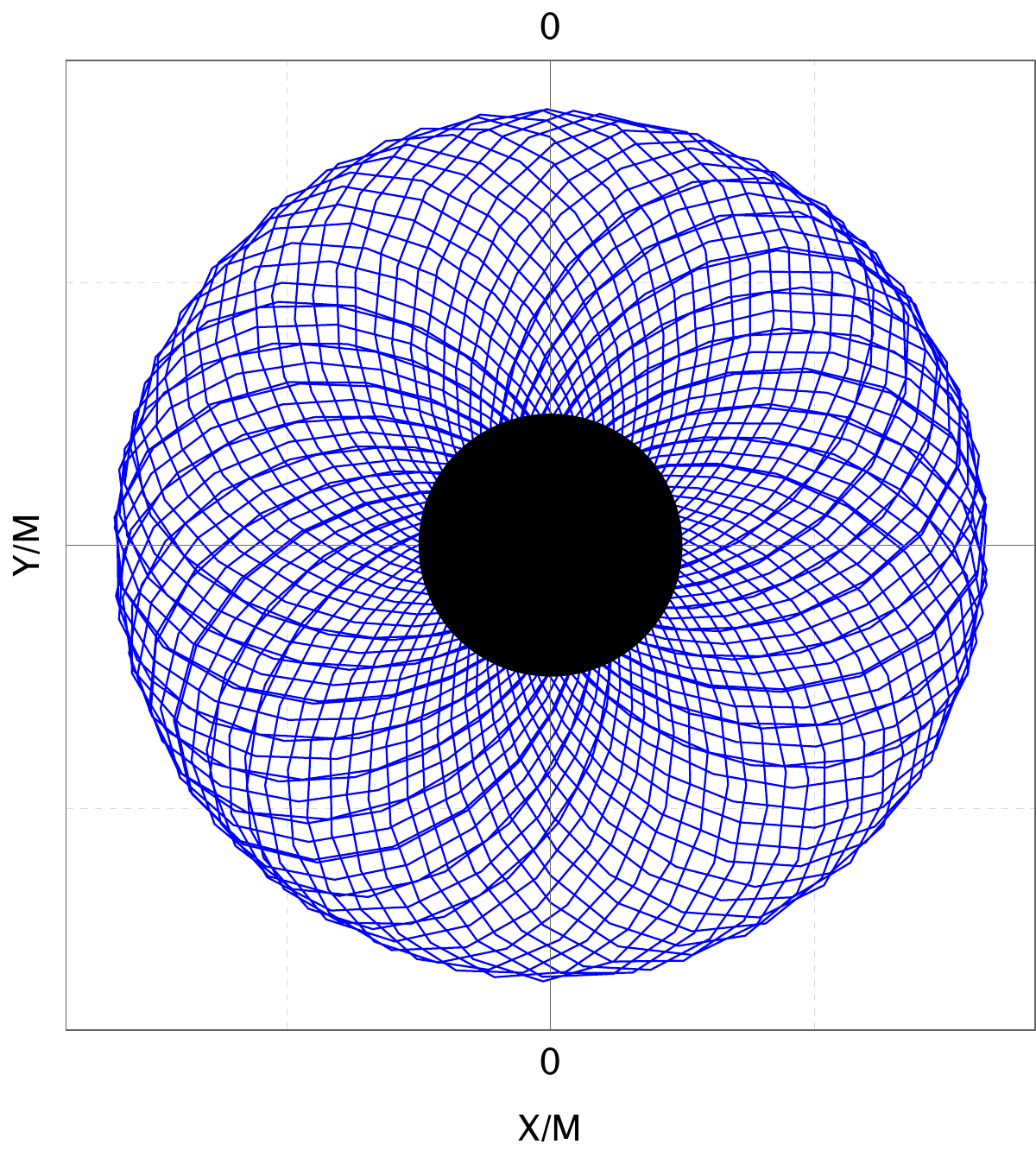}\\
		(i) $\lambda/M=-0.30$ \hspace{4cm} (ii) $\lambda/M=-0.35$ \hspace{4cm} (iii) $\lambda/M=-0.40$\\
		\caption{Photon trajectories in the $(X, Y)$-plane for different values of the PFDM parameter $\lambda$. The black hole parameters are fixed at $\ell=-0.2$ and $Q/M=0.5$. In contrast to Fig.~\ref{fig:trajectory-1}, the deformation of the photon paths shown here is entirely driven by changes in the dark-matter sector. The figure indicates that varying $\lambda$ modifies the bending pattern, shifts the location of unstable circular orbits, and consequently alters the optical appearance of the photon sphere and the black hole shadow.}
		\label{fig:trajectory-2}
	\end{figure}
	
	\begin{center}
		{\bf D.\, Photon Trajectories }
	\end{center}
	
	Photon trajectories describe the paths followed by light in curved spacetime, which are determined by the underlying gravitational field. Their behavior near a black hole reveals features such as the photon sphere, light bending, and the formation of the black hole shadow.
	
	The equation of orbit using Eqs.~(\ref{bb2}) and (\ref{bb4}) is given by
	\begin{equation}
		\left(\frac{1}{r^2}\frac{dr}{d\phi}\right)^2+\frac{1}{ (1-\ell) r^2}=\frac{1}{\beta^2}+\frac{2 M}{r^3}-\frac{Q^2}{(1-\ell)^2 r^4}-\frac{\lambda}{r^3}\ln\!\frac{r}{|\lambda|}.\label{ee1}
	\end{equation}
	Transforming to a new variable via $u(\phi)=\frac{1}{r(\phi)}$, we find
	\begin{equation}
		\left(\frac{du}{d\phi}\right)^2+\frac{u^2}{1-\ell}=\frac{1}{\beta^2}+2 M u^3-\frac{Q^2}{(1-\ell)^2} u^4+\lambda u^3 \ln\!\,(|\lambda| u).\label{ee2}
	\end{equation}
	
	Differentiating both sides w. r. to $\phi$ and after simplification results
	\begin{equation}
		\frac{d^2u}{d\phi^2}+\frac{u}{1-\ell}=\left(3 M+\frac{\lambda}{2}+\frac{3\lambda}{2} \ln(|\lambda| u)\right) u^2-\frac{2 Q^2}{(1-\ell)^2} u^3.\label{ee3}
	\end{equation}
	
	The above second-order differential equation represents the photon trajectory in the gravitational field of a charged black hole in the presence of Lorentz-violating effects and perfect fluid dark matter. It encodes how the combined influence of the electric charge $Q$, Lorentz-violating parameter $\ell$, and PFDM parameter $\lambda$ shapes the motion of photons around the black hole, including the locations of unstable circular orbits that define the photon sphere.
	
	In Fig.~\ref{fig:trajectory-1}, we depict the photon trajectories in the $(X, Y)$-plane for various values of the Lorentz-violating parameter $\ell$, while keeping the black hole parameters fixed at $\lambda/M = -0.1$ and $Q/M = 0.5$. The intricate pattern of the red curves illustrates how the bending of light is affected by changes in the Lorentz-violating effects.  
	
	Similarly, Fig.~\ref{fig:trajectory-2} shows the photon trajectories for different values of the PFDM parameter $\lambda$, with $\ell = -0.2$ and $Q/M = 0.5$ held constant. The complex pattern of the blue curves demonstrates the influence of perfect fluid dark matter on light deflection near the black hole. In both figures, the central black circle represents the black hole shadow, corresponding to photons that are captured by the event horizon.
	
	\section{Epicyclic Frequencies for Neutral Test Particles }
	
	Test particles orbiting a black hole in the equatorial plane ($\theta=\pi/2$) along stable circular trajectories undergo small oscillations when slightly perturbed from equilibrium. Considering small deviations of the form $r = r_c + \delta r$ and $\theta = \pi/2 + \delta \theta$, the motion can be described as harmonic oscillations in the radial and vertical (latitudinal) directions. The corresponding equations of motion take the form \cite{Torok2005,Stuchlik2013}
	\begin{equation}
		\frac{d^2 \delta r}{dt^2} + \omega_r^2 \, \delta r = 0, 
		\qquad 
		\frac{d^2 \delta \theta}{dt^2} + \omega_\theta^2 \, \delta \theta = 0,
		\label{ef1}
	\end{equation}
	where $\omega_r$ and $\omega_\theta$ denote the radial and vertical epicyclic frequencies, respectively.
	
	The orbital (azimuthal) frequency of the circular motion is given by
	\begin{equation}
		\omega_\phi=\omega_K = \dot{\phi} = \frac{\mathcal{L}}{g_{\theta\theta}},
		\label{ef2}
	\end{equation}
	where $\mathcal{L}$ is the conserved angular momentum per unit mass of test particles.
	
	The radial and latitudinal epicyclic frequencies can be derived using the Hamiltonian formalism \cite{Kolos2015,Tursunov2016,Stuchlik2016,Kolos2017,Stuchlik2021,Vrba2021}. The Hamiltonian for a test particle reads
	\begin{equation}
		H = \frac{1}{2} g^{\mu\nu} p_{\mu} p_{\nu} + \frac{m^{2}}{2}.
		\label{ef3}
	\end{equation}
	In our analysis, it is convenient to decompose the Hamiltonian into dynamic and potential parts as
	\begin{equation}
		H = H_{\mathrm{dyn}} + H_{\mathrm{pot}},
		\label{ef4}
	\end{equation}
	where
	\begin{align}
		H_{\mathrm{dyn}} &= \frac{1}{2} \left(f(r) p_{r}^{2} +\frac{p_{\theta}^{2}}{r^2} \right), \label{ef5} \\
		H_{\mathrm{pot}} &= \frac{1}{2} \left(-\frac{\mathcal{E}^{2}}{f(r)}+\frac{\mathcal{L}^{2}}{r^2 \sin^2 \theta} + 1 \right). \label{ef6}
	\end{align}
	
	The potential part of the Hamiltonian effectively determines the behavior of the epicyclic oscillations. The squared radial and latitudinal epicyclic frequencies are then obtained as
	\begin{align}
		\omega_r^{2} &= \frac{1}{g_{rr}} \frac{\partial^{2} H_{\mathrm{pot}}}{\partial r^{2}}\Big{|}_{r=r_c,\theta=\pi/2}, \\
		\omega_{\theta}^{2} &= \frac{1}{g_{\theta\theta}} \frac{\partial^{2} H_{\mathrm{pot}}}{\partial \theta^{2}}\Big{|}_{r=r_c,\theta=\pi/2}.
		\label{ef7}
	\end{align}
	
	For a distant observer, the corresponding observable frequencies are redshifted and can be expressed as \cite{Stella1998,Stella1999,Rezzolla2003}
	\begin{equation}
		\Omega_i = -\frac{\omega_i}{g^{tt}\,\mathcal{E}}=\frac{f(r)}{\mathcal{E}} \omega_i, \qquad (i =K, r, \theta),
		\label{ef8}
	\end{equation}
	where $\mathcal{E}=\frac{dt/d\tau}{f(r)}$ denotes the conserved energy per unit mass of the particles.
	
	In the stable circular orbits of fixed radii $r=r_c$, the specific energy and the specific angular momentum per unit mass of the neutral test particles are given by
	\begin{equation}
		\mathcal{L}_{\rm sp}=\sqrt{\frac{r^3\,f'(r)}{2\,f(r)-r \,f'(r)}},\qquad \mathcal{E}_{\rm sp}=\sqrt{\frac{2\,f^2(r)}{2\,f(r)-r \,f'(r)}}.
	\end{equation}
	
	For the spherically symmetric metric \eqref{metric}, the epicyclic frequencies are given by
	\begin{align}
		\Omega_K^2 &= \frac{f'(r)}{2\,r},\label{ef9}\\[4pt]
		\Omega_r^2 &= \frac{1}{2}\left[f(r)\,f''(r) - 2\left(f'(r)\right)^2 + \frac{3\,f(r)\,f'(r)}{r}\right],\label{ef10}\\[4pt]
		\Omega_\theta^2 &= \Omega_K^2,\label{ef11}
	\end{align}
	where the equality $\Omega_\theta = \Omega_K$ is a direct consequence of spherical symmetry. In physical units, the frequencies measured by a distant observer read
	\begin{equation}
		\nu_i = \frac{1}{2\pi}\,\frac{c^3}{G M_p}\,\Omega_i\;\;[\text{Hz}],\label{nuHz}
	\end{equation}
	where $i \in \{K, r, \theta\}$, $M_P$ is the physical mass related to the solar mass $M_\odot$ by $M_P = 10\,M_\odot$, $c = 3 \times 10^{8}\,\text{m/s}$ is the speed of light, and $G = 6.674 \times 10^{-11}\,\text{N\,m}^2\text{/kg}^2$ is the gravitational constant.
	
	The innermost stable circular orbit is determined by the marginal-stability condition $\Omega_r^2(r_{\rm ISCO})=0$, supplemented by the requirement $\Omega_r^2>0$ outside the orbit. In the present spacetime, $r_{\rm ISCO}$ encodes the combined influence of the charge, the Lorentz-violating KR parameter, and the PFDM contribution. Since the radial epicyclic frequency vanishes at the ISCO and becomes imaginary in the unstable region, the behavior of $\nu_r$ near its first zero provides a direct diagnostic of how the strong-field stability boundary is displaced by the parameters $\{Q,\ell,\lambda\}$.
	
	\begin{figure}[ht!]
		\centering
		\includegraphics[width=0.95\linewidth]{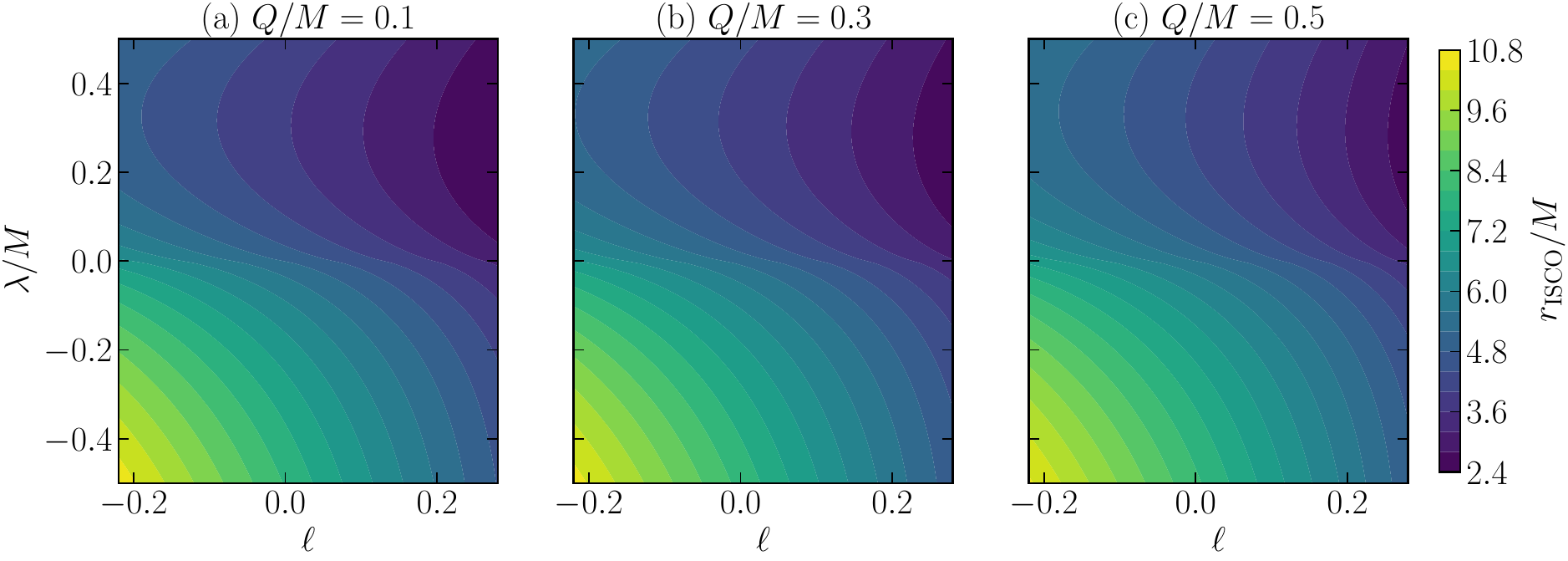}
		\caption{Contour maps of the ISCO radius $r_{\rm ISCO}/M$ in the $(\ell,\lambda/M)$ parameter space for three representative values of the charge, namely $Q/M=0.1$, $0.3$, and $0.5$.}
		\label{fig:isco_map}
	\end{figure}
	
	The contour maps displayed in Fig.~\ref{fig:isco_map} provide a compact global view of the stability boundary in the parameter plane. The figure shows that $r_{\rm ISCO}$ is systematically larger in the region of negative $\lambda$ and negative $\ell$, whereas it decreases as one moves toward positive values of these parameters. Hence, in the plotted domain, the PFDM contribution and the KR-induced Lorentz-violating sector act together to shift the innermost stable orbit inward or outward, depending on the sign of the deformation. The dependence on the electric charge is milder than the one on $\lambda$ and $\ell$, but it remains visible through the gradual reshaping of the contour levels from the left to the right panel. From an astrophysical point of view, this figure is particularly useful because it shows at a glance which sectors of the parameter space favor higher orbital frequencies via smaller ISCO radii, and which sectors suppress them by pushing the stability boundary to larger distances.
	
\begin{figure}[ht!]
		\centering
		\includegraphics[width=5.4cm]{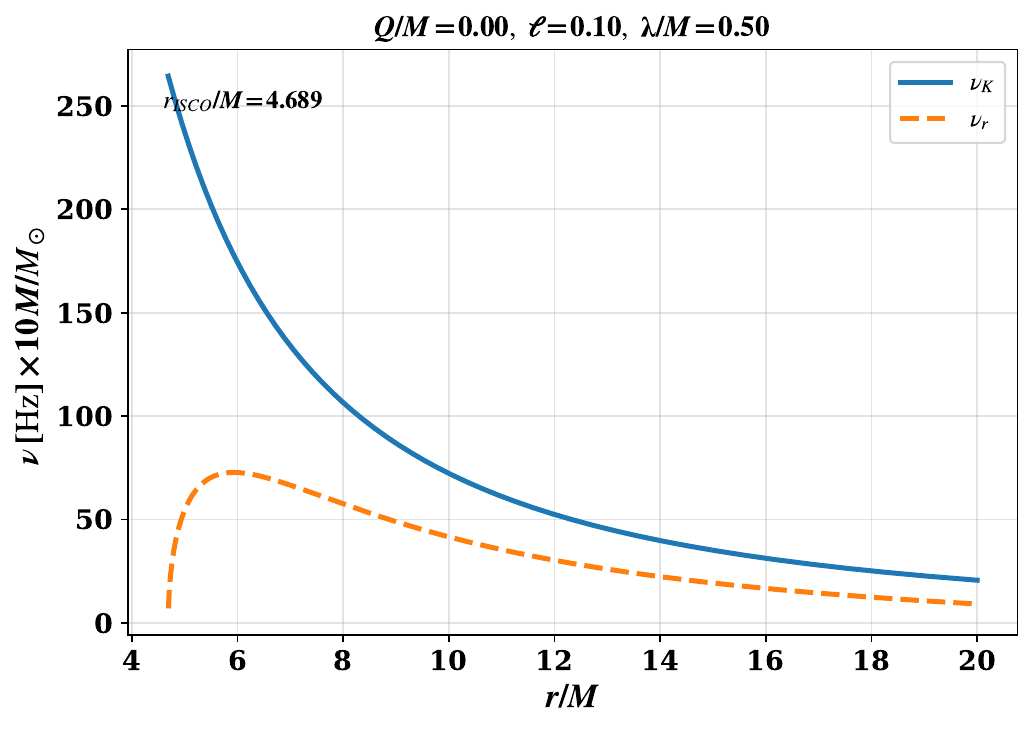}\quad
		\includegraphics[width=5.4cm]{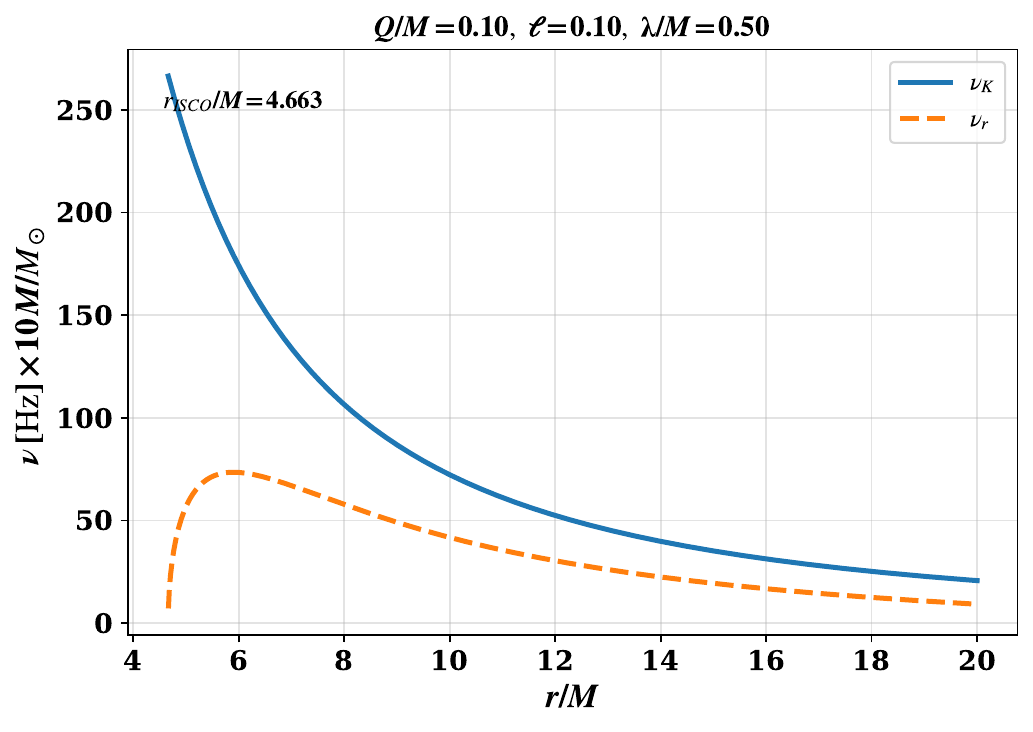}\quad
		\includegraphics[width=5.4cm]{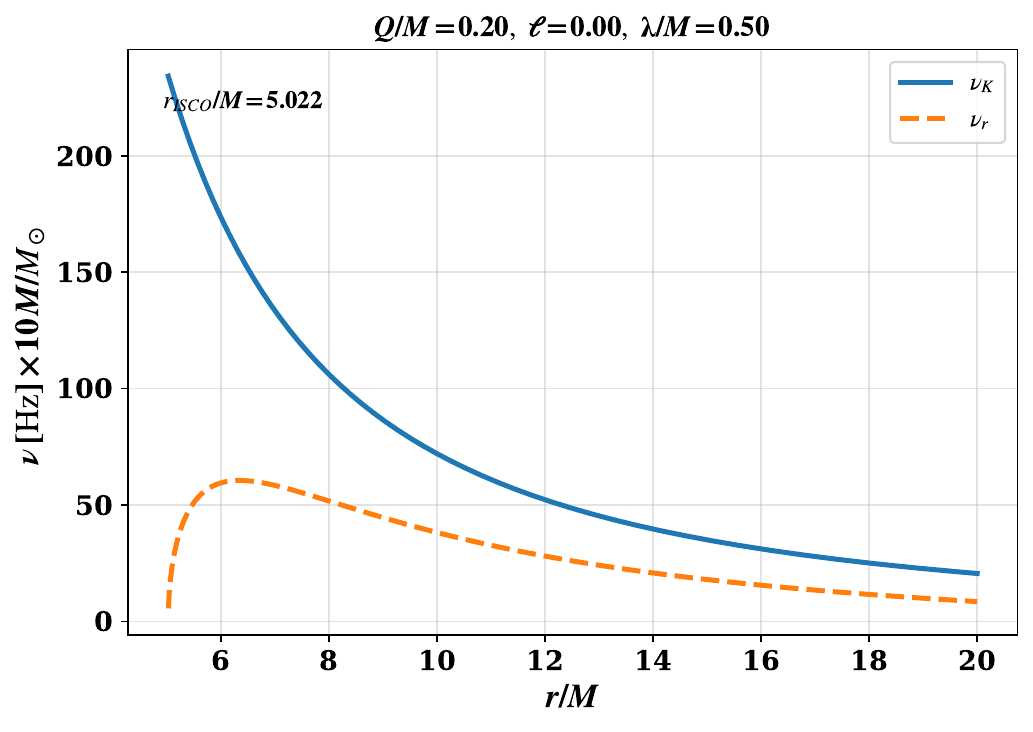}\\
		\includegraphics[width=5.4cm]{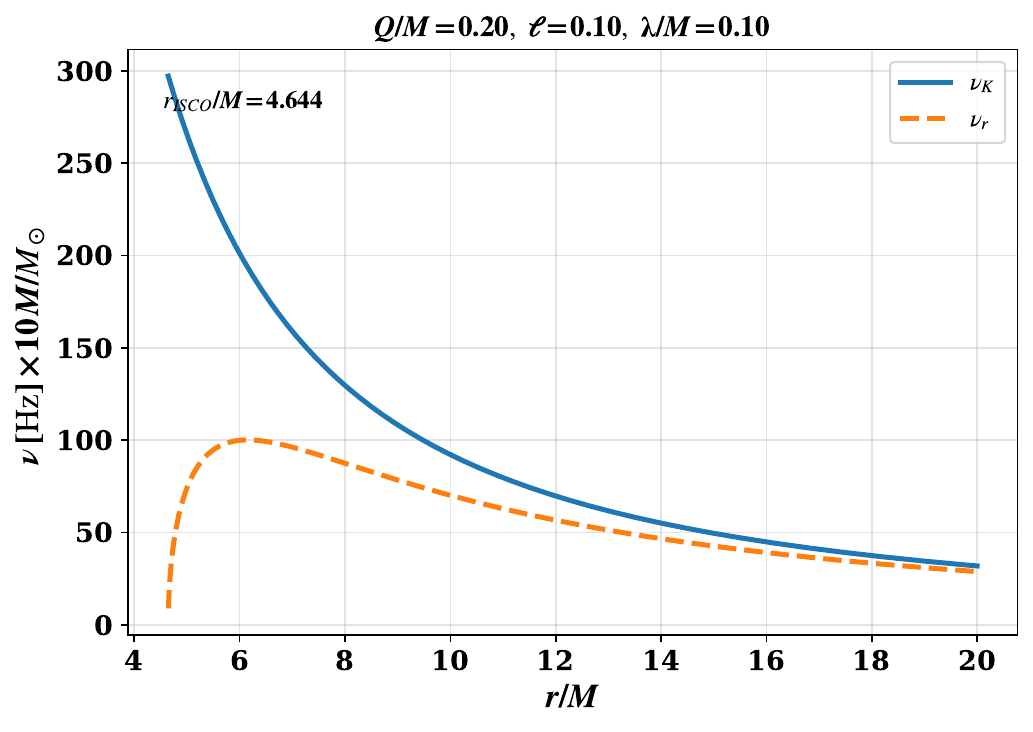}\quad
		\includegraphics[width=5.4cm]{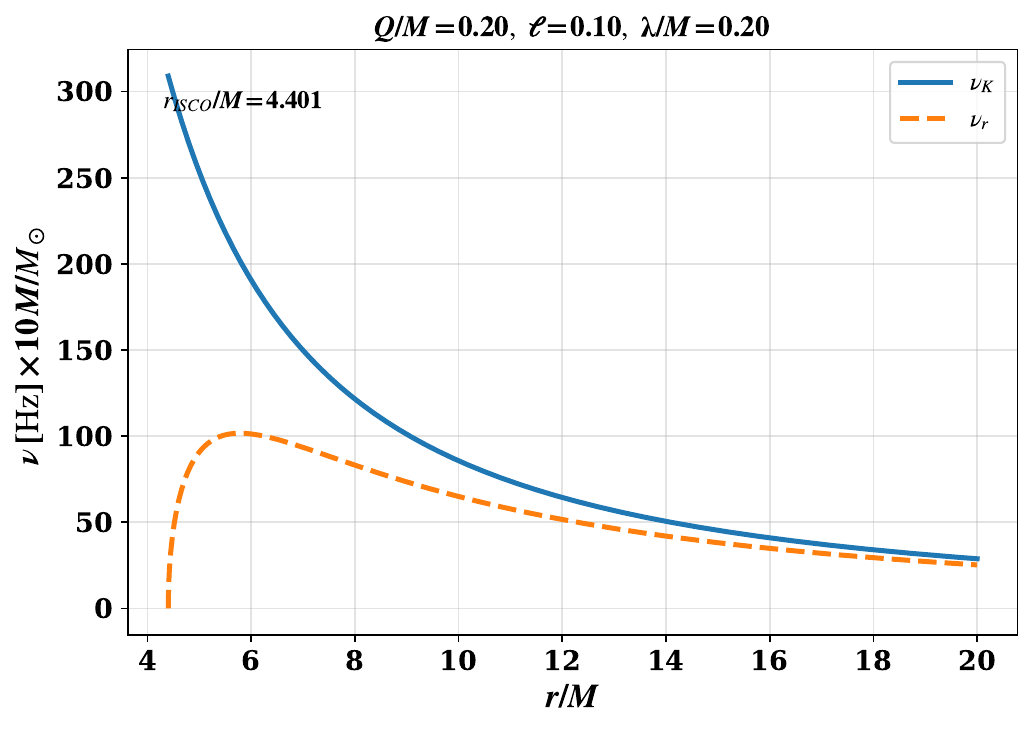}\quad
		\includegraphics[width=5.4cm]{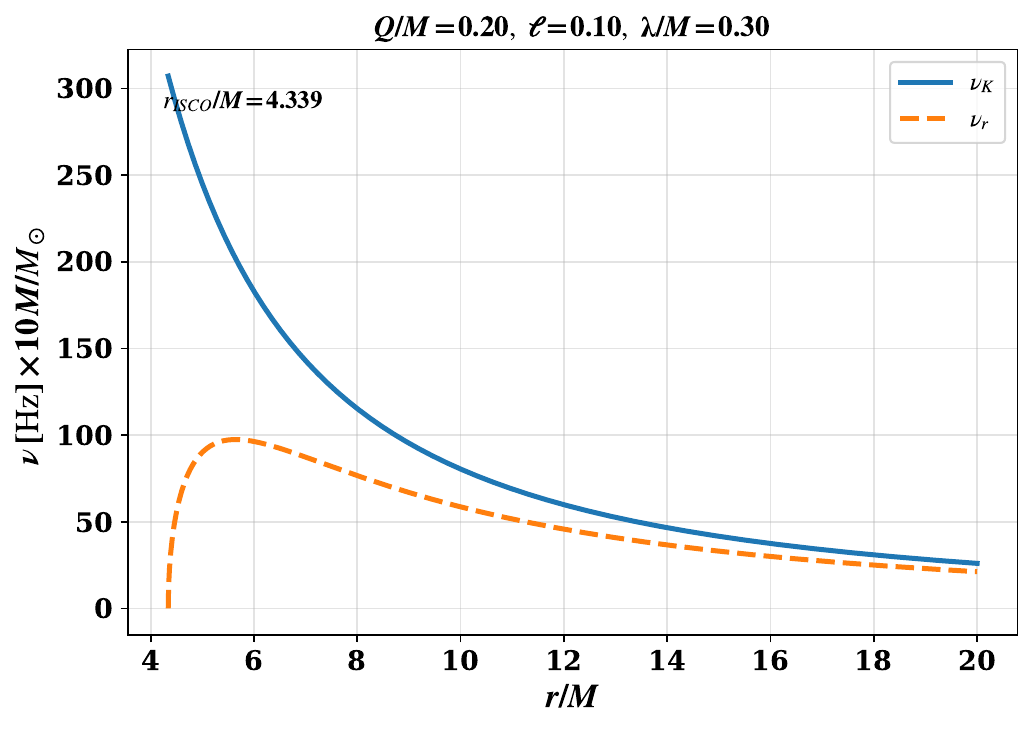}\\
		\includegraphics[width=5.4cm]{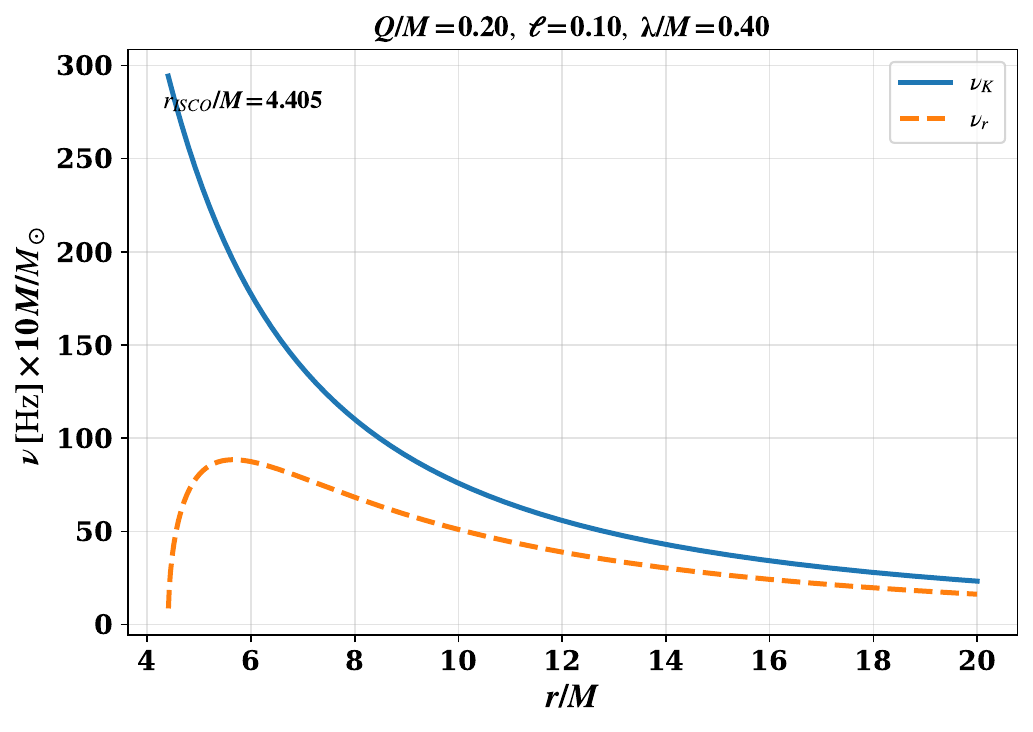}\quad
		\includegraphics[width=5.4cm]{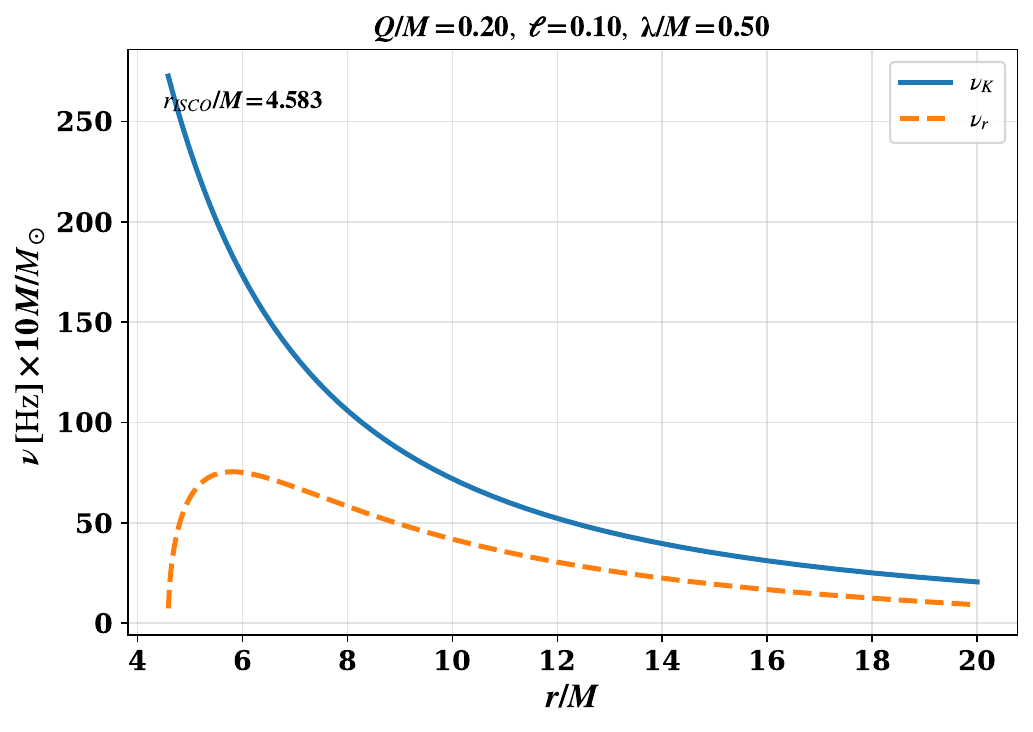}\quad
		\includegraphics[width=5.4cm]{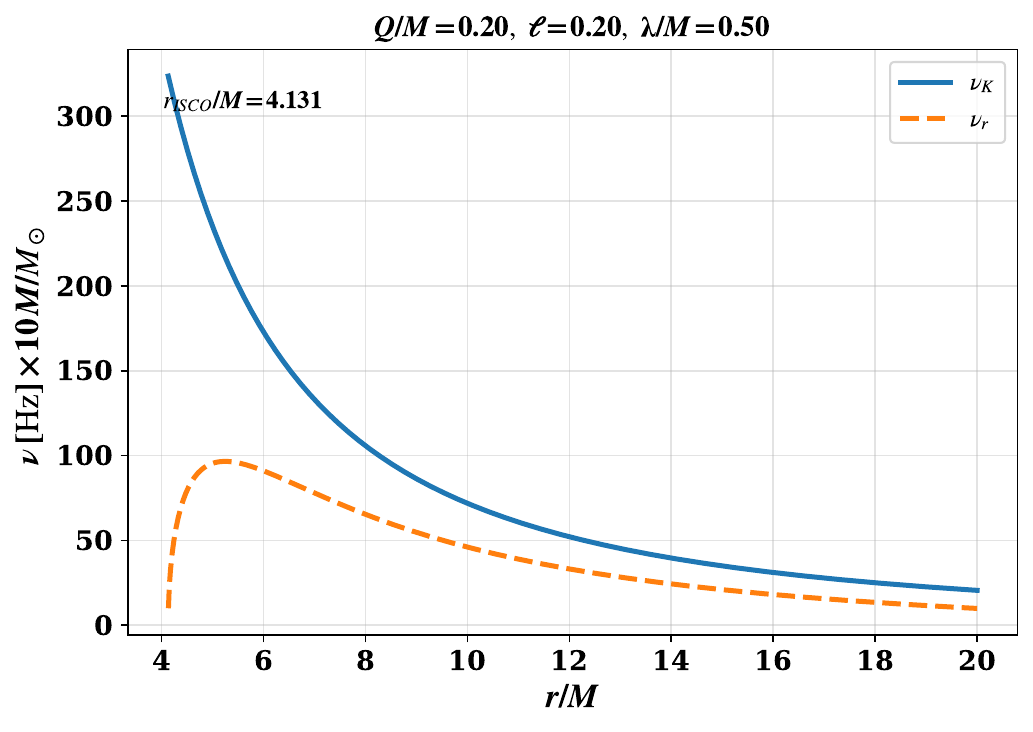}\\
		\includegraphics[width=5.4cm]{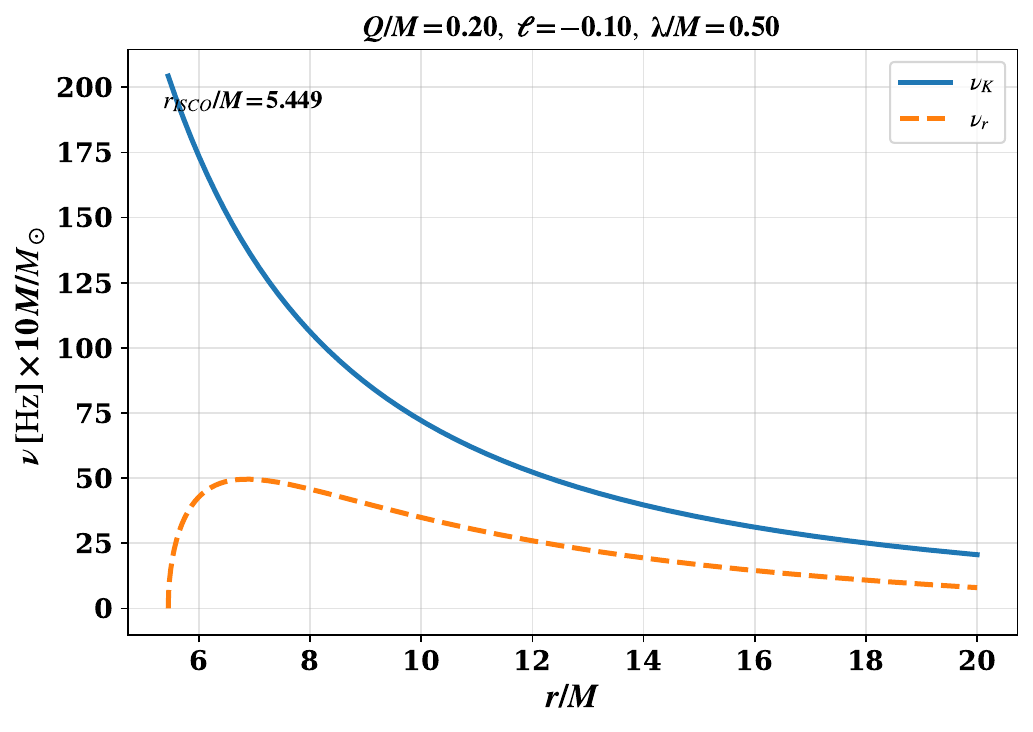}\quad
		\includegraphics[width=5.4cm]{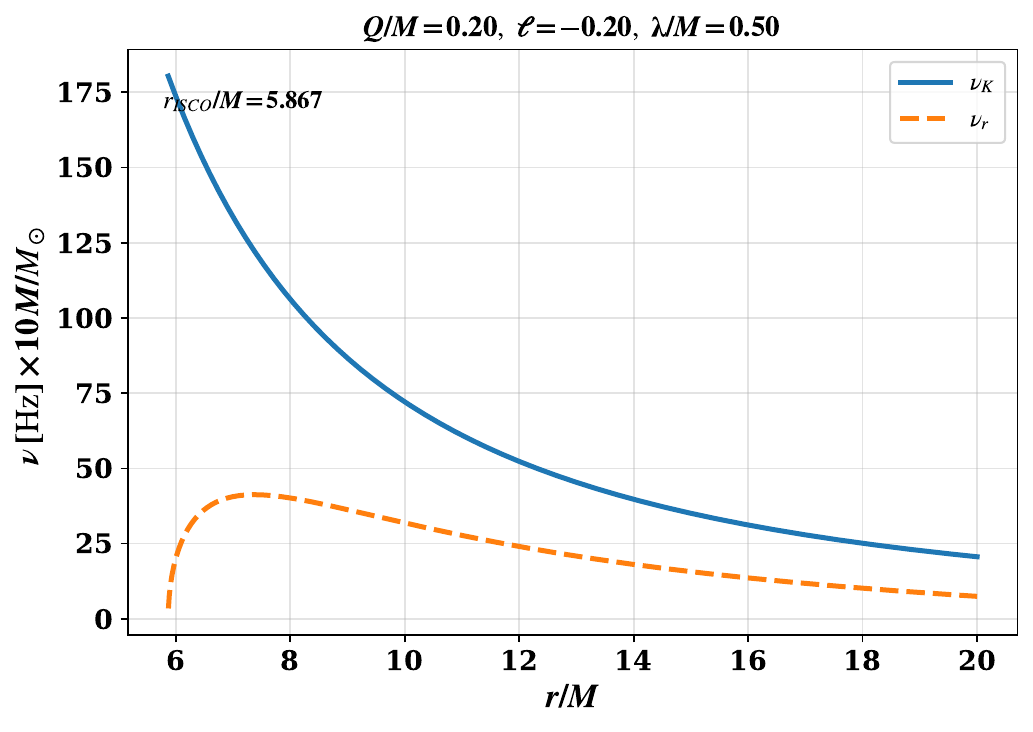}\quad
		\includegraphics[width=5.4cm]{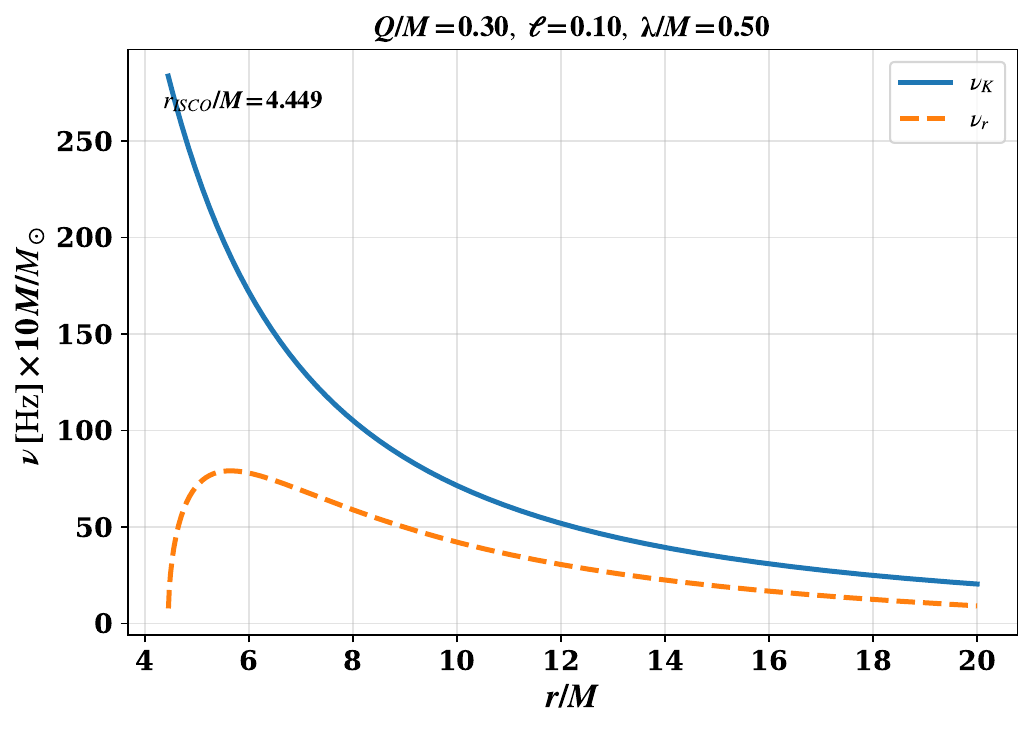}
		\caption{The radial profiles of $\nu_r$ and $\nu_K$, plotted for various values of the spacetime parameters.}
		\label{fig:nu_rnu_K}
	\end{figure}

	In Fig. \ref{fig:nu_rnu_K}, we have plotted the radial profiles of the above epicyclic frequencies for different values of the black hole parameters.
	
	Figure~\ref{fig:nu_rnu_K} shows that the Keplerian frequency $\nu_K$ monotonically decreases with radius, while the radial epicyclic frequency $\nu_r$ rises from zero at the ISCO, reaches a maximum, and then gradually decays at large distances. The displacement of the left endpoint of each curve reflects the sensitivity of the ISCO to the spacetime parameters. In particular, the panels indicate that the combined action of $Q$, $\ell$, and $\lambda$ can shift the stability boundary inward or outward, thereby changing the range of admissible orbital frequencies available to QPO models. This behavior is especially relevant for the phenomenological interpretation of twin-peak QPOs, since the observed upper and lower frequencies are constructed from these fundamental geodesic frequencies.
	
	%%%%%%%%%%%%%
	\subsection{The QPO models}\label{subsec:QPO_models}
	
	The electromagnetic radiation coming from accretion disks around BHs is usually connected with the oscillatory motion of particles in the gravitational background. In particular, charged particles can emit radiation at frequencies matching their characteristic oscillation frequencies around the compact object. Because of this, the dynamics of test particles in BH spacetimes gives a natural setup for explaining QPOs in terms of quasi-harmonic oscillations in the radial and angular directions. Many QPO signals have been detected in such systems, and their frequencies have been measured with high precision. Nevertheless, there is still no single mechanism that is accepted universally as the definite origin of these oscillations. This issue is still under active investigation, especially when one wants to test gravity models or study the position of the inner edge of the accretion disk, which is strongly connected with the ISCO radius \cite{rayimbaev_quasiperiodic_2022,JR2022}. Therefore, the analysis of QPOs can be considered as an important tool for constraining the characteristic radii of the innermost part of the disk, estimating the mass of the central BH, and exploring gravitational effects in the strong-field regime.
	
\begin{figure}[ht!]
		\centering
		\includegraphics[width=1.0\textwidth]{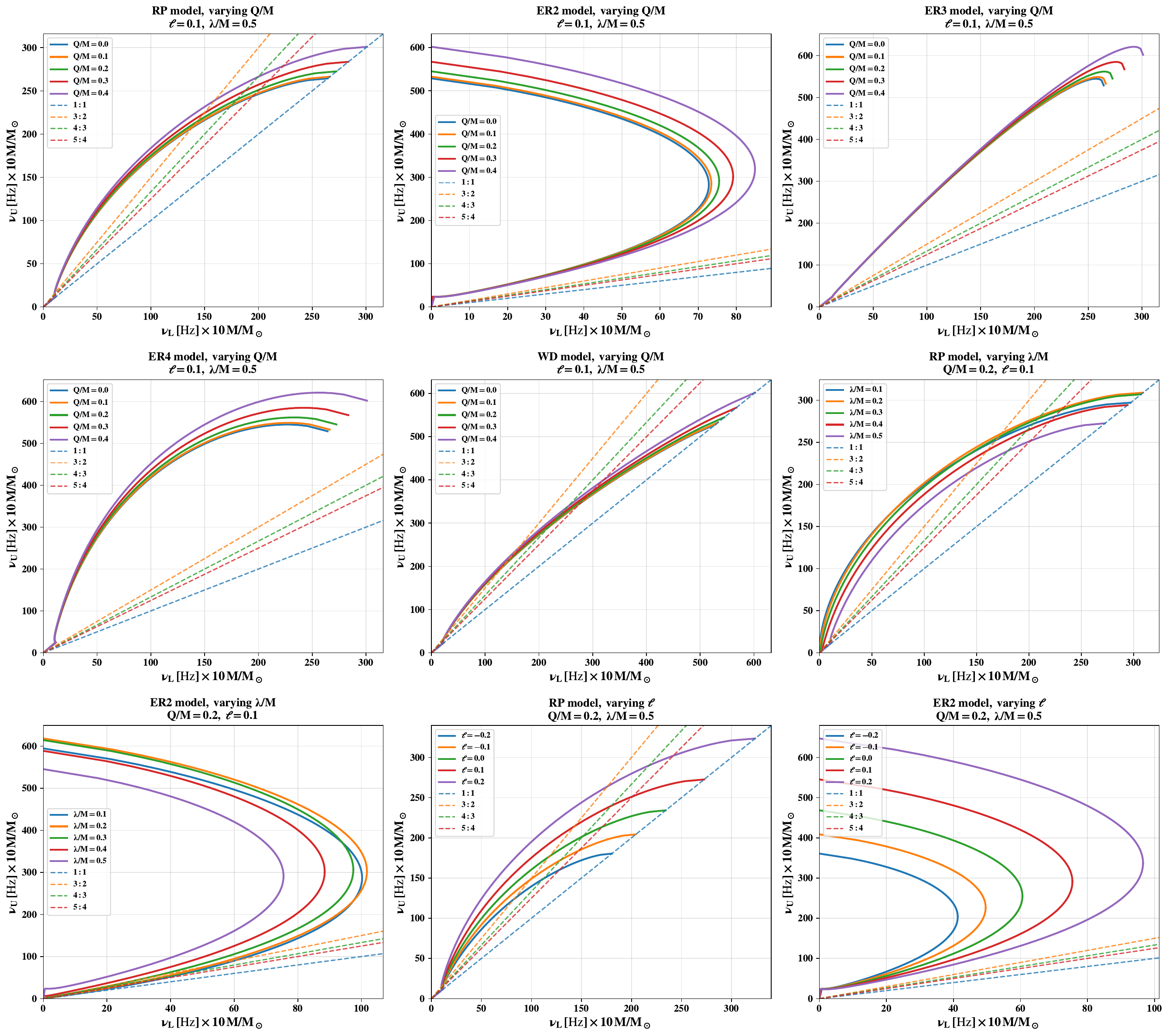}
		\caption{Relations between the upper and lower frequencies of twin-peaked QPOs in the RP, ER2--ER4, and WD models around the black hole, for several representative values of $Q/M$, for fixed $\ell$ and $\lambda/M$ for each case.}
		\label{fig:nuUnuL}
	\end{figure}

	In this subsection, we study the relation between the upper and lower frequencies of twin-peaked QPOs around charged KR black holes in PFDM, and we also compare these results with the corresponding frequency relations obtained for neutral particles orbiting BHs \cite{stuchlik_models_2016}. For this purpose, we use several phenomenological models that can be directly compared with astrophysical observations, following the approaches discussed in Ref.~\cite{shahzadi_epicyclic_2021}. These models give practical ways to interpret QPO data in terms of the fundamental frequencies of motion. More specifically, the models considered include the relativistic precession (RP) model and its invariant version RP$\overline{1,2}$; the epicyclic resonance (ER) models, together with the invariant ER$\overline{1,5}$; and the tidal disruption (TD) and warped disk (WD) models. In this subsection, however, we focus on deriving the upper and lower QPO frequencies in the RP, ER, and WD models. Below, we briefly summarize how these models are written in terms of the fundamental frequencies introduced in the previous subsection, and then we present a graphical discussion of the corresponding upper-lower frequency relations:
	\begin{enumerate}
		
		\item[(i)] In the RP model, the twin-peaked QPO frequencies are related to the orbital motion and the radial epicyclic oscillation. In this case, the upper and lower frequencies are identified as $\nu_U = \nu_K$ and $\nu_L = \nu_K - \nu_r$, respectively \cite{stella_correlations_1999}.
		
		\item[(ii)] In the ER models, the accretion flow is assumed to be thick enough so that resonant oscillations can appear along the geodesic motion of radiating particles. Then, the observed QPOs are interpreted through different combinations of oscillation modes. Here, we take three representative examples, namely ER2, ER3, and ER4, which differ from one another in how the oscillation modes are coupled. For these models, the frequency pairs are given by $(\nu_U, \nu_L) = (2\nu_K - \nu_r,\, \nu_r)$, $(\nu_U, \nu_L) = (2\nu_K + \nu_r,\, \nu_K)$, and $(\nu_U, \nu_L) = (2\nu_K + \nu_r,\, \nu_K - \nu_r)$, respectively \cite{abramowicz_precise_2001}.
		
		\item[(iii)] In the WD model, the QPOs are assumed to come from the oscillatory motion of test particles in a thin accretion disk. In this framework, the upper and lower frequencies take the form $\nu_U = 2\nu_K - \nu_r$ and $\nu_L = 2(\nu_K - \nu_r)$, respectively \cite{kato_resonant_2004,kato_frequency_2008}.
		
	\end{enumerate}
	In Fig.~\ref{fig:nuUnuL}, we show the correlated behavior between the upper and lower frequencies of twin-peaked QPOs in the black hole spacetime. The frequencies are plotted for orbital radii extending from the ISCO up to infinity, and for several representative choices of the black hole parameter $Q$. The plots also contain a set of straight inclined lines corresponding to the specific frequency ratios $\nu_U\!:\!\nu_L = 3\!:\!2$, $4\!:\!3$, $5\!:\!4$, and $1\!:\!1$. These lines represent the possible combinations of upper and lower frequencies that satisfy each ratio \cite{kolos_possible_2017,kolos_quasi-periodic_2020,shahzadi_epicyclic_2021,Shahzadi2024}. In particular, the $1\!:\!1$ line, which is often called the ``graveyard'' of twin-peaked QPOs, indicates the region where the two frequencies become equal, meaning that the system effectively shows only one QPO peak.
	The fact that some of the QPO curves do not intersect the $1\!:\!1$ line is a direct consequence of the kinematical definitions of the corresponding models. Recall that, the $1\!:\!1$ line corresponds to the condition $\nu_U=\nu_L$. For the RP and WD models, this condition reduces to $\nu_r=0$, which is naturally satisfied at the ISCO, so these curves terminate on the $1\!:\!1$ line at a finite radius. By contrast, in the ER2 model, the equality $\nu_U=\nu_L$ implies $\nu_K=\nu_r$, which is approached only asymptotically as $r\to\infty$, so the corresponding curve tends to the $1\!:\!1$ line only at the origin. For the ER3 model, the condition $\nu_U=\nu_L$ gives $\nu_K+\nu_r=0$, while for ER4 it yields $\nu_K+2\nu_r=0$. Since both $\nu_K$ and $\nu_r$ are non-negative in the physically relevant region of stable circular motion, these equations cannot be satisfied at any finite radius. Therefore, the absence of an intersection of the ER3 and ER4 curves with the $1\!:\!1$ line is not a numerical artifact, but rather an intrinsic property of the underlying QPO prescriptions.
	
	From a phenomenological perspective, Fig.~\ref{fig:nuUnuL} makes clear that different sectors of the parameter space lead to systematically different frequency--frequency tracks. This is useful because it shows that the charged KR+PFDM background does not simply rescale the standard geodesic models, but can also reshape the curvature and endpoint structure of the QPO relations. In practice, this means that observational QPO data can be used not only to estimate the characteristic orbital radius but also to constrain the amount of room left for Lorentz-violating and dark-matter effects in the strong-field region.
	
	\subsection{Black hole parameter estimation from QPO data}\label{subsec:MCMC}
	
	In this subsection, we constrain the parameters governing test-particle motion around the charged KR black hole immersed in PFDM by confronting the model with QPO observations from black hole candidates spanning stellar-mass, intermediate-mass, and supermassive scales.
	
	For the stellar-mass class, we consider the compact objects in the microquasars GRO~J1655-40 and XTE~J1550-564. As an intermediate-mass black hole candidate, we adopt M82~X-1, an ultraluminous X-ray source in the galaxy M82 whose estimated mass lies in the range $\sim 10^{2}$--$10^{3}\,M_\odot$ \cite{fiorito_is_2004,torok_possible_2005,stuchlik_mass_2015}. We also include the supermassive black hole Sgr~A$^\ast$ at the Galactic center, which is of particular interest in the context of millihertz QPOs. The QPO data employed in the analysis are summarized in Table~\ref{tab:1}.
	\begin{table*}
		\centering
		\begin{tabular}{c|ccccc}
			source  & $\nu_U$ [Hz] & $\Delta\nu_U$ [Hz] & $\nu_L$ [Hz] & $\Delta\nu_L$ [Hz] & mass [$M_\odot$]\\
			\hline
			XTE J1550-564 \cite{orosz_improved_2011}& 276 & $\pm3$ & 184 & $\pm5$ & $9.1\pm0.61$ \\
			GRO J1655-40 \cite{strohmayer_discovery_2001} & 451 & $\pm5$ & 298  & $\pm4$ & $5.4\pm0.3$ \\
			M82 X-1 \cite{pasham_400-solar-mass_2014} & 5.07 & $\pm0.06$ & 3.32 & $\pm0.06$ & $415\pm63$ \\
			Sgr A* \cite{ghez_measuring_2008} & $1.445\times 10^{-3}$ &  $\pm0.16\times 10^{-3}$ & $0.886\times10^{-3}$ & $\pm0.04\times10^{-3}$ & $(4.1\pm0.6)\times 10^6$ \\
		\end{tabular}
		\caption{Twin-peak QPO frequencies for the black hole candidates considered in this work.}
		\label{tab:1}
	\end{table*}
	To infer the parameters of the charged KR+PFDM black hole, we perform a Markov chain Monte Carlo (MCMC) analysis using the Python package \texttt{emcee} \cite{ascl,zhadyranova_exploring_2024,abdulkhamidov_parameter_2024}. Within the Bayesian framework, the posterior probability distribution is given by \cite{liu_constraints_2023,mitra_charged_2024}
	\begin{equation}
		\mathcal{P}(\theta|\mathcal{M}) = \frac{P(\mathcal{D}|\theta,\mathcal{M})\,\pi(\theta|\mathcal{M})}{P(\mathcal{D}|\mathcal{M})},
		\label{eq:posterior_0}
	\end{equation}
	where $\pi(\theta|\mathcal{M})$ and $P(\mathcal{D}|\theta,\mathcal{M})$ denote the prior and likelihood functions, respectively.
	
	For the parameter vector $\theta_i=\{r/M,\ell,M/M_\odot,Q/M,\lambda/M\}$, 
	the prior for each parameter is taken as
	\begin{equation}
		\pi(\theta_i)\propto \exp\left[-\frac{1}{2}\left(\frac{\theta_i-\theta_{0,i}}{\sigma_i}\right)^2\right],
		\qquad
		\theta_{\mathrm{low},i}<\theta_i<\theta_{\mathrm{high},i},
		\label{eq:pi_i}
	\end{equation}
	where $\theta_{0,i}$ and $\sigma_i$ are the corresponding mean values and standard deviations.
	
	The likelihood function is constructed from two observational inputs associated with the upper and lower QPO frequencies, so that the total log-likelihood reads
	\begin{equation}
		\ln\mathfrak{L} = \ln\mathfrak{L}_U + \ln\mathfrak{L}_L,
		\label{eq:like_0}
	\end{equation}
	with
	\begin{subequations}
		\begin{align}
			\ln\mathfrak{L}_U &= -\frac{1}{2}\sum_j\left(\frac{\nu_{\phi,\mathrm{obs}}^j - \nu_{\phi,\mathrm{theo}}^j}{\sigma_{\phi,\mathrm{obs}}^j}\right)^2,\\
			\ln\mathfrak{L}_L &= -\frac{1}{2}\sum_j\left(\frac{\nu_{\mathrm{per},\mathrm{obs}}^j - \nu_{\mathrm{per},\mathrm{theo}}^j}{\sigma_{\mathrm{per},\mathrm{obs}}^j}\right)^2.
		\end{align}
	\end{subequations}
	Here, $\ln\mathfrak{L}_U$ and $\ln\mathfrak{L}_L$ correspond to the upper and lower QPO likelihoods, associated with the orbital frequency $\nu_{\phi}\equiv\nu_K$ and the periapsis precession frequency $\nu_{\mathrm{per}}=\nu_K-\nu_r$, respectively. The quantities $(\nu_{\phi,\mathrm{obs}}^j,\nu_{\mathrm{per},\mathrm{obs}}^j)$ represent the observed frequencies, whereas $(\nu_{\phi,\mathrm{theo}}^j,\nu_{\mathrm{per},\mathrm{theo}}^j)$ denote the theoretical predictions of the model.
	
	We then estimate the parameter set $\{r/M,\,\ell,\,M/M_\odot,\,Q/M,\,\lambda/M\}$ for the charged KR black hole in PFDM by sampling the parameter space with Gaussian priors constructed from the observational properties of the QPO sources listed in Table~\ref{tab:1}. The analysis employs chains containing of the order of $10^5$ samples, which allows an efficient exploration of the physically admissible region of the parameter space and the identification of the parameter values that best reproduce the observational data. The resulting best-fit values are presented in Table~\ref{tab:posterior_constraints_rp}.
	\begin{table*}[t]
		\centering
		\renewcommand{\arraystretch}{1.15}
		\begin{tabular}{lccccc}
			\hline
			Source & $r/M$ & $\ell$ & $M/M_\odot$ & $Q/M$ & $\lambda/M$ \\
			\hline
			XTE J1550-564 
			& $4.93175^{+0.20914}_{-0.17779}$ 
			& $0.08526^{+0.04837}_{-0.04771}$ 
			& $9.05122^{+0.45136}_{-0.44632}$ 
			& $0.25418^{+0.16773}_{-0.16900}$ 
			& $0.30137^{+0.13308}_{-0.13113}$ \\
			
			GRO J1655-40 
			& $5.02968^{+0.20876}_{-0.18768}$ 
			& $0.07282^{+0.04969}_{-0.04607}$ 
			& $5.38415^{+0.26487}_{-0.27968}$ 
			& $0.24652^{+0.16985}_{-0.16629}$ 
			& $0.30227^{+0.13038}_{-0.12773}$ \\
			
			M82 X-1 
			& $5.51492^{+0.24421}_{-0.21414}$ 
			& $-0.00346^{+0.05449}_{-0.04684}$ 
			& $413.477^{+19.7937}_{-19.9262}$ 
			& $0.24843^{+0.17111}_{-0.17243}$ 
			& $0.30964^{+0.12881}_{-0.14237}$ \\
			
			Sgr A$^\ast$ 
			& $3.59485^{+0.20231}_{-0.14722}$ 
			& $0.18385^{+0.01215}_{-0.02352}$ 
			& $(3.8984^{+0.1980}_{-0.1986})\times 10^{6}$ 
			& $0.45839^{+0.03123}_{-0.06966}$ 
			& $0.27209^{+0.10085}_{-0.09352}$ \\
			\hline
		\end{tabular}
		\caption{Posterior constraints on the charged KR+PFDM black hole parameters obtained from the RP-model MCMC analysis for the four HF QPO sources. The quoted uncertainties correspond to the $16$th and $84$th percentiles of the posterior distributions.}
		\label{tab:posterior_constraints_rp}
	\end{table*}
	It is worth emphasizing that, although Bayesian inference and MCMC techniques are particularly powerful when applied to large observational datasets, the presently available sample of high-frequency QPO measurements remains limited, as noted in earlier studies. Therefore, our goal is not to derive highly restrictive statistical bounds, but rather to identify the regions of parameter space in which the model is capable of reproducing the observed QPO frequencies. The constraints obtained here should thus be regarded as preliminary and model dependent, rather than as definitive statistical determinations. A more robust assessment will require a substantially larger set of observational data in the future; see, for example, Ref.~\cite{Stuchlik2021}.
	
\begin{figure*}[ht!]
		\centering
		
		\begin{minipage}{0.47\textwidth}
			\centering
			\begin{overpic}[width=\linewidth]{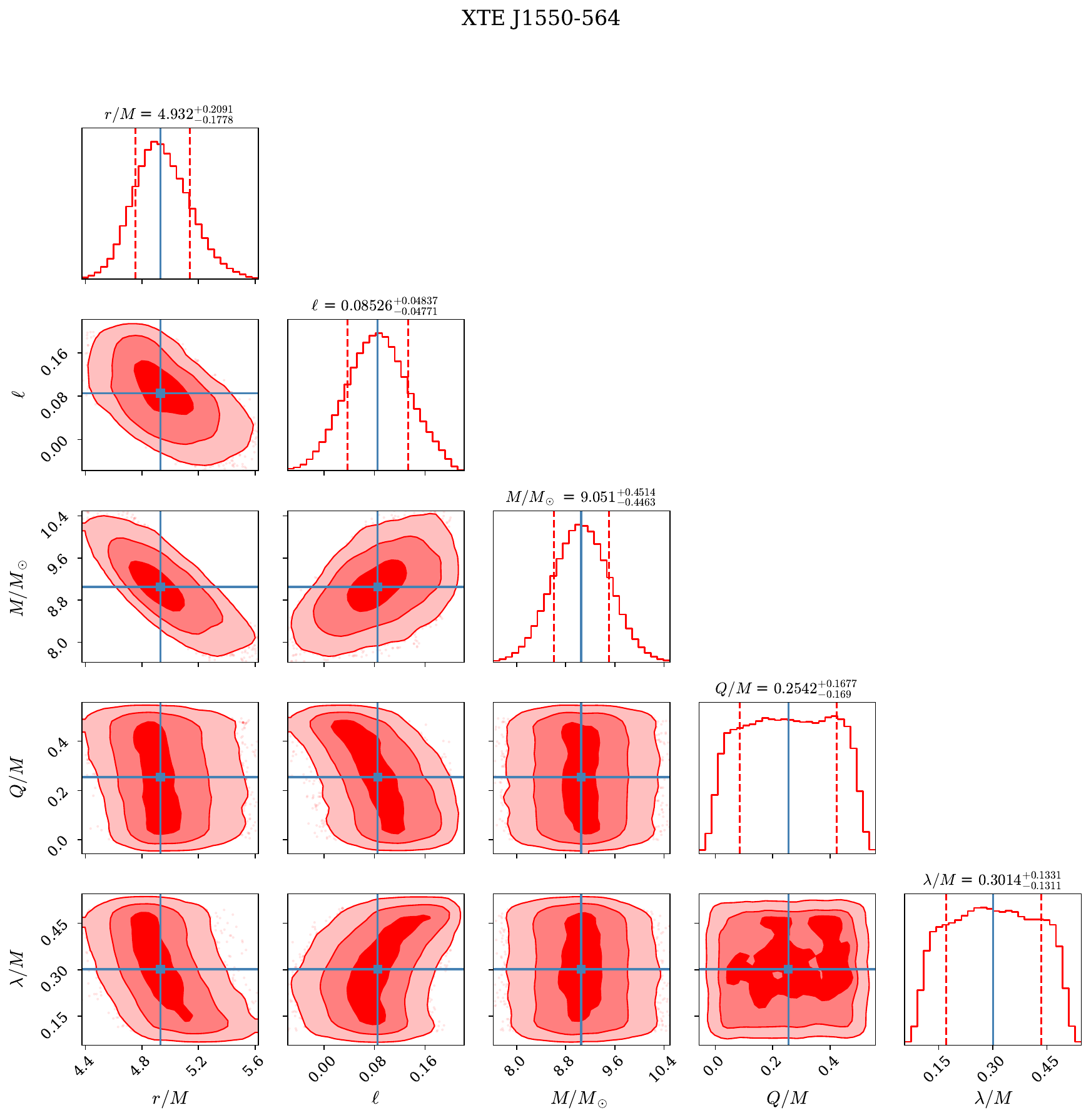}
				\put(4,92){\large \bfseries (a)}
			\end{overpic}
		\end{minipage}
		\hfill
		\begin{minipage}{0.47\textwidth}
			\centering
			\begin{overpic}[width=\linewidth]{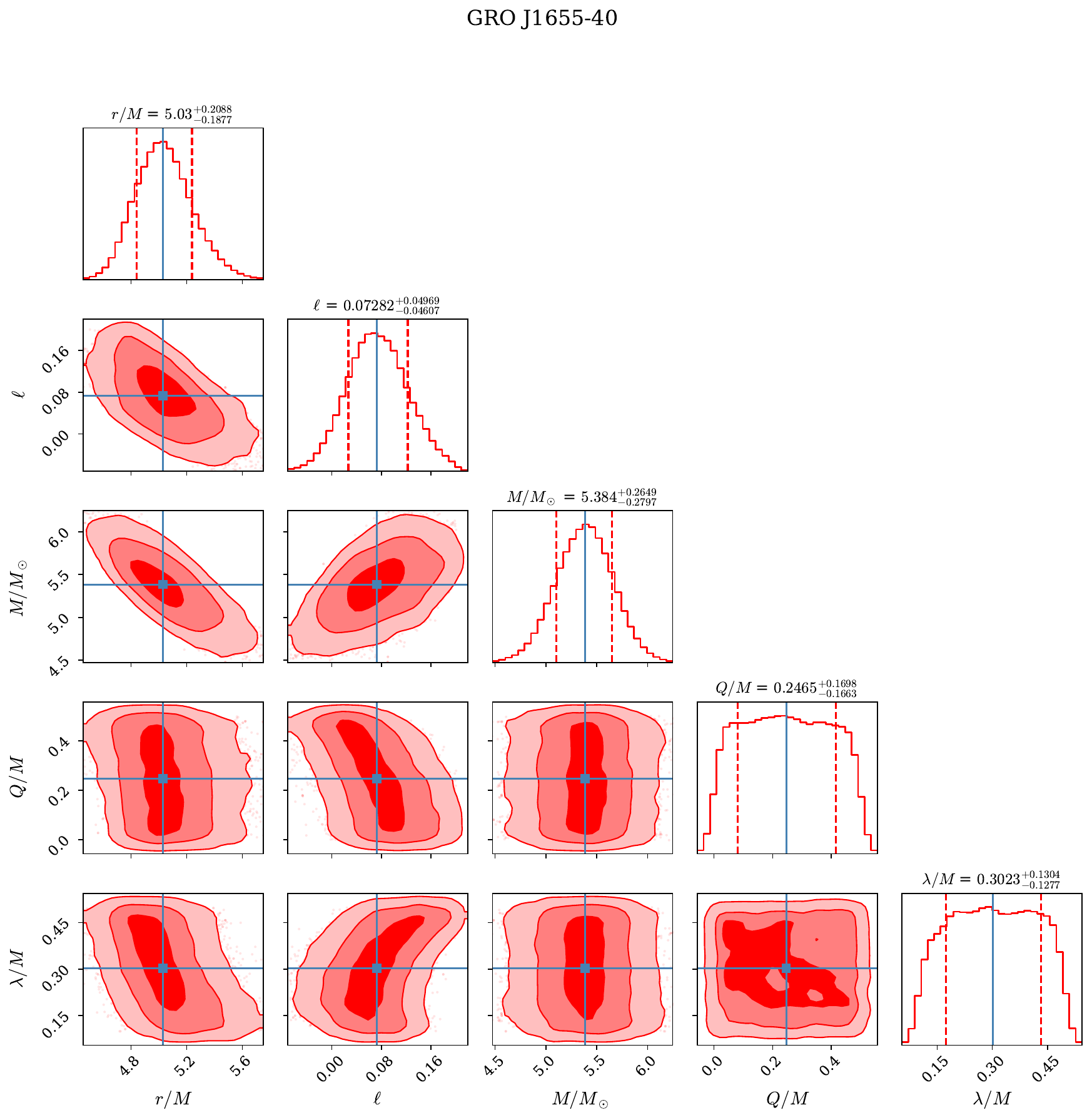}
				\put(4,92){\large \bfseries (b)}
			\end{overpic}
		\end{minipage}
		
		\vspace{0.6cm}
		
		\begin{minipage}{0.47\textwidth}
			\centering
			\begin{overpic}[width=\linewidth]{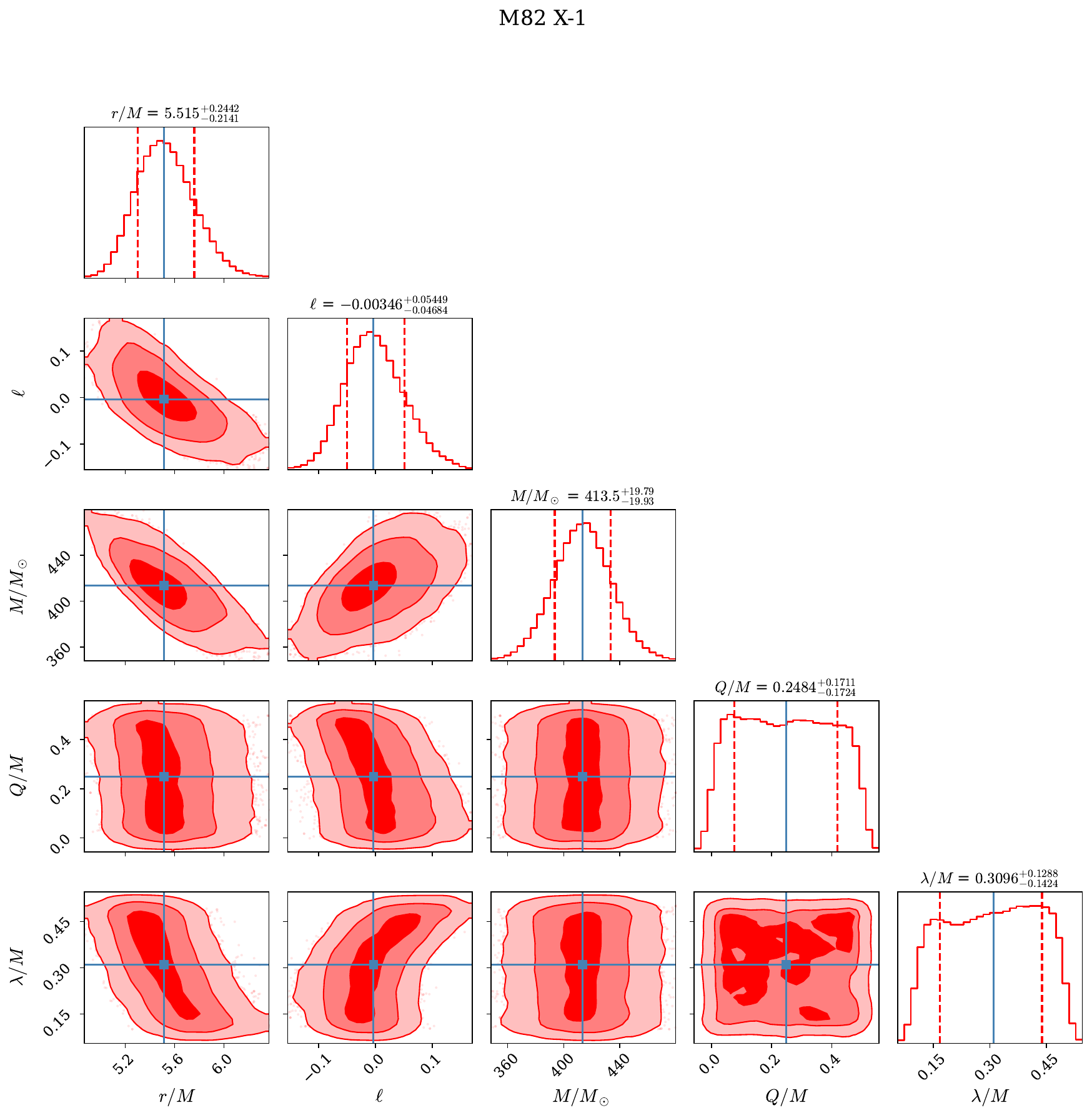}
				\put(4,92){\large \bfseries (c)}
			\end{overpic}
		\end{minipage}
		\hfill
		\begin{minipage}{0.47\textwidth}
			\centering
			\begin{overpic}[width=\linewidth]{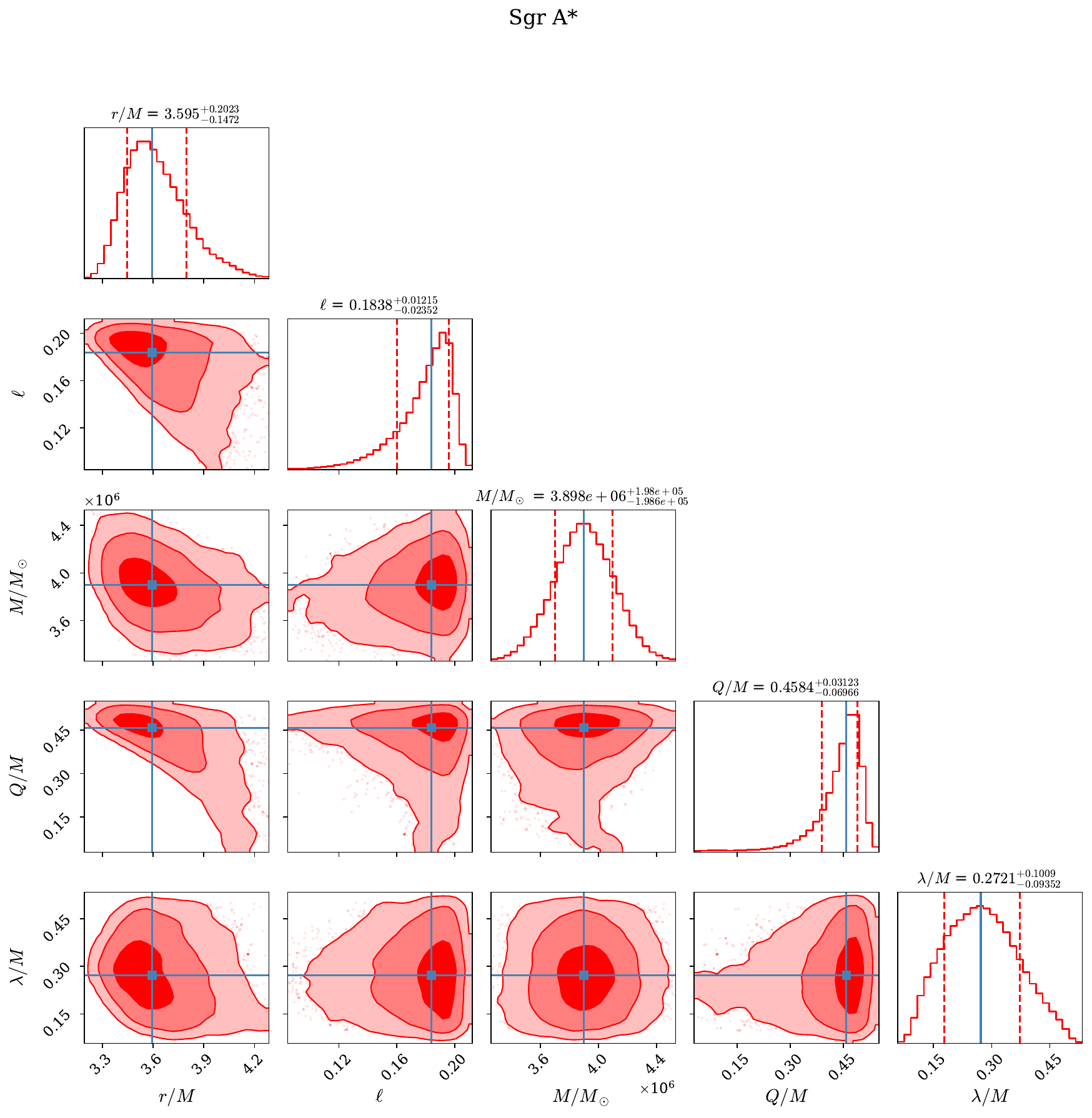}
				\put(4,92){\large \bfseries (d)}
			\end{overpic}
		\end{minipage}
		
		\caption{Constraints on the parameters of the charged KR black hole in PFDM derived from the five-dimensional MCMC analysis of the QPO data. Panels: (a) XTE J1550-564, (b) GRO J1655-40, (c) M82 X-1, and (d) Sgr A*.}
		\label{fig:MCMC}
	\end{figure*}

	The posterior probability distributions of the model parameters are displayed in Fig.~\ref{fig:MCMC}. The contour levels correspond to the $1\sigma$ ($68\%$), $2\sigma$ ($95\%$), and $3\sigma$ ($99\%$) confidence regions in the full five-dimensional parameter space, with the shaded areas indicating the corresponding credibility intervals.
	
	The corner plots exhibit a clear overall pattern. The sources XTE J1550--564, GRO J1655--40, and M82 X$-$1 show very similar posterior structures, favoring values of $r/M$ close to $5$ and small to moderate values of the Lorentz-violating parameter $\ell$. For these three systems, the posterior distributions of $Q/M$ and $\lambda/M$ remain relatively broad, indicating that the present RP-model fit does not strongly constrain them. In contrast, Sgr A* prefers a noticeably smaller radius, a larger positive value of $\ell$, and a higher value of $Q/M$, with a visibly tighter posterior localization in these parameters. In all cases, nontrivial parameter correlations are present, most notably an anticorrelation between $r/M$ and $\ell$, while the mass posterior remains centered around the corresponding observational estimate. Taken together, these results indicate that the first three sources probe a broadly similar region of the KR+PFDM parameter space, whereas Sgr A* favors a more distinct sector associated with stronger effective geometric deviations.
	
	These posterior trends also provide a useful consistency check for the dynamical analysis developed earlier in the paper. The values preferred by the RP-model fits correspond to regions of parameter space where the epicyclic frequencies remain well-behaved and where the ISCO lies within a range compatible with the observed QPO pairs. In this sense, the Bayesian analysis does not stand as an isolated statistical exercise; rather, it reinforces the geometrical picture emerging from the study of circular motion, namely that the charged KR+PFDM background leaves potentially observable imprints on the frequencies generated in the innermost region of the accretion flow.
	
	\begin{figure}[ht!]
		\centering
		\includegraphics[width=0.92\linewidth]{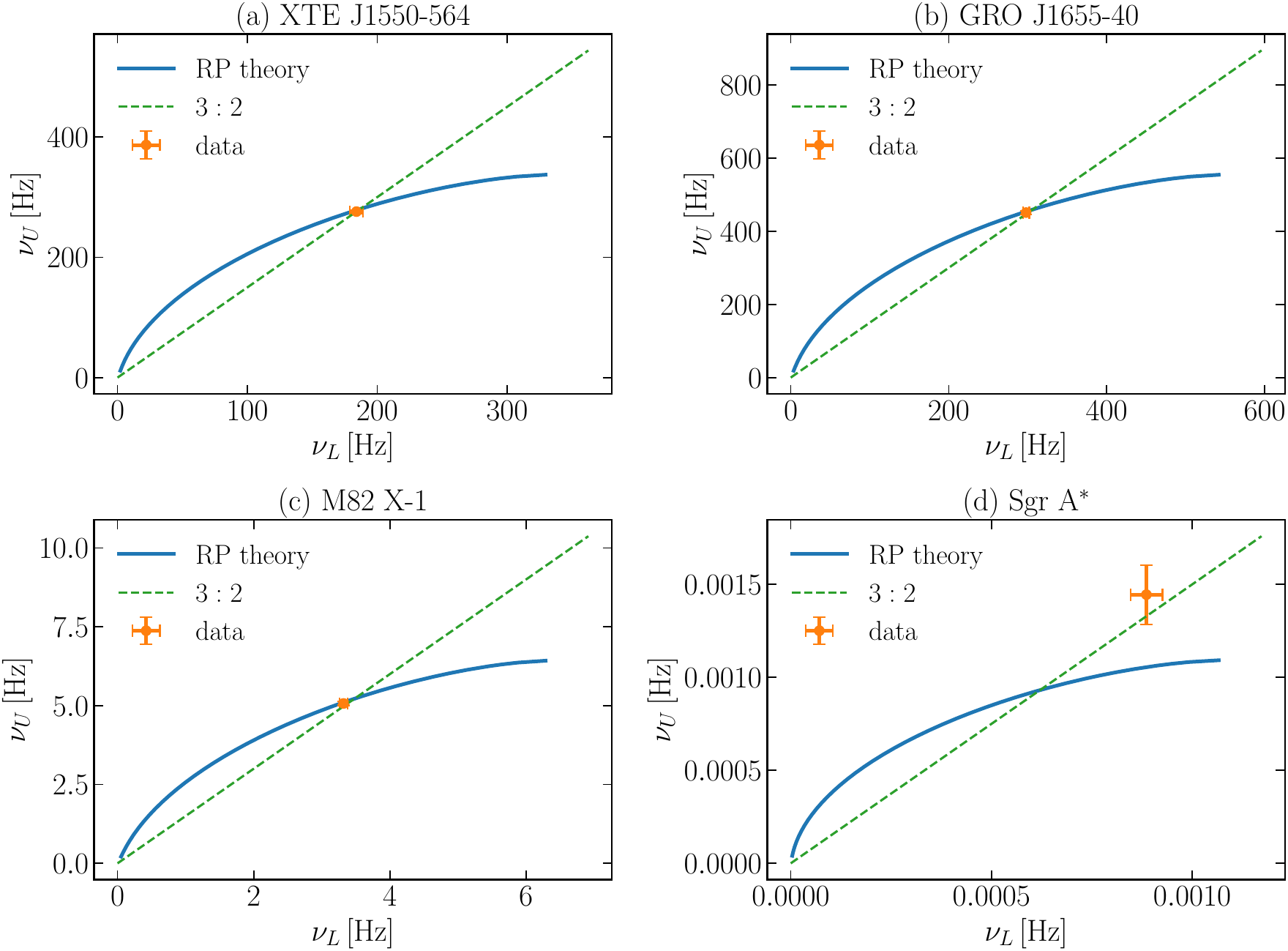}
		\caption{Overlay between the RP-model theoretical curves and the observed twin-peak QPO data for XTE J1550--564, GRO J1655--40, M82 X-1, and Sgr A$^\ast$, using the central values of the inferred parameters. The dashed line indicates the $3:2$ ratio.}
		\label{fig:qpo_overlay_rp}
	\end{figure}
	
	Figure~\ref{fig:qpo_overlay_rp} offers a direct visual assessment of the fit quality by placing the observational points and their uncertainties on top of the RP-model curves built from the inferred central parameters. For XTE J1550--564, GRO J1655--40, and M82 X-1, the data lie very close to the theoretical trajectories and also sit near the $3:2$ reference line, indicating that the fitted KR+PFDM configuration reproduces the measured twin-peak pattern rather naturally. The situation for Sgr A$^\ast$ is more subtle: the observational point remains in the same general sector of the plane, but it is visibly offset from the theoretical curve, suggesting that this source is either more sensitive to the exact parameter values or more affected by model dependence than the stellar- and intermediate-mass cases. This behavior is consistent with the broader discussion of the posterior distributions and reinforces the idea that the first three sources are described more tightly by the present RP implementation, whereas Sgr A$^\ast$ may require either a refined treatment or a broader phenomenological interpretation.
	
	\section{Thermodynamics}
	
	In this section, we investigate the thermodynamic sector of the charged black hole in KR-gravity surrounded by perfect-fluid dark matter (PFDM). Our goal is to clarify how the Lorentz-violating parameter $\ell$, the electric charge $Q$, and the PFDM parameter $\lambda$ jointly affect the horizon structure, the Hawking temperature, the entropy, the heat capacity, and the Gibbs free energy. Since the metric is not asymptotically flat, some thermodynamic definitions require special care, particularly the normalization used in the definition of the surface gravity. For this reason, before discussing the thermal variables themselves, it is useful to first analyze how the event horizon changes across the parameter space, because all subsequent thermodynamic quantities depend explicitly on the horizon radius.
	
	The event horizon is obtained from the largest real root of the lapse function,
	\begin{equation}
		f(r_h)=0.
	\end{equation}
	Because the lapse function contains both the KR correction and the logarithmic PFDM term, a closed analytic expression for $r_h$ is generally not available. Nevertheless, the horizon radius can be determined numerically with high accuracy for any chosen set of parameters. In Tables \ref{tab:horizontal-1} and \ref{tab:horizontal-2}, we list representative values of $r_h/M$ by varying $\ell$ and $\lambda$ for fixed charge $Q/M=0.5$, separating the negative-$\lambda$ and positive-$\lambda$ sectors for convenience.
	
	\begin{table}[h!]
		\centering
		\caption{Numerical values of the event-horizon radius $r_h/M$ for several combinations of the Lorentz-violating parameter $\ell$ and the PFDM parameter $\lambda/M$, with fixed charge $Q/M=0.5$. The negative-$\lambda$ sector shown here illustrates how increasingly negative PFDM contributions enlarge the outer horizon, while larger positive $\ell$ tends to reduce it.}
		\begin{tabular}{|c|c|c|c|c|c|}
			\hline
			$\ell \backslash \lambda/M$ & $-0.1$ & $-0.2$ & $-0.3$ & $-0.4$ & $-0.5$ \\
			\hline
			0.05  & 2.05961 & 2.24179 & 2.37962 & 2.48897 & 2.57671 \\
			0.10  & 1.92144 & 2.08939 & 2.21425 & 2.31129 & 2.38708 \\
			0.15  & 1.77941 & 1.93359 & 2.04577 & 2.13069 & 2.19463 \\
			0.20  & 1.63189 & 1.77286 & 1.87264 & 1.94557 & 1.99760 \\
			\hline
		\end{tabular}
		\label{tab:horizontal-1}
	\end{table} 
	
	\begin{table}[h!]
		\centering
		\caption{Numerical values of the event-horizon radius $r_h/M$ for several combinations of $\ell$ and $\lambda/M$, again with fixed charge $Q/M=0.5$. In the positive-$\lambda$ sector, the horizon radius decreases as $\lambda$ increases, while more negative values of $\ell$ partially compensate this effect by pushing the horizon outward.}
		\begin{tabular}{|c|c|c|c|c|c|}
			\hline
			$\ell \backslash \lambda/M$ & $0.1$ & $0.2$ & $0.3$ & $0.4$ & $0.5$ \\
			\hline
			-0.2  & 1.9368  & 1.75971 & 1.65879 & 1.60359 & 1.5783  \\
			-0.15 & 1.84694 & 1.68104 & 1.58816 & 1.53894 & 1.51818 \\
			-0.1  & 1.75535 & 1.60046 & 1.51552 & 1.47227 & 1.45603 \\
			-0.05 & 1.66162 & 1.51753 & 1.44049 & 1.4032  & 1.39153 \\
			\hline
		\end{tabular}
		\label{tab:horizontal-2}
	\end{table} 
	
	The numerical values in these tables already reveal two important trends. First, for fixed $\ell$, negative values of $\lambda$ tend to increase the event-horizon radius, whereas positive values of $\lambda$ reduce it. Second, for fixed $\lambda$, increasing $\ell$ generally shifts the horizon inward, while more negative values of $\ell$ enlarge the horizon. This indicates that the PFDM contribution and the Lorentz-violating KR sector do not act independently; rather, they combine to determine whether the black hole becomes geometrically more compact or more extended.
	
	\begin{figure}[ht!]
		\centering
		\includegraphics[width=0.95\linewidth]{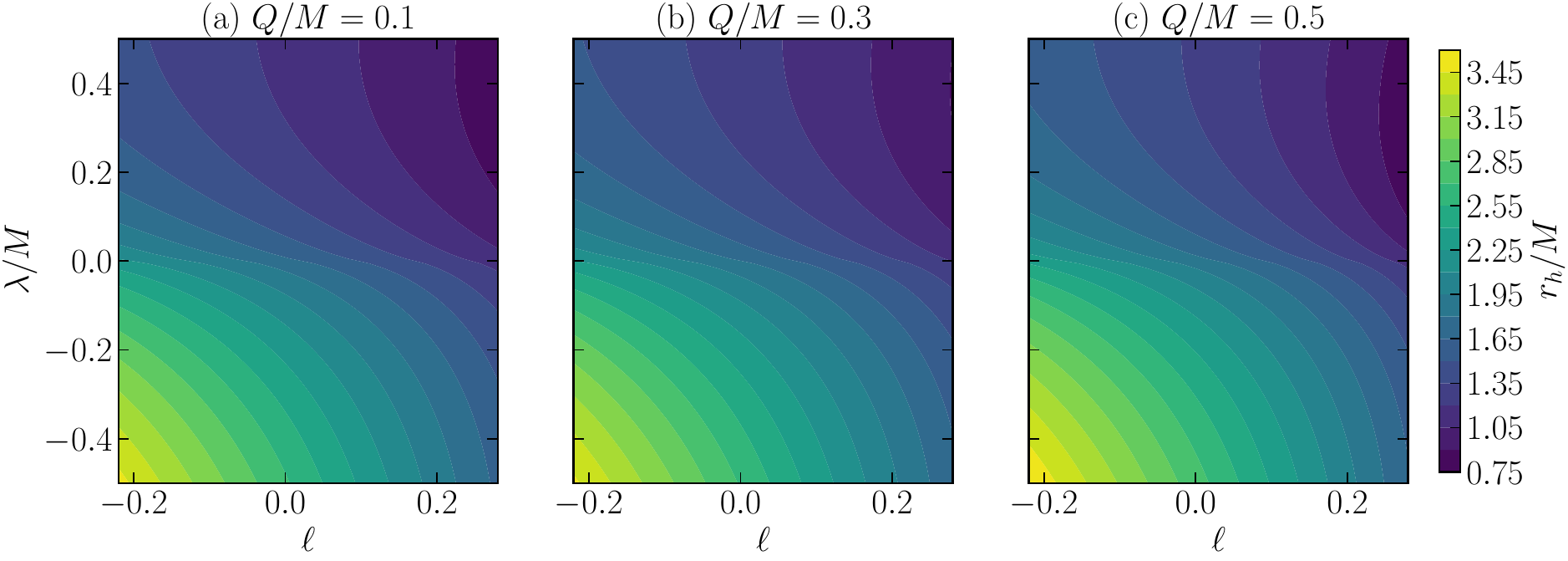}
		\caption{Contour maps of the event-horizon radius $r_h/M$ in the $(\ell,\lambda/M)$ plane for three representative charge values, $Q/M=0.1$, $0.3$, and $0.5$. The maps provide a global view of the horizon structure and show that the largest horizons occur in the sector of negative $\lambda$ and negative $\ell$, whereas the most compact configurations are concentrated toward positive $\lambda$ and positive $\ell$. The charge dependence is present in all panels, but the strongest gradients are primarily driven by the PFDM and KR parameters.}
		\label{fig:horizon_map}
	\end{figure}
	
	The global behavior of the horizon is summarized in Fig.~\ref{fig:horizon_map}. The contour maps make clear that the event horizon becomes larger in the region of negative $\lambda$ and negative $\ell$, while the smallest values of $r_h$ occur for positive $\lambda$ and positive $\ell$. Thus, within the plotted domain, both the PFDM sector and the Lorentz-violating KR background can either enlarge or shrink the outer horizon depending on the direction in which the parameter space is explored. The comparison among the three panels further shows that the electric charge modifies the contour pattern smoothly, while the dominant deformation is controlled by $\lambda$ and $\ell$. This figure is particularly useful as a bridge to the thermodynamic analysis, because all thermal quantities inherit their parameter dependence through the horizon radius.
	
	Once the event horizon is known, the black hole mass can be expressed in terms of $r_h$ by imposing the horizon condition $f(r_h)=0$. Solving this condition for $M$, one obtains
	\begin{equation}
		M=\frac{r_h}{2}\left[\frac{1}{1-\ell}+\frac{Q^2}{(1-\ell)^2 r_h^2} + \frac{\lambda }{r_h} \ln \frac{r_h}{|\lambda |}\right].
		\label{ss1}
	\end{equation}
	This relation shows explicitly how the thermodynamic mass receives three distinct contributions: the geometric KR correction through the factor $(1-\ell)^{-1}$, the usual Reissner--Nordstr\"om-like electrostatic term proportional to $Q^2$, and the logarithmic PFDM term proportional to $\lambda$. In this sense, Eq.~(\ref{ss1}) provides the starting point for the entire thermodynamic description, since the remaining quantities can be obtained from suitable derivatives of $M(r_h,Q,\lambda,\ell)$.
	
	\begin{figure}[ht!]
		\centering
		\includegraphics[width=0.45\linewidth]{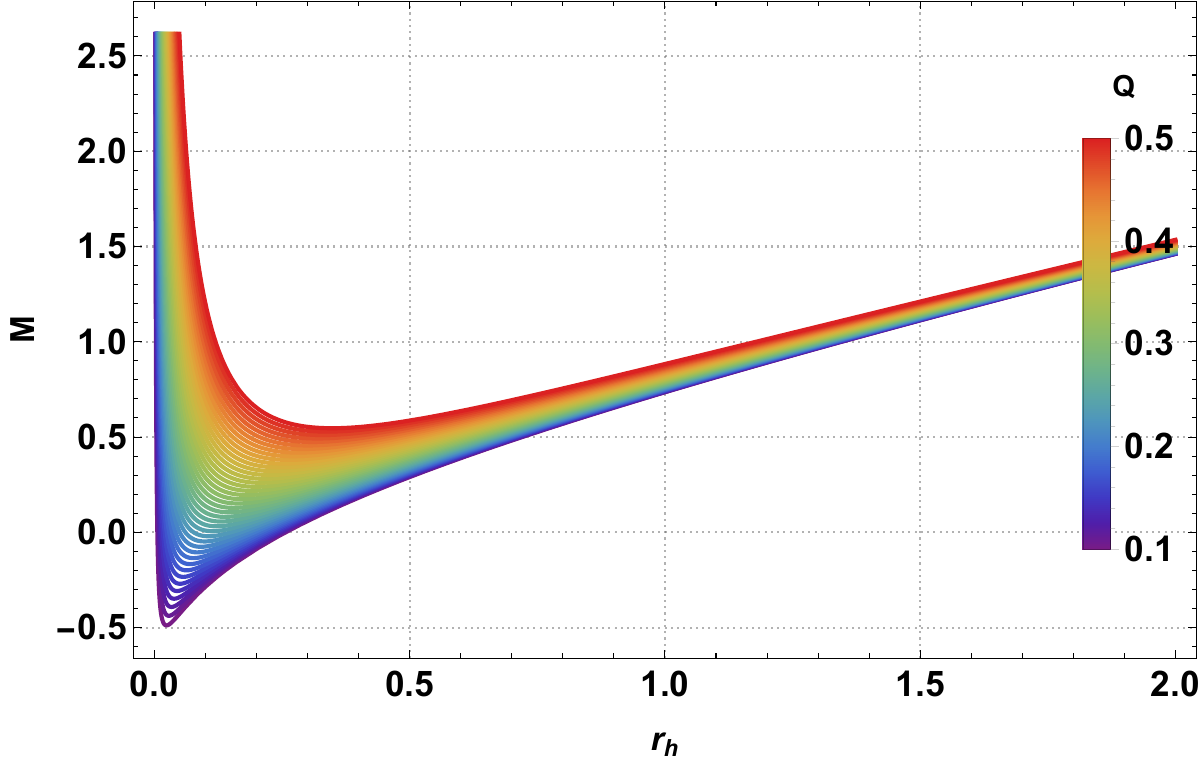}\qquad
		\includegraphics[width=0.45\linewidth]{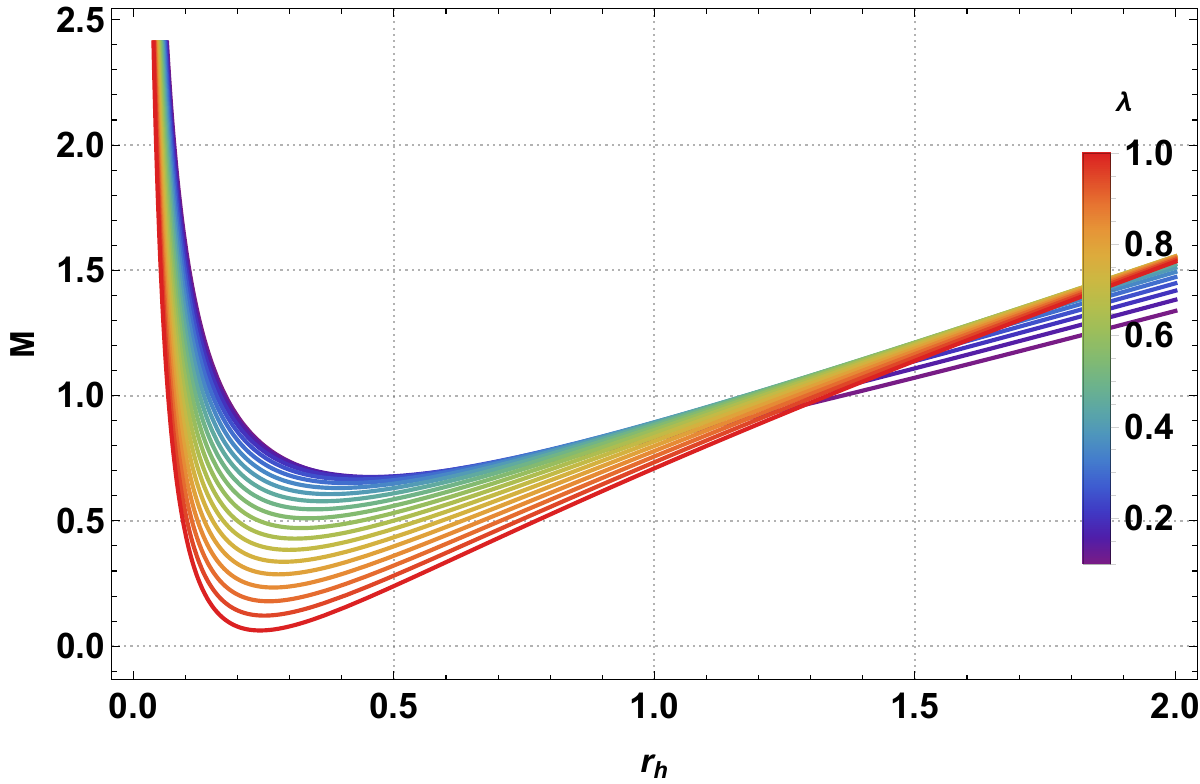}\\
		(i) $\lambda=0.2,\,\ell=0.1$ \hspace{6cm} (ii) $Q=0.5,\,\ell=0.1$\\
		\hfill\\
		\includegraphics[width=0.45\linewidth]{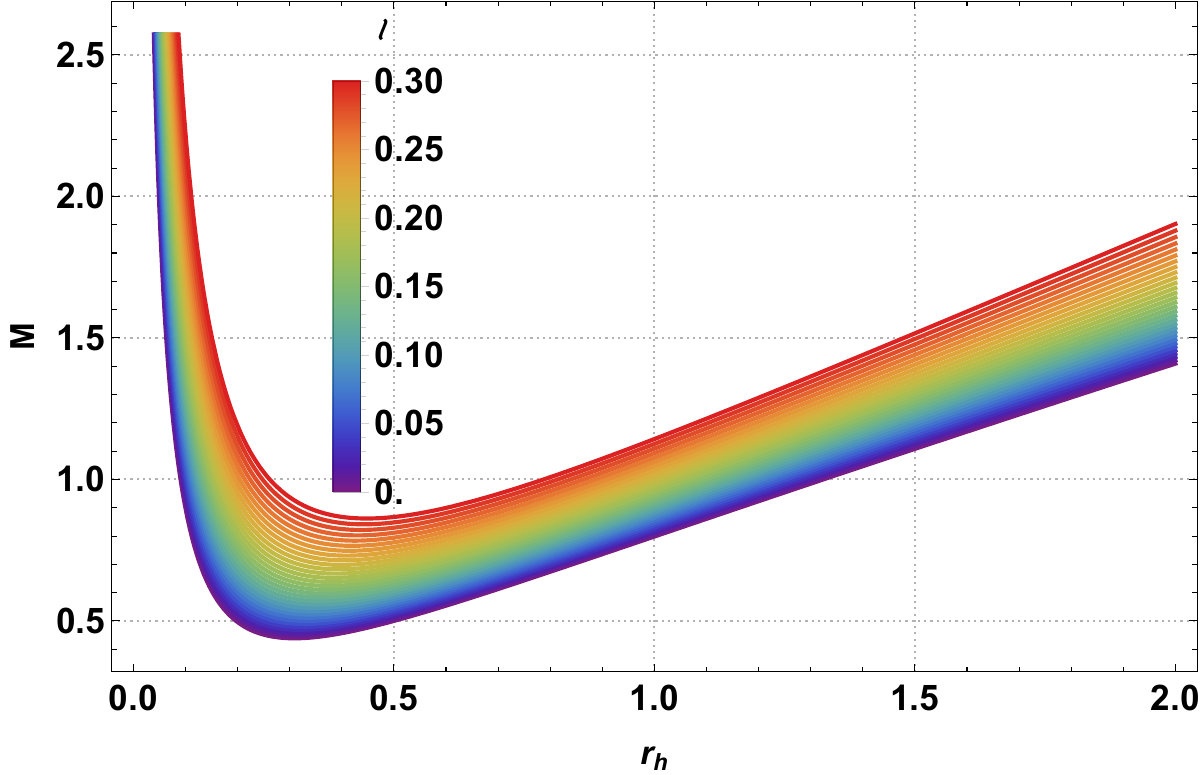}\\
		(iii) $Q=0.5=\lambda$
		\caption{Behavior of the mass function $M(r_h)$ for different values of the electric charge $Q$, PFDM parameter $\lambda$, and Lorentz-violating parameter $\ell$. Panel (i) isolates the effect of the charge for fixed PFDM and KR background; panel (ii) emphasizes the role of the PFDM contribution for fixed charge and KR parameter; and panel (iii) shows how varying $\ell$ shifts the full mass profile when the charge and PFDM strength are kept equal. The figure illustrates how the three sectors modify the amount of mass required to sustain a horizon of given radius.}
		\label{fig:mass}
	\end{figure}
	
	Figure~\ref{fig:mass} illustrates the behavior of the mass function as the parameters are varied. The main message of this plot is that, for a given horizon radius, the required black hole mass is not fixed solely by the horizon scale, but also by the way in which the charge, the PFDM environment, and the Lorentz-violating KR background distribute the effective gravitational contribution. Therefore, the same horizon radius can correspond to different mass values depending on the microscopic and environmental content of the spacetime.
	
	We now turn to the Hawking temperature. In asymptotically flat black hole spacetimes, the temperature is usually obtained directly from the surface gravity using the standard normalization of the timelike Killing vector at spatial infinity. In the present case, however, this procedure must be modified because the metric does not satisfy the usual asymptotic condition. In fact, from Eq.~(\ref{condition}) one has
	\begin{equation}
		\lim_{r\to\infty} f(r)=\frac{1}{1-\ell}\neq 1,
	\end{equation}
	so the normalization of the Killing field at infinity differs from the conventional one. For this reason, the correct temperature must be computed with the properly normalized surface gravity.
	
	The surface gravity is defined by
	\begin{equation}
		\kappa=-\lim_{r \to r_h}\,\frac{\mathcal{C}}{2}\frac{\partial_r g_{tt}}{\sqrt{-g_{tt}\,g_{rr}}},
		\label{ss3}
	\end{equation}
	where the normalization factor $\mathcal{C}$ is determined from the asymptotic behavior of the metric as
	\begin{equation}
		\mathcal{C}=1/\sqrt{-\lim_{r \to \infty} g_{tt}}.
		\label{ss4}
	\end{equation}
	For the spacetime under consideration, these expressions yield
	\begin{equation}
		\kappa=\frac{\mathcal{C}}{2} f'(r_h),\qquad \mathcal{C}=\sqrt{1-\ell}.
		\label{ss5}
	\end{equation}
	Hence, the Hawking temperature is
	\begin{equation}
		T_H=\frac{\kappa}{2 \pi}=\frac{\sqrt{1-\ell}}{4 \pi r_h}\left[\frac{1}{1-\ell}- \frac{Q^2}{(1-\ell)^2\,r_h^2}+ \frac{\lambda}{r_h}\right].
		\label{ss6}
	\end{equation}
	
	Equation~(\ref{ss6}) makes the physical content of the temperature especially transparent. The factor $\sqrt{1-\ell}$ originates from the nonstandard normalization associated with the non-asymptotically flat background, while the bracketed term contains the effective competition among the geometric KR contribution, the electric charge, and the PFDM term. The charge contribution tends to reduce the temperature, in analogy with the standard charged-black-hole case, because it weakens the surface gravity near extremality. By contrast, the PFDM contribution may either increase or decrease the temperature depending on the sign of $\lambda$ and on the value of the horizon radius. Thus, the thermal response of the system is controlled not by a single deformation parameter, but by the combined geometry of the KR+PFDM background.
	
	\begin{figure}[ht!]
		\centering
		\includegraphics[width=0.45\linewidth]{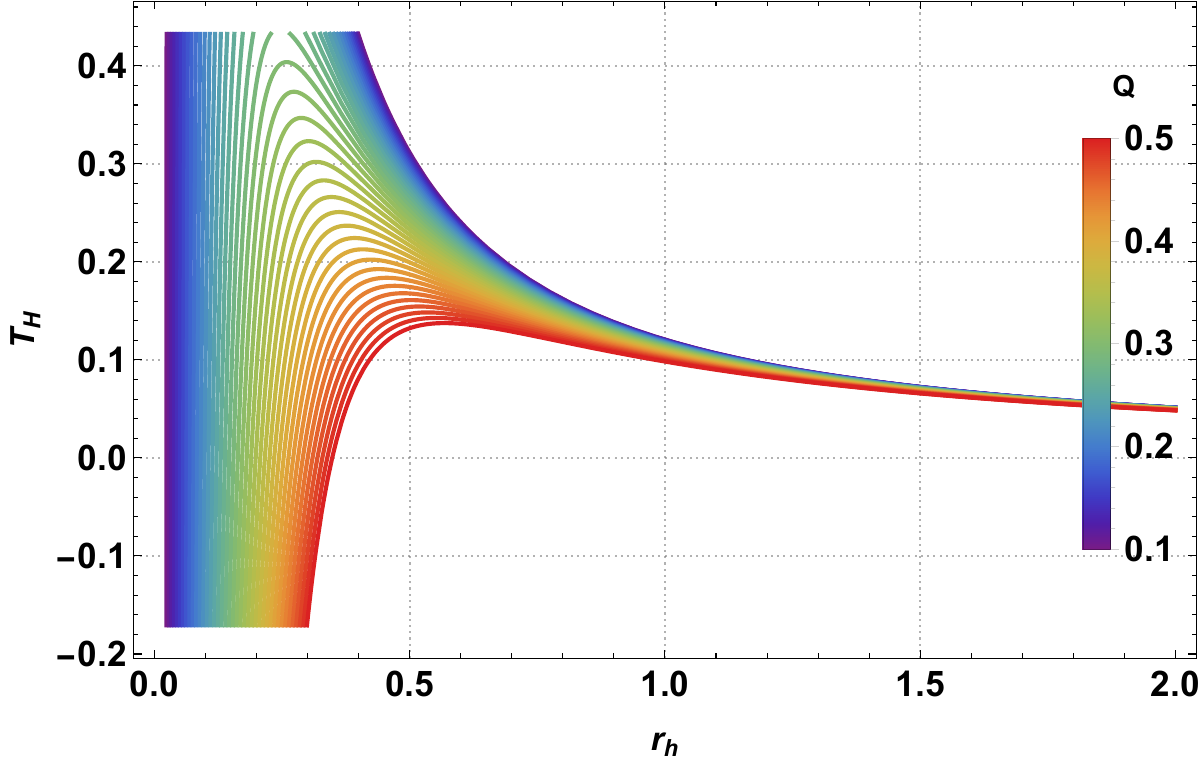}\qquad
		\includegraphics[width=0.45\linewidth]{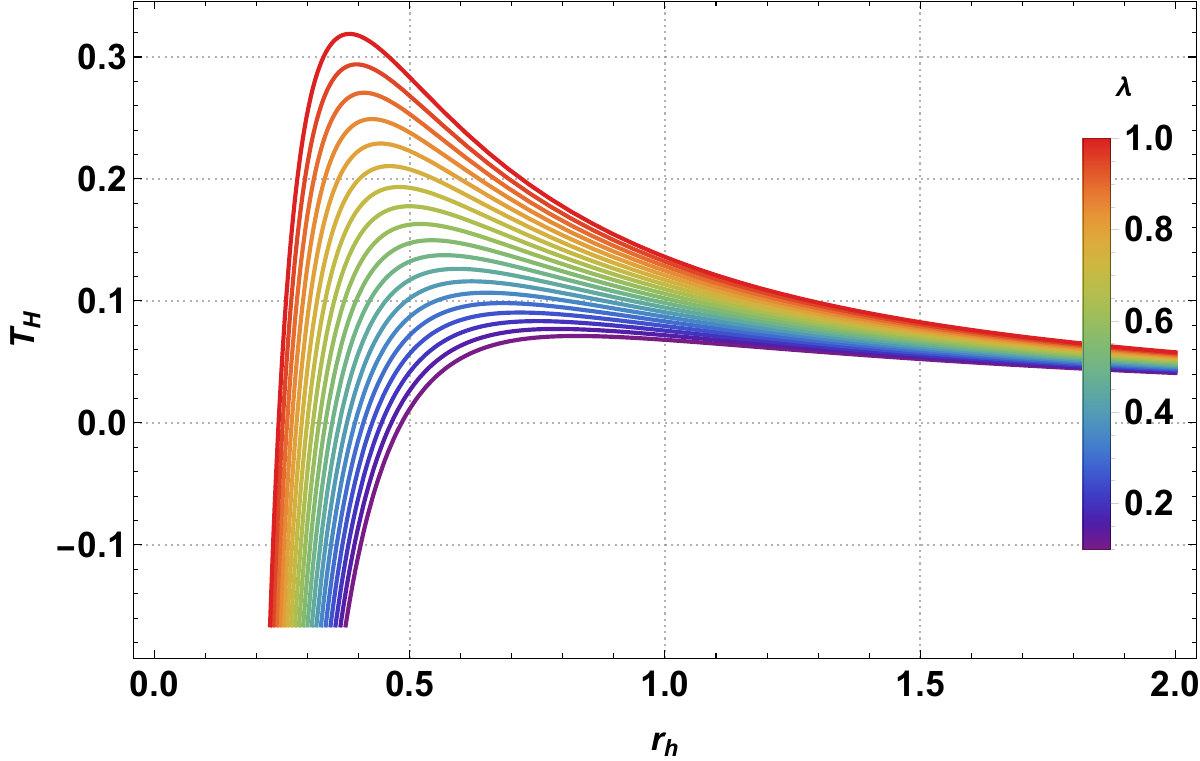}\\
		(i) $\lambda=0.2,\,\ell=0.1$ \hspace{6cm} (ii) $Q=0.5,\,\ell=0.1$\\
		\hfill\\
		\includegraphics[width=0.45\linewidth]{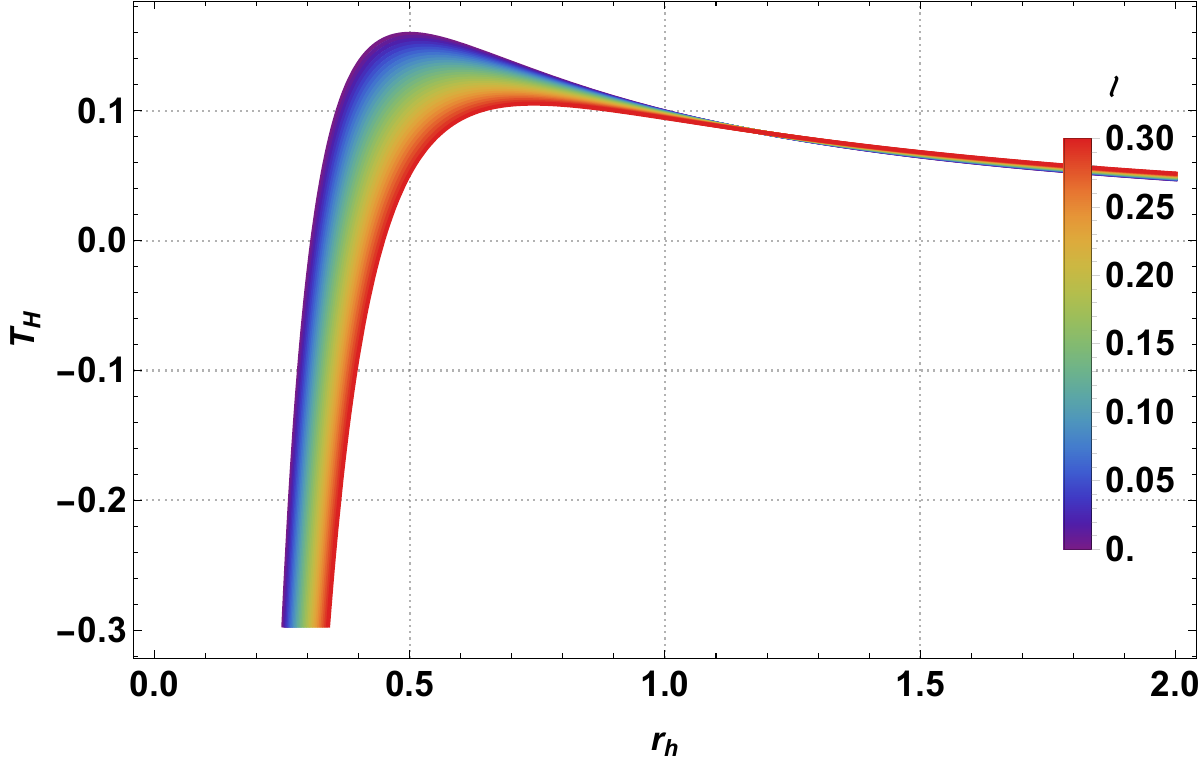}\\
		(iii) $Q=0.5=\lambda$
		\caption{Hawking temperature $T_H$ as a function of the horizon radius for different values of $Q$, $\lambda$, and $\ell$. The three panels disentangle the role of each sector: panel (i) shows the suppression induced by the electric charge, panel (ii) displays the shift produced by the PFDM contribution, and panel (iii) highlights the effect of the Lorentz-violating KR parameter on both the amplitude and the position of the thermal extrema. These profiles are useful for identifying the parameter regions associated with hotter and colder black hole states.}
		\label{fig:temperature}
	\end{figure}
	
	The behavior of the Hawking temperature is shown in Fig.~\ref{fig:temperature}. The figure reveals how the temperature profile is deformed when each parameter is varied separately. In particular, the presence of a maximum in the temperature curve is thermodynamically relevant, because it often signals the transition between small-black-hole and large-black-hole branches with qualitatively different thermal behavior. More specifically, the rising portion of the curve corresponds to a regime in which the temperature increases with the horizon radius, whereas the descending part corresponds to the more familiar regime in which larger horizons are colder. The position of this turnover is shifted by $Q$, $\lambda$, and $\ell$, showing that the KR-induced Lorentz violation and the PFDM environment modify the effective thermal scale of the black hole.
	
	The entropy of the system can be derived from the first law of thermodynamics. Integrating $dS=dM/T_H$, one finds
	\begin{equation}
		S=\int \frac{dM}{T_H}=\frac{\pi r_h^2}{\sqrt{1-\ell}}.
		\label{ss7}
	\end{equation}
	This result deserves emphasis. In standard Einstein gravity for asymptotically flat black holes, one recovers the Bekenstein--Hawking area law $S=A/4=\pi r_h^2$ (in natural units). Here, however, the entropy acquires an additional factor $1/\sqrt{1-\ell}$, so that the KR background changes the proportionality between entropy and area. The departure from the standard area law is therefore controlled entirely by the Lorentz-violating parameter, even though the PFDM sector still affects the numerical value of the entropy indirectly through its impact on the horizon radius.
	
	\begin{figure}[ht!]
		\centering
		\includegraphics[width=0.90\linewidth]{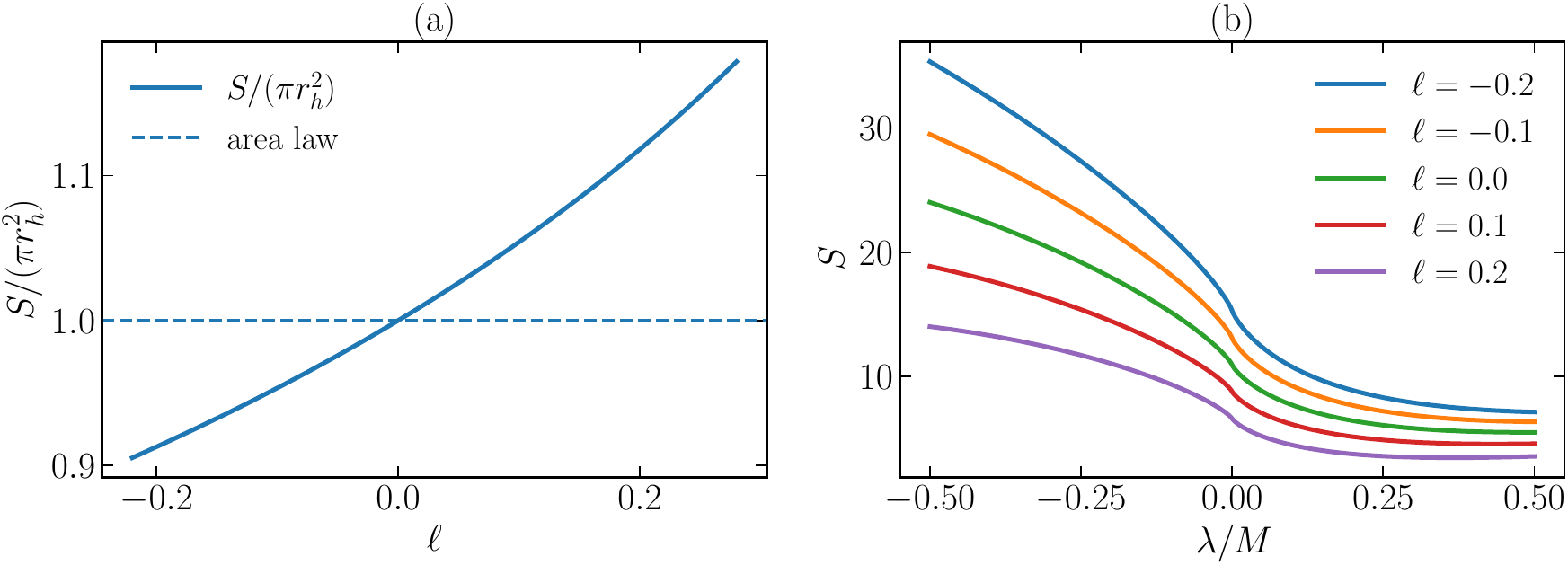}
		\caption{Entropy diagnostics for the charged KR+PFDM black hole. The left panel shows the ratio $S/(\pi r_h^2)=1/\sqrt{1-\ell}$ as a function of the Lorentz-violating parameter, explicitly displaying the deviation from the standard Bekenstein--Hawking area law. The right panel presents the entropy as a function of $\lambda/M$ for several representative values of $\ell$, thereby showing how the PFDM environment modifies the entropy through the corresponding change in the horizon radius.}
		\label{fig:entropy_extra}
	\end{figure}
	
	Figure~\ref{fig:entropy_extra} makes this deviation especially transparent. The left panel shows that the ratio $S/(\pi r_h^2)$ equals unity only at $\ell=0$, as expected from the standard area law. For negative $\ell$, the ratio drops below unity, indicating an entropy smaller than the usual geometric value, while for positive $\ell$ it becomes larger than unity. Thus, the KR-induced Lorentz-violating parameter controls whether the entropy is suppressed or enhanced relative to the standard Bekenstein--Hawking prescription. The right panel complements this picture by displaying the entropy as a function of $\lambda/M$ for several values of $\ell$. Even though $\lambda$ does not enter Eq.~(\ref{ss7}) explicitly as a prefactor, it changes the horizon radius and therefore modifies the entropy indirectly. This figure shows that the PFDM environment leaves a clear and systematic imprint on the entropy landscape of the solution.
	
	A central diagnostic of local thermodynamic stability is the specific heat capacity. It measures the response of the black hole mass to a change in temperature and determines whether the thermal branch is stable under small fluctuations. Using the definitions above, the specific heat is
	\begin{align}
		C=\frac{\partial M}{\partial T_H}=\frac{2\pi r_h^2}{\sqrt{1-\ell}}\, 
		\frac{\left[\dfrac{1}{1-\ell} - \dfrac{Q^2}{(1-\ell)^2 r_h^2} + \dfrac{\lambda}{r_h}\right]}
		{\left[-\dfrac{1}{1-\ell} + \dfrac{3 Q^2}{(1-\ell)^2 r_h^2} - \dfrac{2 \lambda}{r_h} \right]}.
		\label{ss8}
	\end{align}
	The sign of $C$ determines the local thermodynamic character of the black hole. When $C>0$, the system is locally stable in the canonical ensemble, because an increase in temperature corresponds to an increase in mass in a controlled way. When $C<0$, the system is locally unstable, as happens for the Schwarzschild black hole in asymptotically flat spacetime. Therefore, the denominator of Eq.~(\ref{ss8}) is particularly important: its zeros signal the points at which the heat capacity diverges and the system changes from one thermodynamic branch to another.
	
	\begin{figure}[ht!]
		\centering
		\includegraphics[width=0.45\linewidth]{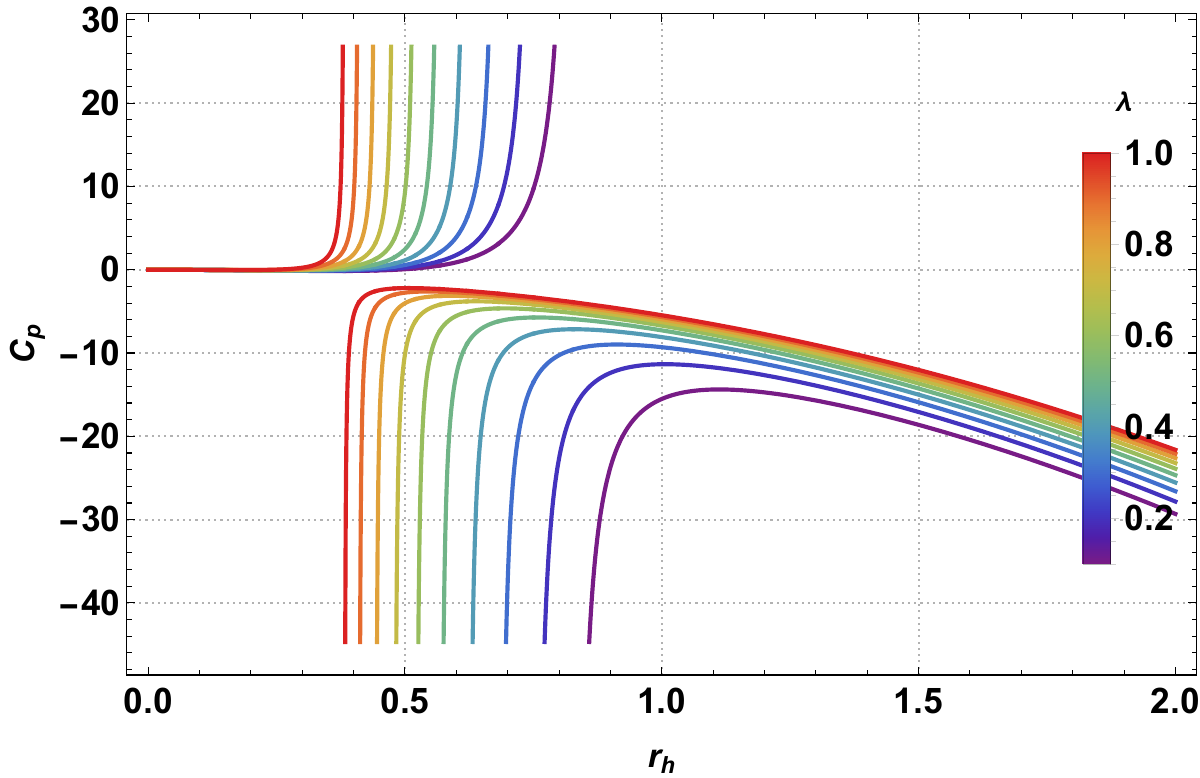}\quad
		\includegraphics[width=0.45\linewidth]{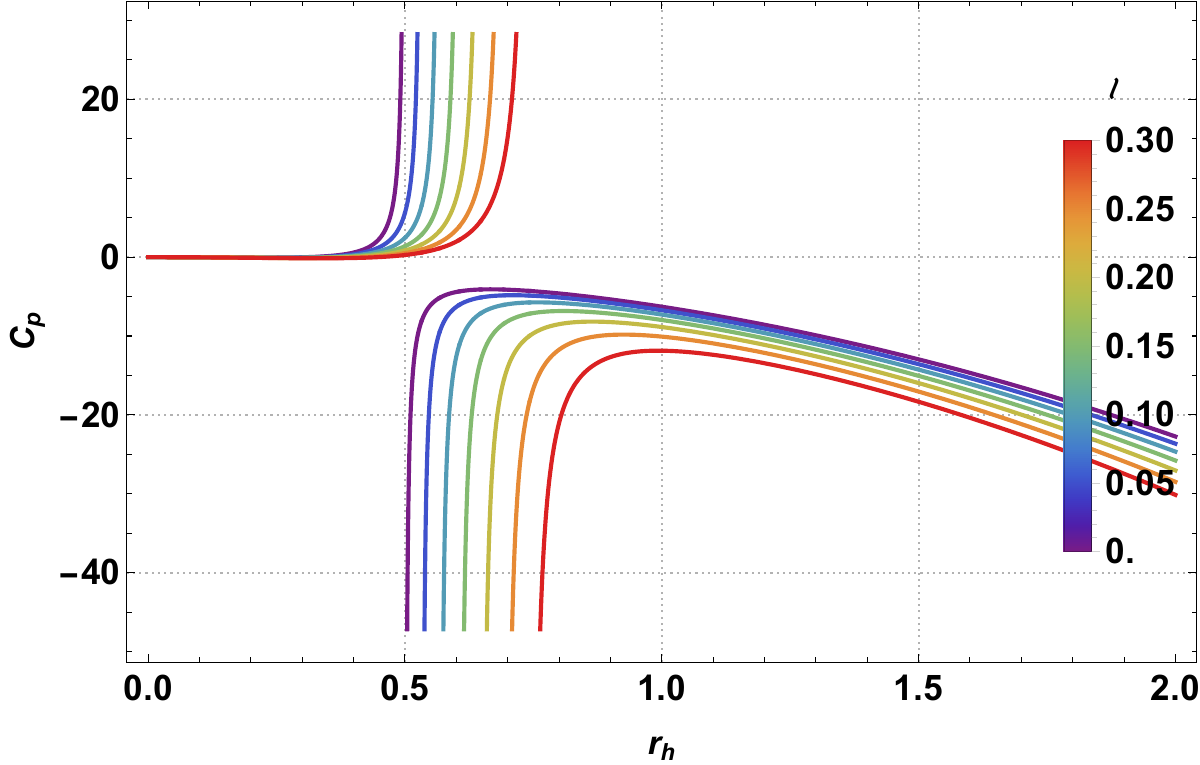}\\
		(i) $\ell=0.1$ \hspace{6cm} (ii) $\lambda=0.5$
		\caption{Specific heat capacity $C$ for the charged KR+PFDM black hole with fixed $Q=0.5$. Panel (i) shows how the PFDM parameter $\lambda$ shifts the divergence point and changes the extent of the stable and unstable branches for fixed Lorentz violation, while panel (ii) displays the corresponding effect of varying $\ell$ for fixed PFDM strength. The sign change across the divergence identifies the second-order phase-transition point separating locally unstable and locally stable thermodynamic sectors.}
		\label{fig:capacity}
	\end{figure}
	
	This behavior is illustrated in Fig.~\ref{fig:capacity}. The divergence of the specific heat separates the unstable and stable branches and therefore marks a second-order phase-transition point in the canonical description. Physically, the KR parameter and the PFDM contribution do more than merely shift the temperature curve; they also control the location of the thermodynamic critical radius that divides the negative-$C$ and positive-$C$ regions. In this way, the geometry directly determines the stability structure of the black hole.
	
	The divergence of the specific heat occurs when the denominator of Eq.~(\ref{ss8}) vanishes, that is, when $r_h=r_*$ satisfies
	\begin{equation}
		\dfrac{r_*^2}{1-\ell} - \dfrac{3 Q^2}{(1-\ell)^2} + 2 \lambda r_*=0.
	\end{equation}
	Solving for $r_*$, one finds
	\begin{equation}
		r_* = -\lambda (1-\ell) + \sqrt{ \lambda^2 (1-\ell)^2 + \frac{3 Q^2}{1-\ell} }.
	\end{equation}
	This expression shows explicitly that the critical radius depends on the interplay between the KR correction, the PFDM background, and the electric charge. Thus, the position of the thermodynamic phase transition is not universal: it is displaced across the parameter space according to the underlying deformation of the geometry.
	
	We next discuss the first law and the corresponding Smarr relation. Treating the black hole mass $M$ as the enthalpy of the thermodynamic system, the modified first law can be written as
	\begin{equation}
		dM=T dS+\Phi dQ_{\rm eff}+\psi d\lambda,
		\label{ss9}
	\end{equation}
	where $T$ is the Hawking temperature, $\Phi$ is the potential conjugate to the effective charge, and $\psi$ is the thermodynamic quantity conjugate to the PFDM parameter. These quantities are given by
	\begin{align}
		T&=\left(\frac{\partial M}{\partial S}\right)_{Q,\lambda}=T_H,\nonumber\\
		\Phi&=\left(\frac{\partial M}{\partial Q_{\rm eff}}\right)_{S,\lambda}=\frac{Q_{\rm eff}}{r_h}=\frac{Q}{(1-\ell) r_h},\nonumber\\
		\psi&=\left(\frac{\partial M}{\partial \lambda}\right)_{S, Q}=\frac{1}{2} \ln \frac{r_h}{|\lambda|} - \frac{1}{2}.
		\label{ss10}
	\end{align}
	Here, the effective charge
	\begin{equation}
		Q_{\rm eff}=\frac{Q}{1-\ell}
	\end{equation}
	makes explicit that the KR background rescales the electrostatic sector. In addition, the conjugate quantity $\psi$ shows that the PFDM parameter enters the thermodynamic description in a nontrivial way through the logarithmic structure inherited from the metric function.
	
	Using the relations above, one verifies that the corresponding Smarr formula reads
	\begin{equation}
		M = 2 T S + \Phi Q_{\rm eff} + \lambda \psi.
		\label{ss11}
	\end{equation}
	Therefore, although the geometry is modified by both Lorentz-violating and PFDM contributions, the structure of the first law and the Smarr relation remains formally consistent once the appropriate effective variables are identified. This is a nontrivial consistency check of the thermodynamic framework.
	
	To complement the local analysis based on the heat capacity, we also examine the Gibbs free energy in the canonical ensemble. It is defined as
	\begin{equation}
		G=M- T_H S=\frac{r_h}{4} \left[\frac{1}{1-\ell} + \frac{3 Q^2}{(1-\ell)^2 r_h^2} + \frac{\lambda}{r_h} \left(2\ln\! \frac{r_h}{|\lambda|} -1\right)\right].
	\end{equation}
	Unlike the specific heat, which determines local stability under small fluctuations, the Gibbs free energy identifies the globally preferred thermodynamic branch. Lower values of $G$ correspond to states that are more favorable in the canonical ensemble.
	
	\begin{figure}[ht!]
		\centering
		\includegraphics[width=0.45\linewidth]{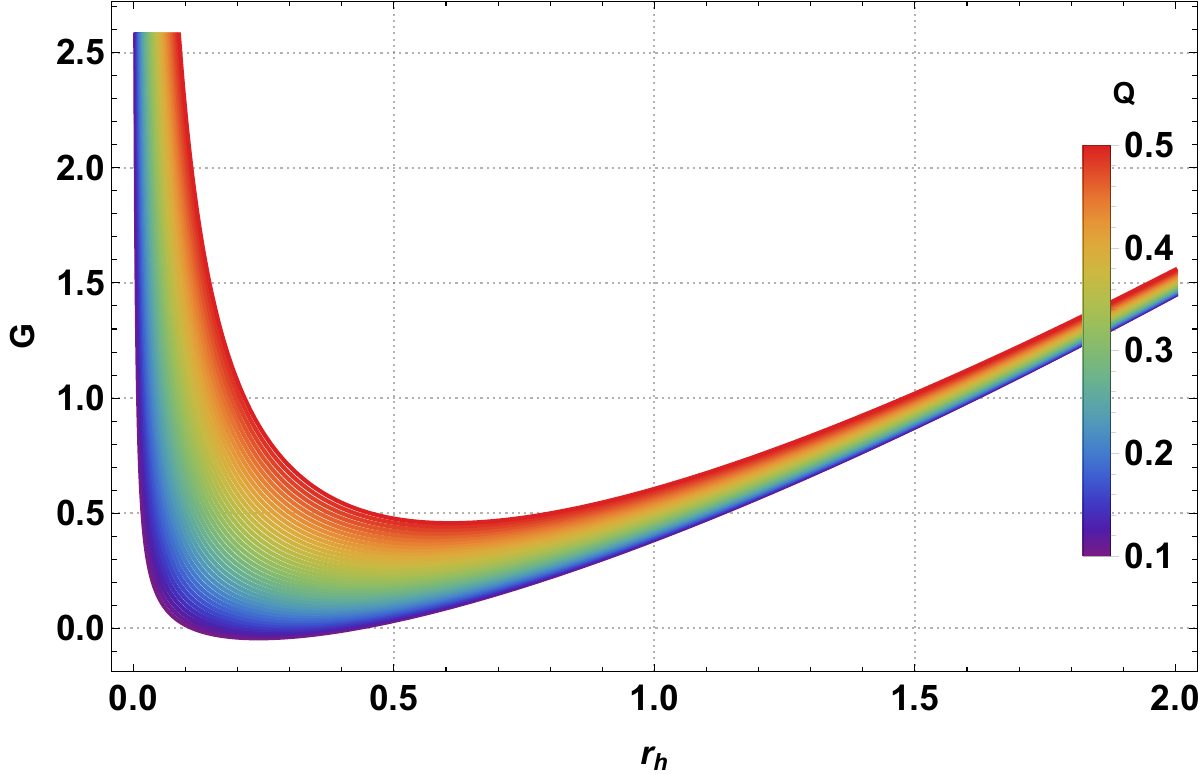}\qquad
		\includegraphics[width=0.45\linewidth]{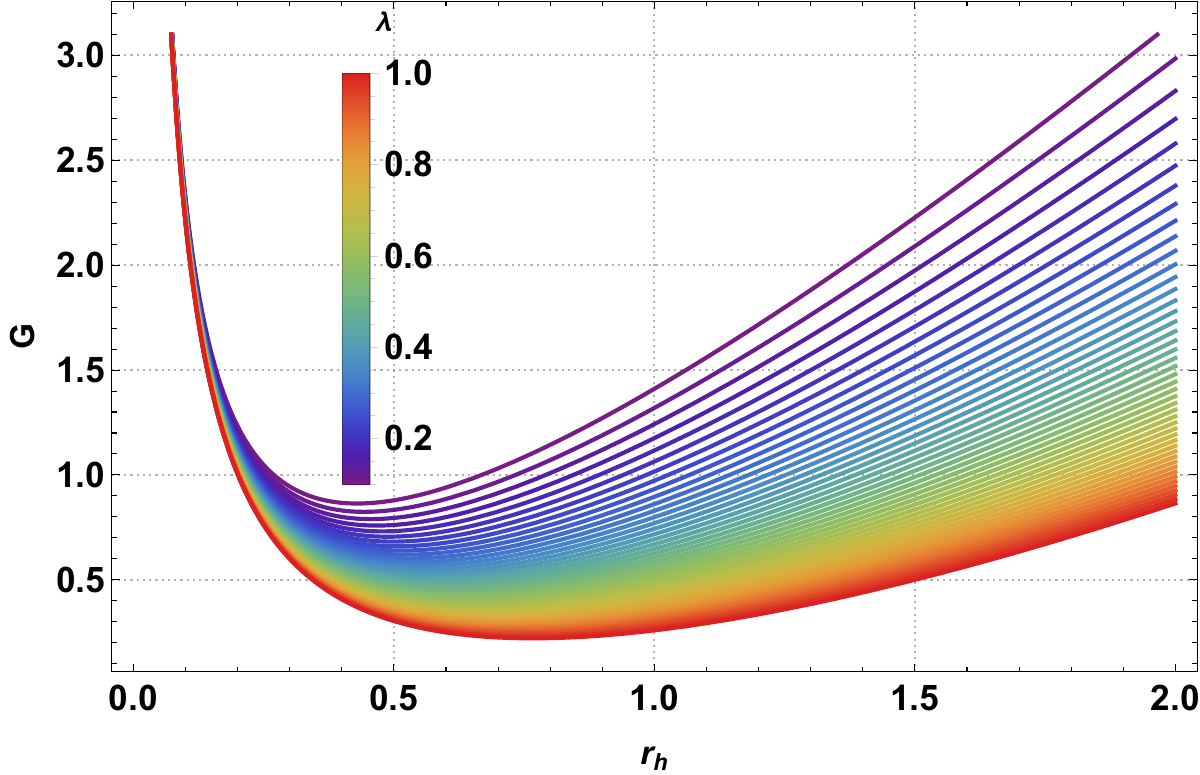}\\
		(i) $\lambda=0.2,\,\ell=0.1$ \hspace{6cm} (ii) $Q=0.5,\,\ell=0.1$\\
		\hfill\\
		\includegraphics[width=0.45\linewidth]{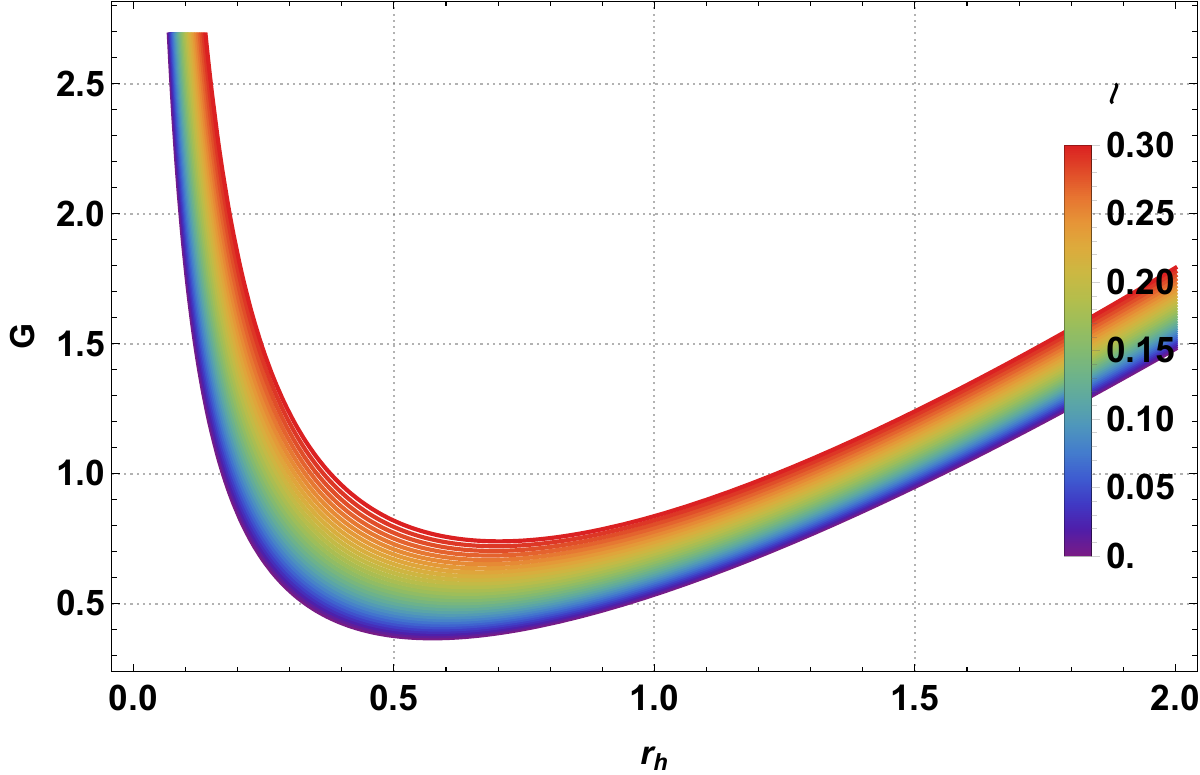}\\
		(iii) $Q=0.5=\lambda$
		\caption{Gibbs free energy $G$ of the charged KR+PFDM black hole for different values of $Q$, $\lambda$, and $\ell$. The three panels show how each parameter deforms the global thermodynamic landscape. Regions where the curves move downward or approach zero correspond to thermodynamically favored branches in the canonical ensemble, while changes in the shape of the curves indicate how the KR and PFDM sectors modify the competition between distinct black hole states.}
		\label{fig:gibbs}
	\end{figure}
	
	The behavior of the Gibbs free energy is presented in Fig.~\ref{fig:gibbs}. Together with the heat-capacity analysis, this figure provides a fuller thermodynamic picture. While $C$ distinguishes locally stable from unstable branches, $G$ shows which of those branches is globally preferred. In particular, parameter domains for which the Gibbs free energy decreases or approaches zero correspond to configurations that are thermodynamically favored relative to hotter competing states in the canonical ensemble. Hence, the KR parameter $\ell$ and the PFDM parameter $\lambda$ influence not only the local response of the black hole to thermal fluctuations, but also its global phase preference.
	
	To summarize, the thermodynamic properties of the charged KR+PFDM black hole are governed by a rich interplay between the Lorentz-violating background, the electric charge, and the surrounding dark matter. The horizon radius sets the geometric scale of the problem; the Hawking temperature inherits a nontrivial normalization due to the non-asymptotically flat character of the metric; the entropy departs from the usual area law through a KR-dependent prefactor; the specific heat reveals a second-order phase transition separating locally stable and unstable branches; and the Gibbs free energy identifies the globally favored branch in the canonical ensemble. Altogether, these results show that the KR and PFDM sectors produce sizable and structured deformations of the thermal behavior of the black hole, thereby enriching the phenomenology well beyond the standard Reissner--Nordstr\"om case.
	
	\section{Sparsity of Hawking Radiation}
	
	Although the Hawking spectrum is thermal, the emission does not proceed continuously. Individual quanta are emitted intermittently, and the radiation profile is inherently sparse. Following~\cite{Hawking1975,Page1976,Gray2016}, the degree of sparsity can be quantified by comparing the characteristic thermal wavelength $\lambda_t = 2\pi/T_H$ with the effective emitting area $\mathcal{A}_{\rm eff}$. The dimensionless sparsity parameter is
	\begin{equation}
		\eta = \frac{\mathcal{C}}{\tilde{g}}\left(\frac{\lambda_t^2}{\mathcal{A}_{\rm eff}}\right),\label{Sparsity-1}
	\end{equation}
	where $\mathcal{C}$ is a dimensionless constant, $\tilde{g}$ the spin degeneracy of the emitted quanta, and $\mathcal{A}_{\rm eff} = \frac{27}{4}\,A_{\rm BH}$ is the effective area.
	
	For the considered spacetime, $\mathcal{A}_{\rm eff} = 27\pi r_h^2$ and the thermal wavelength reads
	\begin{equation}
		\lambda_t = 8\pi^2 r_h\,(1-\ell)^{-1/2}\left(\frac{1}{1-\ell}- \frac{Q^2}{(1-\ell)^2\,r_h^2}+ \frac{\lambda}{r_h}\right)^{-1}.\label{sparsity-2}
	\end{equation}
	Substituting this into Eq.~(\ref{Sparsity-1}), we obtain
	\begin{equation}
		\eta =(1-\ell)^{-1}\left(\frac{1}{1-\ell}- \frac{Q^2}{(1-\ell)^2\,r_h^2}+ \frac{\lambda}{r_h}\right)^{-2} \eta_{\rm Sch.},
		\label{sparsity-3}
	\end{equation}
	where $\eta_{\rm Sch.} = 64\pi^3/27$ is the Schwarzschild sparsity.
	
	\begin{table}[h!]
		\centering
		\begin{tabular}{|c|c|c|c|c|c|}
			\hline
			$\ell \backslash \lambda$ & -0.5 & -0.4 & -0.3 & -0.2 & -0.1 \\
			\hline
			0.05 & 1.57753 & 1.46655 & 1.3666  & 1.27591 & 1.1944  \\
			0.10 & 1.54701 & 1.43392 & 1.33391 & 1.24503 & 1.1677  \\
			0.15 & 1.53032 & 1.41285 & 1.31143 & 1.22376 & 1.15103 \\
			0.20 & 1.53703 & 1.41104 & 1.30578 & 1.21835 & 1.15118 \\
			\hline
		\end{tabular}
		\caption{Numerical values of sparsity ratio $\eta/\eta_{\rm sch}$ for $Q=0.5$, for different $\ell$ and $\lambda$.}
		\label{tab:sparsity-1}
	\end{table}
	
	\begin{table}[h!]
		\centering
		\begin{tabular}{|c|c|c|c|c|c|}
			\hline
			$\ell \backslash \lambda$ & 0.1 & 0.2 & 0.3 & 0.4 & 0.5 \\
			\hline
			-0.20 & 1.18474 & 1.04988 & 0.92124 & 0.80847 & 0.713874 \\
			-0.15 & 1.15337 & 1.0237  & 0.898958 & 0.789223 & 0.697076 \\
			-0.10 & 1.12482 & 1.00014 & 0.878803 & 0.771576 & 0.681426 \\
			-0.05 & 1.10012 & 0.980199 & 0.861599 & 0.756153 & 0.667377 \\
			\hline
		\end{tabular}
		\caption{Numerical values of sparsity ratio $\eta/\eta_{\rm sch}$ for $Q=0.5$, for different $\ell$ and $\lambda$.}
		\label{tab:sparsity-2}
	\end{table}
	
	From the above expression, we have seen that the sparsity of the radiation emitted by the black hole depends on the Lorentz-violating parameter $\ell$, the PFDM parameter $\lambda$, and the charge $Q$. In Tables \ref{tab:sparsity-1}-\ref{tab:sparsity-2}, we presented the numerical values of the sparsity parameters for different values of $\ell$ and $\lambda$.
	
	\begin{figure}[ht!]
		\centering
		\includegraphics[width=0.95\linewidth]{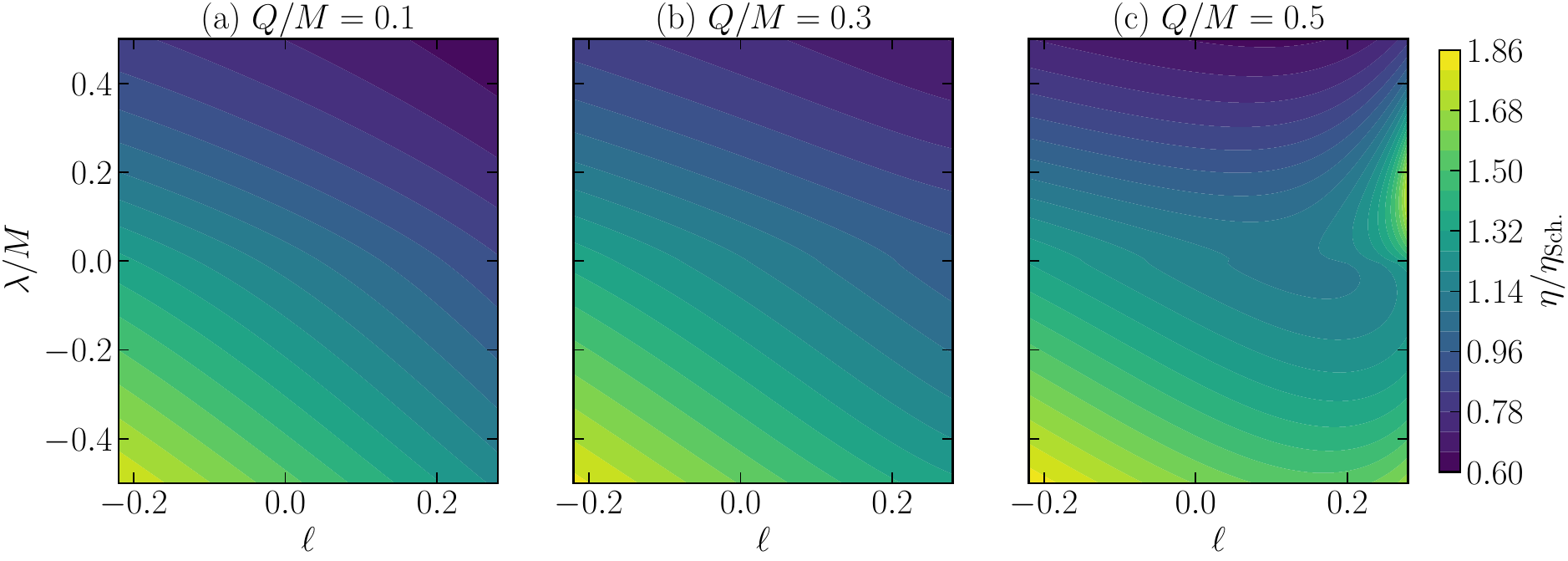}
		\caption{Contour maps of the sparsity ratio $\eta/\eta_{\rm Sch.}$ in the $(\ell,\lambda/M)$ parameter space for $Q/M=0.1$, $0.3$, and $0.5$.}
		\label{fig:sparsity_map}
	\end{figure}
	
	Figure~\ref{fig:sparsity_map} extends the tabulated information by showing how the sparsity ratio varies continuously over the parameter plane. For small and intermediate values of the charge, the contour structure is smooth and largely monotonic: the emission becomes more sparse in the region of negative $\lambda$ and negative $\ell$, whereas it is less sparse toward positive values of these parameters. In the $Q/M=0.5$ panel, however, the contour pattern develops a more localized feature in the positive-$\ell$, mildly positive-$\lambda$ sector, indicating that larger charges can produce a more intricate interplay between the horizon scale and the effective temperature entering the sparsity parameter. This figure is particularly valuable because it shows that the quantum-emission phenomenology is not simply inherited from the classical thermodynamic sector in a trivial way; rather, it exhibits a structured dependence on the Lorentz-violating background, the dark-matter environment, and the electric charge.
	
	The numerical values indicate that the emission can become either sparser or denser than in the Schwarzschild case, depending on the sign and magnitude of the PFDM parameter, as well as on the value of $\ell$. When $\eta/\eta_{\rm Sch.}>1$, the Hawking quanta are emitted more intermittently than in the Schwarzschild background, whereas $\eta/\eta_{\rm Sch.}<1$ signals a comparatively less sparse emission channel. This result is physically relevant because it shows that Lorentz-violating effects and the surrounding dark-matter environment modify not only the equilibrium thermodynamic quantities, but also the temporal character of the quantum emission process.
	
	\section{Conclusion}\label{sec:5}
	
	In this work, we investigated a charged black hole solution in Lorentz-violating KR-gravity surrounded by perfect fluid dark matter, with the goal of understanding how the parameters $Q$, $\ell$, and $\lambda$ modify the optical, dynamical, and thermodynamic properties of the spacetime. Starting from the null geodesic sector, we derived the effective potential for photons and analyzed the associated photon sphere, shadow radius, effective radial force, and photon trajectories. Our results show that the electric charge and the PFDM parameter tend to reduce both the photon-sphere radius and the shadow size, whereas decreasing the Lorentz-violating parameter enlarges these observables. The trajectory plots further illustrate that both the KR background and the dark-matter environment leave visible imprints on the bending and capture of light near the black hole.
	
	We then examined the motion of neutral test particles through the Keplerian and radial epicyclic frequencies, emphasizing the role of the ISCO as the boundary of stable circular motion. The behavior of these frequencies demonstrates that the strong-field orbital structure is highly sensitive to the combined action of $Q$, $\ell$, and $\lambda$. Building on this result, we constructed several twin-peak QPO relations within the RP, ER, and WD models and showed that the corresponding frequency--frequency tracks are systematically deformed by the modified geometry. The subsequent MCMC analysis using observational data from XTE J1550--564, GRO J1655--40, M82 X-1, and Sgr A$^\ast$ indicates that the model can reproduce the measured QPO pairs in physically admissible regions of the parameter space and reveals source-dependent correlations among the fitted parameters.
	
	On the thermodynamic side, we derived the horizon mass relation, the modified Hawking temperature appropriate for the non-asymptotically flat background, the entropy, the specific heat, and the Gibbs free energy. The entropy differs from the standard Bekenstein--Hawking area law by a factor controlled by the Lorentz-violating parameter, while the divergence of the specific heat identifies the critical radius that separates locally stable and unstable thermodynamic branches. The Gibbs free energy provides complementary information about global stability in the canonical ensemble and shows that the KR and PFDM sectors can substantially alter the preferred thermodynamic branch of the solution.
	
	Finally, we studied the sparsity of Hawking radiation and found that the temporal character of the emission is also modified by the same parameters that govern the geometry. Depending on the region of the parameter space, the radiation may become more intermittent or less intermittent than in the Schwarzschild case. Altogether, these results indicate that charged KR black holes immersed in PFDM constitute a rich phenomenological framework in which Lorentz-violating physics, environmental matter effects, and black-hole observables can be investigated simultaneously. In the next step, it will be natural to complement the present analytical study with additional graphical diagnostics and, eventually, with extensions including plasma effects, rotating generalizations, and further observational consistency tests.
	
	\section*{Acknowledgments}
	
	F.A. acknowledges the Inter University Centre for Astronomy and Astrophysics (IUCAA), Pune, India for granting visiting associateship. M.F. acknowledges financial support from Agencia Nacional de Investigación y Desarrollo (ANID) through the FONDECYT postdoctoral Grant No. 3260029. E. O. Silva acknowledges the support from Conselho Nacional de Desenvolvimento Cient\'{i}fico e Tecnol\'{o}gico (CNPq) (grants 306308/2022-3), Funda\c c\~ao de Amparo \`{a} Pesquisa e ao Desenvolvimento Cient\'{i}fico e Tecnol\'{o}gico do Maranh\~ao (FAPEMA) (grants UNIVERSAL-06395/22), and Coordena\c c\~ao de Aperfei\c coamento de Pessoal de N\'{i}vel Superior (CAPES) - Brazil (Code 001).

\end{document}